\documentclass{aa}

\usepackage{graphicx}
\usepackage[english]{babel}
\usepackage{natbib}
\usepackage{subfig}
\usepackage{amssymb}
\usepackage{wrapfig}
\usepackage{pdflscape}
\usepackage{lscape}
\usepackage{longtable}
\usepackage{appendix}
\usepackage{textcomp}
\usepackage{multirow}
\bibpunct{(}{)}{;}{a}{}{,}
\usepackage{array}
\usepackage{txfonts}

\def\setsymbol#1#2{\expandafter\def\csname #1\endcsname{#2}}
\def\getsymbol#1{\csname #1\endcsname}

\def\Planck{\textit{Planck}}

\newbox\tablebox    \newdimen\tablewidth
\def\leaderfil{\leaders\hbox to 5pt{\hss.\hss}\hfil}
\def\tablenote#1 #2\par{\begingroup \parindent=0.8em
    \abovedisplayshortskip=0pt\belowdisplayshortskip=0pt
    \noindent
    $$\hss\vbox{\hsize\tablewidth \hangindent=\parindent \hangafter=1 \noindent
    \hbox to \parindent{$^#1$\hss}\strut#2\strut\par}\hss$$
    \endgroup}

\def\L2{\ifmmode L_2\else $L_2$\fi}

\def\DeltaT{\ifmmode \Delta T\else $\Delta T$\fi}
\def\deltat{\ifmmode \Delta t\else $\Delta t$\fi}
\def\fknee{\ifmmode f_{\rm knee}\else $f_{\rm knee}$\fi}
\def\Fmax{\ifmmode F_{\rm max}\else $F_{\rm max}$\fi}
\def\solar{\ifmmode{\rm M}_{\mathord\odot}\else${\rm M}_{\mathord\odot}$\fi}
\def\Msolar{\ifmmode{\rm M}_{\mathord\odot}\else${\rm M}_{\mathord\odot}$\fi}
\def\Lsolar{\ifmmode{\rm L}_{\mathord\odot}\else${\rm L}_{\mathord\odot}$\fi}
\def\inv{\ifmmode^{-1}\else$^{-1}$\fi}
\def\mo{\ifmmode^{-1}\else$^{-1}$\fi}
\def\sup#1{\ifmmode ^{\rm #1}\else $^{\rm #1}$\fi}
\def\expo#1{\ifmmode \times 10^{#1}\else $\times 10^{#1}$\fi}
\def\,{\thinspace}
\def\lsim{\mathrel{\raise .4ex\hbox{\rlap{$<$}\lower 1.2ex\hbox{$\sim$}}}}
\def\gsim{\mathrel{\raise .4ex\hbox{\rlap{$>$}\lower 1.2ex\hbox{$\sim$}}}}

\def\simprop{\mathrel{\raise .4ex\hbox{\rlap{$\propto$}\lower 1.2ex\hbox{$\sim$}}}}
\def\deg{\ifmmode^\circ\else$^\circ$\fi}
\def\pdeg{\ifmmode $\setbox0=\hbox{$^{\circ}$}\rlap{\hskip.11\wd0 .}$^{\circ}
          \else \setbox0=\hbox{$^{\circ}$}\rlap{\hskip.11\wd0 .}$^{\circ}$\fi}
\def\arcs{\ifmmode {^{\scriptstyle\prime\prime}}
          \else $^{\scriptstyle\prime\prime}$\fi}
\def\arcm{\ifmmode {^{\scriptstyle\prime}}
          \else $^{\scriptstyle\prime}$\fi}
\newdimen\sa  \newdimen\sb
\def\parcs{\sa=.07em \sb=.03em
     \ifmmode \hbox{\rlap{.}}^{\scriptstyle\prime\kern -\sb\prime}\hbox{\kern -\sa}
     \else \rlap{.}$^{\scriptstyle\prime\kern -\sb\prime}$\kern -\sa\fi}
\def\parcm{\sa=.08em \sb=.03em
     \ifmmode \hbox{\rlap{.}\kern\sa}^{\scriptstyle\prime}\hbox{\kern-\sb}
     \else \rlap{.}\kern\sa$^{\scriptstyle\prime}$\kern-\sb\fi}
\def\ra[#1 #2 #3.#4]{#1\sup{h}#2\sup{m}#3\sup{s}\llap.#4}
\def\dec[#1 #2 #3.#4]{#1\deg#2\arcm#3\arcs\llap.#4}
\def\deco[#1 #2 #3]{#1\deg#2\arcm#3\arcs}
\def\rra[#1 #2]{#1\sup{h}#2\sup{m}}

\def\dots{\relax\ifmmode \ldots\else $\ldots$\fi}
\def\WHzsr{\ifmmode $W\,Hz\mo\,sr\mo$\else W\,Hz\mo\,sr\mo\fi}
\def\mHz{\ifmmode $\,mHz$\else \,mHz\fi}
\def\GHz{\ifmmode $\,GHz$\else \,GHz\fi}
\def\mKs{\ifmmode $\,mK\,s$^{1/2}\else \,mK\,s$^{1/2}$\fi}
\def\muKs{\ifmmode \,\mu$K\,s$^{1/2}\else \,$\mu$K\,s$^{1/2}$\fi}
\def\muKRJs{\ifmmode \,\mu$K$_{\rm RJ}$\,s$^{1/2}\else \,$\mu$K$_{\rm RJ}$\,s$^{1/2}$\fi}
\def\muKHz{\ifmmode \,\mu$K\,Hz$^{-1/2}\else \,$\mu$K\,Hz$^{-1/2}$\fi}
\def\MJysr{\ifmmode \,$MJy\,sr\mo$\else \,MJy\,sr\mo\fi}
\def\MJysrmK{\ifmmode \,$MJy\,sr\mo$\,mK$_{\rm CMB}\mo\else \,MJy\,sr\mo\,mK$_{\rm CMB}\mo$\fi}
\def\microns{\ifmmode \,\mu$m$\else \,$\mu$m\fi}

\def\muK{\ifmmode \,\mu$K$\else \,$\mu$\hbox{K}\fi}
\def\microK{\ifmmode \,\mu$K$\else \,$\mu$\hbox{K}\fi}
\def\muW{\ifmmode \,\mu$W$\else \,$\mu$\hbox{W}\fi}
\def\kms{\ifmmode $\,km\,s$^{-1}\else \,km\,s$^{-1}$\fi}
\def\kmsMpc{\ifmmode $\,\kms\,Mpc\mo$\else \,\kms\,Mpc\mo\fi}
\providecommand{\sorthelp}[1]{}

\begin{document} 

\title{37 GHz observations of narrow-line Seyfert 1 galaxies \thanks{Table~\ref{tab:data1h0323} is only available in electronic form at the CDS via anonymous ftp to cdsarc.u-strasbg.fr (130.79.128.5) or via http://cdsweb.u-strasbg.fr/cgi-bin/qcat?J/A+A/.}}


   \author{A. L\"{a}hteenm\"{a}ki\inst{1}\fnmsep\inst{2}, E. J\"{a}rvel\"{a}\inst{1}\fnmsep\inst{2}, T. Hovatta\inst{1}\fnmsep\inst{2},  M. Tornikoski\inst{1}, 
D. L. Harrison\inst{3}\fnmsep\inst{4}, M. L\'{o}pez-Caniego\inst{5}\fnmsep, W. Max-Moerbeck\inst{6}, M. Mingaliev\inst{7}\fnmsep\inst{8}, 
T. J. Pearson\inst{9}, V. Ramakrishnan\inst{1}, A. C. S. Readhead\inst{9}, R. A. Reeves\inst{10}, J. L. Richards\inst{9}, Y. Sotnikova\inst{8}, J. Tammi\inst{1}}

   \institute{Aalto University Mets\"{a}hovi Radio Observatory, Mets\"{a}hovintie 114, Kylm\"{a}l\"{a}, FI-02540, Finland
         \and
             Aalto University Department of Radio Science and Engineering, P.O. BOX 13000, FI-00076 AALTO, Finland
\and
Institute of Astronomy, University of Cambridge, Madingley Road, Cambridge CB3 0HA, U.K.
\and
Kavli Institute for Cosmology Cambridge, Madingley Road, Cambridge, CB3 0HA, U.K.
                \and
European Space Agency, ESAC, Planck Science Office, Camino bajo del Castillo, s/n, Urbanizaci\'{o}n Villafranca del Castillo, Villanueva de la Ca\~{n}ada, 
Madrid, Spain
\and
Max-Planck-Institut f\"{u}r Radioastronomie, Auf dem H\"{u}gel 69, 53121, Bonn, Germany
\and
Special Astrophysical Observatory, Russian Academy of Sciences, Nizhnij Arkhyz, 369167 Russia 
               \and
               Kazan Federal University, 18 Kremlyovskaya St. Kazan, 420008 Russia
\and
Cahill Center for Astronomy and Astrophysics, California Institute of Technology, Pasadena CA,  91125, USA
\and
CePIA, Departamento de Astronom\'{i}a, Universidad de Concepci\'{o}n, Casilla 160-C, Concepci\'{o}n, Chile
                  }

   \date{Received ; accepted }


   \abstract {Observations at 37~GHz, performed at Mets\"{a}hovi Radio Observatory, are presented for a sample of 78 radio-loud and radio-quiet narrow-line Seyfert 1 (NLS1) galaxies, together with additional lower
     and higher frequency radio data from RATAN-600, Owens Valley Radio Observatory, and the \textit{Planck} satellite. Most of the data have been gathered between February
     2012 and April 2015 but for some sources even longer lightcurves exist.
The detection rate at 37~GHz is around 19\%, comparable to other populations of active galactic nuclei presumed 
to be faint at radio frequencies, such as BL Lac objects. Variability and spectral indices are determined for sources with enough detections. Based on the radio 
data, many NLS1 galaxies show a blazar-like radio spectra exhibiting significant variability. The spectra at a given time are often inverted or convex. The source of the high-frequency radio emission in NLS1 galaxies, detected at 37~GHz, is most probably a relativistic jet rather than star formation.
Jets in NLS1 galaxies are therefore expected to be a much more common phenomenon than earlier assumed.}


\titlerunning{37 GHz observations of Narrow-line Seyfert 1 galaxies}
\authorrunning{A. L\"{a}hteenm\"{a}ki et al.} 

   \maketitle


\section{Introduction}
\label{intro}

Narrow-line Seyfert 1 (NLS1) galaxies are a distinctive class of gamma-ray emitting active galactic nuclei (AGN). They were first described in 1985 by \citet{1985osterbrock1} and are
characterized by the properties of their optical spectra; in NLS1 galaxies also permitted emission lines are narrow 
\citep[{by definition FWHM(H$\beta$) $< 2000$ km s$^{-1}$},][]{1989goodrich1} and [O~III]/H$\beta <$ 3, 
with exceptions allowed if there are strong [Fe~VIII] and [Fe~X] present \citep{1985osterbrock1}. Many NLS1 galaxies show strong Fe~II emission \citep{1985osterbrock1}. 

Some NLS1 sources show rapid, high-amplitude variability at X-rays at very short time scales \citep[{e.g.},][]{2000boller1, 2013fabian1}. They show a strong soft X-ray excess, and have 
more diverse soft X-ray (0.1--2.5 keV) photon indices ($\Gamma \approx$ 1--5) than those of Type 1 Seyfert galaxies ($\Gamma \approx$ 2) \citep{1996boller1}. NLS1 sources have on average 
a steep X-ray spectrum that steepens the narrower the H$\beta$ is \citep{1992puchnarewicz1, 1996boller1}.

Most NLS1 sources, particularly the radio-quiet variety, are hosted by late-type galaxies, but a few of them are also found in peculiar, interacting or E/S0 systems \citep{2007ohta1, 2014leontavares1}. 
Black hole masses in NLS1 galaxies are low or intermediate \citep[{$M_{\text{BH}} < 10^{8}$ $M_{\sun}$},][]{2000peterson1} and accrete at high rates 
\citep[{0.1--1 Eddington rate or even above},][]{1992boroson1}. Some studies suggest that they tend to lie below the normal stellar velocity dispersion of the bulge $M_{\text{BH}}$ -- $\sigma_{\ast}$ and luminosity of the bulge $M_{\text{BH}}$ -- $L_{\text{bulge}}$ relations \citep{2001mathur1}, whereas other studies claim that this is not the case \citep{2015woo1}.
Based on these properties NLS1 galaxies are believed to be rather young active galactic nuclei in the early stages of their evolution \citep{2001mathur1}.

Only about 7\% of NLS1 galaxies are radio-loud ($RL=S_{\text{radio}} / S_{\text{optical}}>10$) , and 2\% -- 3\% very radio loud ($RL>100$) \citep{2006komossa1}. These generally appear to have a very compact radio morphology without extended radio emission, but evidence of kiloparsec-scale structures has been found in several radio-loud NLS1 galaxies \citep{2010gliozzi1, 2012doi1, 2015richards1, 2015gu1}. Some radio-quiet NLS1 
galaxies also show parsec-scale radio structures associated with non-thermal
processes and indicating the presence of a jet-producing central engine \citep{2013doi1,2015doi1,2015richards,2016lister}. Sub-luminal and superluminal speeds
have been measured in some NLS1 galaxies, suggesting Lorentz factors and viewing angles similar to BL Lac objects (BLOs) and flat-spectrum radio quasars \citep{2016lister}.
It is noteworthy that most of the sources with radio structures harbour, on average, more massive black holes than those without radio structures, i.e. $M_{\text{BH}} > 10^7 M_{\sun}$ \citep{2012doi1,2015jarvela1,2015foschini1}.
Extended radio structures were indeed expected after Large Area Telescope (LAT) onboard \textit{Fermi Gamma-ray Space Telescope} detected 
gamma-ray emission in NLS1 galaxies \citep[{e.g.},][]{2009abdo2}, thus confirming the presence of powerful relativistic jets.

NLS1 galaxies defy our current knowledge of AGN and relativistic jet systems. They are intrinsically different compared
to other gamma-ray emitting AGN, i.e., blazars and radio galaxies, even though their observational properties often resemble them; their host galaxies, black hole masses, accretion rates and radio morphologies are distinct.
Moreover, it is unclear at the moment whether they form a homogeneous class, or, for example, if radio-quiet, radio-loud, and radio-silent NLS1 sources have
disparate parent populations. Evidently all NLS1 sources are not intrinsically similar \citep{2014caccianiga1, 2015berton1}.

Radio observations are crucial for understanding these sources because the origin of the radio emission could be the relativistic jets or, alternatively,
 star formation processes in the galaxy itself \citep[{see, e.g.},][]{2015caccianiga1}. However, due to their faintness in low radio frequency surveys,
such as the VLA FIRST 1.4~GHz survey, NLS1 galaxies have been scarcely observed at higher radio frequencies; most observing programmes concentrate only
on the brightest individuals \citep[{see, e.g.},][for a comprehensive variability and radio spectra analysis of four radio-loud and gamma-ray-detected NLS1 galaxies]{angelakis2015}. 
To study the properties of NLS1 galaxies as a class, we need observations of larger and more diverse samples.

In this paper we publish the first results of an extensive observing programme of NLS1 galaxies at 37~GHz, launched at Mets\"{a}hovi Radio Observatory, and 
combine them with additional quasi-simultaneous radio data both at lower and higher frequencies from Owens Valley Radio Observatory (OVRO), RATAN-600 and the 
\textit{Planck} satellite, to examine their radio spectra. The so-called ``Mets\"{a}hovi NLS1 pilot survey'' consists of at least three-epoch 
observations of 78 NLS1 sources at 37~GHz. Basic statistical and variability analyses, 
flux density curves, and radio spectra are presented. One of our goals is to also define a set of sources for 
long-term monitoring and multifrequency studies.

We define radio-loudness (\textit{RL}) as $\textit{RL} = S_{1.4~\text{GHz}} / S_{400~\text{nm}}$, where the flux densities have been K-corrected
\citep[{K-correction was done as suggested in}][]{2011foschini1}.
In this paper a source is called radio-quiet if \textit{RL} $< 10$, radio-loud if \textit{RL} $> 10$, and very radio-loud if \textit{RL} $> 100$. A source is called radio-silent
if no radio emission has been observed from it. Throughout the paper we assume the convention $S_\nu \propto \nu^{\alpha}$, where $S_\nu$ is the flux density and $\alpha$ is the spectral index, and 
the cosmology with H$_{0}$ = 73 km s$^{-1}$ Mpc$^{-1}$, $\Omega_{\text{matter}}$ = 0.27 and $\Omega_{\text{vacuum}}$ = 0.73 \citep{2007spergel1}.


\section{Sample selection}
\label{sec:sample}

The Mets\"{a}hovi NLS1 observing programme at 22 and 37~GHz currently consists of 160 NLS1 galaxies, divided into several samples based on different selection criteria.
The sources are diverse, but all of them have properties that make them good candidates for higher frequency radio observations. Our objective was to characterise the radio properties, 
for example, detection rate and variability, of a NLS1 sample of statistically significant size and with varying properties, and to compile a sample of detected sources for future 
multiwavelength observations. In this paper our pilot survey containing the first two NLS1 samples (78 sources) is presented. These two samples contain the radio-loudest and most interesting targets that were introduced to the observing programme first, to test its feasiblity. The detection limit of the Mets\"{a}hovi radio telescope is relatively high (see Sect.~\ref{sec:mh}), and we wanted to first check how many of the sources we could detect. Our samples do not form  a complete sample, nor did we aim at one. Therefore the source selection was driven by the detection probability and how interesting the sources in general are. Observations of sample 1 were started in February 2012, and the second sample was added in November 2013. Other samples were added to the programme after the feasibility was confirmed with the first two samples, and their observations are on-going.

\subsection{Sample 1}

The Mets\"{a}hovi NLS1 sample 1 consists of 45 sources. They were mostly selected from \citet{2011foschini1} with the addition of two sources from \citet{2006komossa1}. 
The sample includes many sources that have recently also appeared in \citet{2015foschini1} and \citet{2015berton1}. Because the main purpose at the
start of the programme was to chart how many of these sources we could detect, some of the radio-loudest known NLS1 sources are included in this sample as well as most of those detected at gamma-rays.
The list of sources, their basic data and statistics of observations at Mets\"{a}hovi are shown in Table~\ref{tab:data1-stat}. Column 1 gives the name of the source. Cols. 2, 3, and 4 list
the redshift and coordinates (in J2000.0 epoch) of the sources. The number of detections $N_{\text{det}}$ and the total number of observations $N_{\text{obs}}$, the detection rate, 
and maximum flux at 37~GHz are given in Cols. 5, 6, and 7, respectively. 1.4~GHz flux densities from the VLA FIRST and NRAO VLA Sky Survey \citep[{NVSS};][]{condon1998} surveys are 
listed in column 8. Radio-loudness and black hole masses are given in Cols.~9 and 10.

\begin{table*}[ht]
\caption[]{Mets\"{a}hovi NLS1 sample 1: basic data and statistics of the 37~GHz observations.}
\centering
\begin{tabular}{l l l l l l l l l l}
\hline\hline
Source                     &  $z$  & RA            & Dec              & $N_{\textrm{Det}}$/$N_{\textrm{Obs}}$ & Det  & $S_{\text{37GHz, max}}$ & $S_{\text{1.4GHz}}$      & \textit{RL}  & log $M_{\text{BH}}$ \\
                           &       & (hh mm ss.ss) & (dd mm ss.ss)    &                                       & (\%) & (Jy)    &  (mJy)         &         & ($M_{\sun}$)   \\ \hline
1H 0323$+$342              & 0.061 & 03 24 41.16   & $+$34 10 45.86   &  14/22                                & 64  & 1.05     &  614.3            & 318\tablefootmark{1}    & 7.6        \\
FBQS J0713$+$3820          & 0.123 & 07 13 40.28   & $+$38 20 39.91   &  0/7                                  & 0   & --     &  10.3             & 20\tablefootmark{1}     & 8.3        \\
FBQS J0744$+$5149          & 0.460 & 07 44 02.28   & $+$51 49 17.50   &  0/3                                  & 0   & --     &  11.9             & 59\tablefootmark{1}     & 8.4        \\
SDSS J075800.05$+$392029.0 & 0.096 & 07 58 00.05   & $+$39 20 29.09   &  0/4                                  & 0   & --     &  11.6             & 90\tablefootmark{1}     & \dots        \\
SDSS J080409.23$+$385348.8 & 0.212 & 08 04 09.24   & $+$38 53 48.82   &  0/5                                  & 0   & --     &  2.7              & 9    & 7.8        \\
RGB J0806$+$728            & 0.098 & 08 06 38.97   & $+$72 48 20.60   &  0/6                                  & 0   & --     &  50.1             & 41\tablefootmark{1}     & 6.7\tablefootmark{2}        \\
SDSS J081432.11$+$560956.6 & 0.510 & 08 14 32.12   & $+$56 09 56.68   &  0/5                                  & 0   & --     &  69.2             & 368   & 8.5        \\
SDSS J084957.97$+$510829.0 & 0.585 & 08 49 57.98   & $+$51 08 29.00   & 67/138                                & 49  & 1.18     &  344.1            & 4162  & 7.5       \\
SDSS J085001.17$+$462600.5 & 0.524 & 08 50 01.17   & $+$46 26 00.54   & 0/5                                   & 0   & --     &  16.4             & 296   & 7.2\tablefootmark{3}       \\
SDSS J090227.16$+$044309.5 & 0.533 & 09 02 27.16   & $+$04 43 09.60   & 0/7                                   & 0   & --     &  156.6            & 1756  & 7.7        \\
SDSS J093703.02$+$361537.1 & 0.180 & 09 37 03.03   & $+$36 15 37.18   & 0/5                                   & 0   & --     &  3.6              & 12\tablefootmark{1}     & 7.3        \\
IRAS 09426$+$1929          & 0.149 & 09 45 29.30   & $+$19 15 45.00   & 0/10                                  & 0   & --     &  17.2             & \dots   & 7.9        \\
SDSS J094857.31$+$002225.4 & 0.585 & 09 48 57.30   & $+$00 22 26.00   & 95/132                                & 72  & 1.13     &  107.5            & 780   & 7.9        \\
SDSS J095317.09$+$283601.5 & 0.659 & 09 53 17.10   & $+$28 36 01.62   & 0/4                                   & 0   & --     &  44.6             & 613   & 8.2        \\
SDSS J103123.73$+$423439.3 & 0.377 & 10 31 23.73   & $+$42 34 39.31   & 1/9                                   & 11  & 0.30     &  16.6             & 239   & 8.4        \\
SDSS J103727.45$+$003635.6 & 0.596 & 10 37 27.45   & $+$00 36 35.60   & 0/7                                   & 0   & --     &  27.2             & 546   & 7.3        \\
SDSS J103859.58$+$422742.2 & 0.221 & 10 38 59.58   & $+$42 27 42.26   & 0/4                                   & 0   & --     &  2.8              & 10\tablefootmark{1}     & 7.8        \\
SDSS J104732.68$+$472532.0 & 0.799 & 10 47 32.66   & $+$47 25 32.11   & 0/3                                   & 0   & --     &  734.0            & 10346 & 7.9        \\
SDSS J104816.58$+$222239.0 & 0.330 & 10 48 16.58   & $+$22 22 39.00   & 0/5                                   & 0   & --     &  1.2              & 10\tablefootmark{1}     & 7.8        \\
SDSS J110223.38$+$223920.7 & 0.453 & 11 02 23.39   & $+$22 39 20.72   & 0/7                                   & 0   & --     &  2.0              & 32\tablefootmark{1}     & 8.1        \\
SDSS J111005.03$+$365336.3 & 0.630 & 11 10 05.04   & $+$36 53 36.13   & 0/4                                   & 0   & --     &  18.6             & 1251  & 7.4        \\
SDSS J111438.89$+$324133.4 & 0.189 & 11 14 38.91   & $+$32 41 33.34   & 0/6                                   & 0   & --     &  110.4            & 1986\tablefootmark{1}   & \dots         \\
SDSS J113824.54$+$365327.1 & 0.357 & 11 38 24.54   & $+$36 53 26.99   & 0/6                                   & 0   & --     &  12.5             & 272   & 7.4        \\
SDSS J114654.28$+$323652.3 & 0.466 & 11 46 54.29   & $+$32 36 52.35   & 0/7                                   & 0   & --     &  14.7             & 140   & 8.0        \\
SDSS J115917.32$+$283814.5 & 0.210 & 11 59 17.32   & $+$28 38 14.56   & 0/5                                   & 0   & --     &  2.2              & 20\tablefootmark{1}     & 7.1        \\
SDSS J122749.14$+$321458.9 & 0.137 & 12 27 49.15   & $+$32 14 59.04   & 0/4                                   & 0   & --     &  6.5              & 91\tablefootmark{1}     & 6.7        \\
SDSS J123852.12$+$394227.8 & 0.623 & 12 38 52.15   & $+$39 42 27.59   & 0/5                                   & 0   & --     &  10.4             & 252   & 7.2       \\
SDSS J124634.65$+$023809.0 & 0.363 & 12 46 34.65   & $+$02 38 09.02   & 0/5                                   & 0   & --     &  37.0             & 1     & 7.8        \\
SDSS J130522.75$+$511640.3 & 0.788 & 13 05 22.75   & $+$51 16 40.26   & 0/4                                   & 0   & --     &  86.9             & 250   & 8.2\tablefootmark{3}        \\
SDSS J134634.97$+$312133.7 & 0.246 & 13 46 34.97   & $+$31 21 33.79   & 0/4                                   & 0   & --     &  1.2              & 11\tablefootmark{1}     & 7.1        \\
SDSS J135845.38$+$265808.5 & 0.331 & 13 58 45.38   & $+$26 58 08.50   & 0/4                                   & 0   & --     &  1.8              & 11\tablefootmark{1}     & 7.8        \\
SDSS J142114.05$+$282452.8 & 0.549 & 14 21 14.07   & $+$28 24 52.18   & 0/7                                   & 0   & --     &  46.8             & 204\tablefootmark{1}    & 8.0        \\
SDSS J143509.49$+$313147.8 & 0.502 & 14 35 09.52   & $+$31 31 48.29   & 0/10                                  & 0   & --     &  44.7             & 855   & 7.5\tablefootmark{2}       \\
SDSS J144318.56$+$472556.7 & 0.705 & 14 43 18.57   & $+$47 25 56.30   & 0/7                                   & 0   & --     &  171.1            & 1017  & 7.4\tablefootmark{2}       \\
SDSS J150506.47$+$032630.8 & 0.408 & 15 05 06.46   & $+$03 26 30.30   & 18/24                                 & 75  & 0.67     &  365.4            & 2924  & 7.3        \\
SDSS J154817.92$+$351128.0 & 0.479 & 15 48 17.92   & $+$35 11 28.10   & 1/13                                  & 8   & 0.33     &  140.9            & 677   & 7.9        \\
SDSS J161259.83$+$421940.3 & 0.233 & 16 12 59.84   & $+$42 19 40.32   & 5/16                                  & 31  & 0.46     &  3.4              & 24\tablefootmark{1}     & 6.9        \\
SDSS J162901.30$+$400759.9 & 0.272 & 16 29 01.31   & $+$40 07 59.91   & 1/10                                  & 10  & 0.35     &  12.0             & 45    & 7.5        \\
SDSS J163323.58$+$471858.9 & 0.116 & 16 33 23.58   & $+$47 18 58.93   & 0/8                                   & 0   & --     &  62.6             & 144   & 6.9        \\
SDSS J163401.94$+$480940.2 & 0.495 & 16 34 01.94   & $+$48 09 40.22   & 0/4                                   & 0   & --     &  7.5              & 169   & 7.8        \\
SDSS J164442.53$+$261913.2 & 0.145 & 16 44 42.53   & $+$26 19 13.29   & 4/15                                  & 27  & 0.38     &  87.5             & 320   & 7.1        \\
SDSS J170330.38$+$454047.1 & 0.060 & 17 03 30.38   & $+$45 40 47.17   & 0/10                                  & 0   & --     &  121.8            & 102\tablefootmark{1}    & 6.5\tablefootmark{2}      \\
FBQS J1713$+$3523          & 0.083 & 17 13 04.46   & $+$35 23 33.65   & 0/8                                   & 0   & --     &  12.0             & 10\tablefootmark{1}     & 7.1\tablefootmark{2}     \\
SDSS J172206.03$+$565451.6 & 0.426 & 17 22 06.03   & $+$56 54 51.63   & 0/4                                   & 0   & --     &  39.8             & 285   & 7.2\tablefootmark{3}        \\
RX J2314.9$+$2243          & 0.169 & 23 14 55.89   & $+$22 43 22.69   & 0/7                                   & 0   & --     &  18.7             & \dots   & 7.9\tablefootmark{2}       \\ \hline

\end{tabular}
\tablefoot{\textit{RL} and $M_{\text{BH}}$ values are taken from \citet{2015jarvela1} and \citet{2015foschini1}, respectively, unless otherwise indicated.}
\tablebib{
(1)~\citet{2011foschini1}; (2) \citet{2015berton1}; (3) \citet{2015jarvela1}.
}
\label{tab:data1-stat}
\end{table*}

\subsection{Sample 2}

Sample 2 includes additional 33 sources. These sources were mostly selected from \citet{2015jarvela1} and \citet{2011foschini1}. Sources selected from \citet{2015jarvela1} 
have radio-loudness larger than 100. One source was added from \citet{2006komossa1} and one from \citet{2006whalen}. The list of sources, their basic data, and 
statistics of observations at Mets\"{a}hovi are shown in Table~\ref{tab:data2-stat}. The columns are as in Table~\ref{tab:data1-stat}. 

\begin{table*}[ht]
\caption[]{Mets\"{a}hovi NLS1 sample 2: basic data and statistics of the 37~GHz observations.}
\centering
\begin{tabular}{l l l l l l l l l l }
\hline\hline
Source                      & $z$   & RA            & Dec             & $N_{\textrm{Det}}$/$N_{\textrm{Obs}}$ & Det & $S_{\text{37GHz, max}}$ & $S_{\text{1.4GHz}}$ & \textit{RL}        & log $M_{\text{BH}}$ \\
                            &       & (hh mm ss.ss) & (dd mm ss.ss)   &                                       & (\%)&    (Jy)      &  (mJy)              &                    & ($M_{\sun}$)   \\ \hline
FBQS J0100$-$0200           & 0.227 & 01 00 32.22   & $-$02 00 46.00  &  0/4                                  & 0   & --           &  6.4                & 8\tablefootmark{1} & 7.6\tablefootmark{2}       \\
SDSS J024225.87$-$004142.6  & 0.383 & 02 42 25.87   & $-$00 41 42.71  &  0/14                                 & 0   & --   &  29.0        & 1224 & 6.9       \\
FBQS J0706$+$3901           & 0.086 & 07 06 25.12   & $+$39 01 51.55  &  0/4                                  & 0   & --   &  5.6         & 7\tablefootmark{1}     & 7.3\tablefootmark{2}        \\
SDSS J080535.17$+$302201.7  & 0.552 & 08 05 35.18   & $+$30 22 01.65  &  1/5\tablefootmark{*}                                  & 20  & 0.26 &  60.8        & 638  & 7.1        \\
SDSS J082244.88$+$460318.1  & 0.351 & 08 22 44.89   & $+$46 03 18.10  &  0/5                                  & 0   & --   &  21.6        & 177  & 7.1        \\
SDSS J082700.24$+$374822.1  & 0.660 & 08 27 00.23   & $+$37 48 22.06  &  0/11                                 & 0   & --   &  1.64        & 1    & 7.4?      \\
SDSS J090157.12$+$063734.6  & 0.530 & 09 01 57.12   & $+$06 37 34.50  & 0/5                                   & 0   & --   &  7.6         & 203  & 7.2        \\
SDSS J100633.91$+$430923.4  & 0.605 & 10 06 33.91   & $+$43 09 23.37  & 0/5                                   & 0   & --   &  12.3        & 130  & 7.7        \\
SDSS J101435.46$+$433056.5  & 0.556 & 10 14 35.46   & $+$43 30 56.43  & 0/3                                   & 0   & --   &  20.0        & 334  & 7.6        \\
SDSS J103128.98$+$091607.2  & 0.636 & 10 31 28.98   & $+$09 16 07.14  & 0/9                                   & 0   & --   &  9.9         & 148  & 7.5        \\
SDSS J103430.53$+$470820.1  & 0.782 & 10 34 30.53   & $+$47 08 20.01  & 0/4\tablefootmark{*}                                   & 0   & --   &  53.8        & 283  & 7.9        \\
SDSS J110542.72$+$020250.9  & 0.455 & 11 05 42.72   & $+$02 02 50.93  & 2/4\tablefootmark{*}                                   & 50  & 0.37 &  212.8       & 2012 & 7.1        \\
SDSS J111756.86$-$000220.6  & 0.457 & 11 17 56.86   & $-$00 02 20.54  & 0/11                                  & 0   & --   &  15.1        & 215  & 7.2        \\
SDSS J112016.17$+$491428.8  & 0.150 & 11 20 16.17   & $+$49 14 28.81  & 0/6                                   & 0   & --   &  12.2        & 176  & 6.6        \\
SDSS J112521.60$+$052358.2  & 0.424 & 11 25 21.60   & $+$05 23 58.14  & 0/5                                   & 0   & --   &  33.2        & 442  & 6.8        \\
SDSS J112702.72$+$030152.0  & 0.416 & 11 27 02.72   & $+$03 01 52.00  & 0/11                                  & 0   & --   &  9.2         & 123  & 7.5        \\
SDSS J120014.08$-$004638.7  & 0.179 & 12 00 14.08   & $-$00 46 38.74  & 0/5                                   & 0   & --   &  21.8        & 73   & 7.2        \\
SDSS J125635.89$+$500852.4  & 0.245 & 12 56 35.87   & $+$50 08 52.54  & 1/6\tablefootmark{*}                                   & 17  & 0.38   &  209.1       & 3203 & 6.9        \\
SDSS J132447.10$+$530257.7  & 0.292 & 13 24 47.10   & $+$53 02 57.73  & 0/7                                   & 0   & --   &  29.2        & 697  & 7.1        \\
SDSS J133345.47$+$414127.7  & 0.225 & 13 33 45.47   & $+$41 41 27.66  & 1/7                                   & 14  & 0.32   &  2.5         & 9\tablefootmark{1}     & 7.7\tablefootmark{2}        \\
SDSS J134834.28$+$262205.9  & 0.917 & 13 48 34.28   & $+$26 22 05.96  & 0/5                                   & 0   & --   &  1.6         & \dots  & 7.7\tablefootmark{2}       \\
SDSS J144848.67$+$372935.7  & 0.243 & 14 48 48.67   & $+$37 29 35.74  & 0/3                                   & 0   & --   &  37.9        & 599  & 6.7        \\
SDSS J145041.93$+$591936.9  & 0.202 & 14 50 41.94   & $+$59 19 37.11  & 2/4                                   & 50  & 0.36   &  3.4         & 22   & 6.6        \\
SDSS J150832.91$+$583422.5  & 0.502 & 15 08 32.91   & $+$58 34 22.55  & 1/5\tablefootmark{*}                                   & 20  & 0.55   &  4.4         & 159  & 7.3        \\
SDSS J151617.16$+$472805.1  & 0.198 & 15 16 17.17   & $+$47 28 05.00  & 0/4                                   & 0   & --   &  25.2        & 163  & 6.8        \\
SBS 1517$+$520              & 0.371 & 15 18 32.86   & $+$51 54 56.74  & 0/5                                   & 0   & --   & 5.9          & 28   & \dots       \\
SDSS J160518.50$+$375653.4  & 0.201 & 16 05 18.49   & $+$37 56 53.42  & 0/3                                   & 0   & --   &  121.8       & 257  & 7.3        \\
SDSS J162543.14$+$490059.0  & 0.545 & 16 25 43.15   & $+$49 00 59.04  & 0/5                                   & 0   & --   &  6.8         & 133  & 7.3        \\
SDSS J162902.05$+$263845.2  & 0.628 & 16 29 02.05   & $+$26 38 45.12  & 0/7                                   & 0   & --   &  22.8        & 164  & 7.5        \\
NVSS J170957$+$234845       & 0.254 & 17 09 57.30   & $+$23 48 47.00  & 0/5                                   & 0   & --   &  4.4         & \dots  & \dots       \\
SDSS J232104.68$-$082537.4  & 0.452 & 23 21 04.68   & $-$08 25 37.40  & 0/4                                   & 0   & --   &  5.0         & 119  & 7.1        \\
SDSS J233903.82$-$091221.3  & 0.660 & 23 39 03.83   & $-$09 12 21.28  & 0/3                                   & 0   & --   &  4.1         & 429  & 7.9        \\
SDSS J234018.85$-$011027.3  & 0.552 & 23 40 18.86   & $-$01 10 27.25  & 0/6                                   & 0   & --   &  12.0        & 161  & 7.5        \\

\end{tabular}

\tablefoot{\textit{RL} and $M_{\text{BH}}$ values are taken from \citet{2015jarvela1} unless otherwise indicated. The five targets with known radio sources nearby are marked with asterisks in the
  $N_{\textrm{Det}}$/$N_{\textrm{Obs}}$ column (see Sect.~\ref{sec:metsahovi-obs} for details).}
\tablebib{
(1)~\citet{2011foschini1}; (2) \citet{2015foschini1}.
}
\label{tab:data2-stat}
\end{table*}

\section{Data}
\label{sec:data}

\subsection{Mets\"{a}hovi Radio Observatory}
\label{sec:mh}
The 13.7-metre radio telescope at Aalto University Mets\"{a}hovi Radio Observatory in Finland is used for monitoring large samples of AGN at 22 and 37~GHz. The measurements
included in this study are made with a 1~GHz-band dual beam 
receiver centered at 36.8~GHz. The observations are on--on observations, alternating the source and the sky in each feed horn. A typical integration time to obtain one flux density data point of a faint source 
is between 1600 and 1800~s. The sensitivity is limited by skynoise due to the location of the telecope, and it has been experimentally shown that the results do not significantly improve after the used maximum integration time of 1800~s. The detection limit of our telescope 
at 37~GHz is on the order of 0.2~Jy under optimal conditions. Data points with a signal-to-noise ratio <~4 are handled as non-detections. The flux density scale is set by observations of DR~21. Sources 
NGC~7027, 3C~274, and 3C~84 are used as secondary calibrators. A detailed description of the data reduction and analysis is given in \citet{terasranta98}. The error estimate in the flux density includes 
the contribution from the measurement rms and the uncertainty of the absolute calibration.

Flux density curves of the detected sources are shown in Figs.~\ref{fig:lc1H0323}--\ref{fig:lcJ164442}. For Mets\"{a}hovi data also non-detections are shown in the curves, denoted as red
diamonds. Non-detections may occur either because the source is too faint or, for example, because of non-ideal weather. The flux levels of the non-detections have been set to an identical
but arbitrary, non-zero value to allow for easier inspection.

\subsection{RATAN-600}
\label{sec:ratan}

Four-frequency broadband radio spectra were obtained with the RATAN-600 radio telescope in transit mode by observing simultaneously
at 4.8, 8.2, 11.2, and 21.7~GHz. The observations were carried out during October in 2013, March, April, October, and November in 2014, and January 2015.
The parameters of the antenna and receivers are listed in Table~\ref{tab:ratan-par}, where $f_{\rm c}$ is the central frequency,
$\Delta f$ is the bandwidth, $\Delta F$ is the flux density detection limit per beam, and BW is the beam width (full width at half-maximum in right 
ascension). The detection limit for the RATAN single sector is approximately 8~mJy (over a 3~s integration) under good conditions at the frequency of 4.8~GHz 
and at an average antenna elevation of $42^{\circ}$. 

Data were reduced using the RATAN standard software FADPS (Flexible Astronomical Data Processing
System) reduction package \citep{1997ASPC..125...46V}. The flux density measurement procedure is described in \citet{2001A&A...370...78M,2012A&A...544A..25M}.
The following flux density calibrators were applied to obtain the calibration coefficients in the scale by \citet{baars77}: 3C~48,
3C~147, 3C~161, 3C~286, 3C~295, 3C~309.1, and NGC~7027. We also used the traditional RATAN flux density calibrators: J0237$-$23, 3C~138, J1154$-$35, and 
J1347$+$12. The measurements of some of the calibrators were corrected for angular size and linear polarization following the data from 
\citet{1994A&A...284..331O} and \citet{1980A&AS...39..379T}. The total error in the flux density includes the uncertainty of the RATAN calibration curve and the error in the 
antenna temperature measurement. The systematic uncertainty of the absolute flux density scale (3--10{\%} at different RATAN frequencies) is also included in the flux density error. 
Finally, the data were averaged over 2--25 days in order to get reliable values of the flux densities. 

RATAN-600 observed 34 sources from sample 1; 32 were detected at least at one frequency. They are shown in Table~\ref{tab:otherdata}. Flux density curves are shown in Figs.~\ref{fig:lc1H0323}--\ref{fig:lcJ164442}.

\begin{table}[!]
\caption{\label{tab:radiometers}Parameters of the RATAN-600 antenna and radiometers.}
\centering
\begin{tabular}{rlcr}
\hline\hline
 $f_{\rm c}$ & $\Delta f$ & $\Delta F$ &  BW \\
  GHz    &   GHz       &  mJy beam$^{-1}$ &  (arcsec) \\
\hline
 $21.7$ & $2.5$  &  $70$ & 11 \\
 $11.2$ & $1.4$  &  $20$ & 16 \\
 $7.7$  & $1.0$  &  $25$ & 22 \\
 $4.8$  & $0.9$  &  $8$  & 36 \\
 $2.3$  & $0.4$  &  $30$ & 80 \\
 $1.1$  & $0.12$ &  $160$& 170 \\
\hline
\end{tabular}
\label{tab:ratan-par}
\end{table}

\subsection{Owens Valley Radio Observatory}
\label{sec:ovro}

The 15\,GHz observations were carried out as part of a high-cadence blazar monitoring programme using the OVRO
40~m telescope \citep{richards11}. 
It uses off-axis dual-beam optics and a cryogenic receiver with a 15.0~GHz center frequency and 3~GHz bandwidth. The two sky beams are 
Dicke switched using the off-source beam as a reference, and the source is alternated between the two beams in an on-on fashion to remove 
atmospheric and ground contamination. In May 2014 a new pseudo-correlation receiver was installed on the 40~m telescope and the
fast gain variations are corrected using a 180 degree phase switch instead of a Dicke switch.
The performance of the new receiver is very similar to the old one and no discontinuity is seen in the light curves.
Calibration is achieved using a temperature-stable diode noise source to remove receiver gain drifts and the flux density scale 
is derived from observations of 3C~286 assuming the \citet{baars77} value of 3.44\,Jy at 15.0\,GHz. The systematic uncertainty of about 5\% in the flux 
density scale is included in the error bars. Complete details of the reduction and calibration procedure are found in \citet{richards11}. 

The 15 sources observed at OVRO, shown in Table~\ref{tab:otherdata}, were originally selected for observations with Very Long
Baseline Array (VLBA), and parsec-scale structures were found in all of them \citep{2015richards}. They are all included in Mets\"{a}hovi
sample 1, and seven of them have been detected also at 37~GHz. We have OVRO data between January 2008 and April 2015, however,
for most sources data have been gathered since mid-2013. The number of detections and the total number of observations, the detection rates, 
and maximum flux at 15~GHz are shown in Table~\ref{tab:ovrosources}. We used S/N$>$4 as a detection limit.
Flux density curves at 15~GHz are shown in Figs.~\ref{fig:lc1H0323}--\ref{fig:lcJ164442}.

\begin{table*}[ht]
\caption[]{Statistics of OVRO observations.}
\centering
\begin{tabular}{l l l l}
\hline\hline
Source                 & $N_{\textrm{Det}}$/$N_{\textrm{Obs}}$ & Det (\%) & $S_{\text{15GHz, max}}$ (Jy)\\ \hline
1H 0323$+$342                & 465/468     &  99  & 0.92  \\  
SDSS J081432.11$+$560956.6   & 44/71       &  62  & 0.03 \\         
SDSS J084957.97$+$510829.0   & 351/360     &  97  & 0.57 \\
SDSS J090227.16$+$044309.5   & 71/72       &  99  & 0.09 \\
SDSS J094857.31$+$002225.4   & 341/345     &  99  & 0.94 \\
SDSS J095317.09$+$283601.5   & 11/62       &  18  & 0.06 \\
SDSS J104732.68$+$472532.0   & 64/64       &  100 & 0.21 \\ 
SDSS J124634.65$+$023809.0   & 7/72        &  10  & 0.03  \\
SDSS J143509.49$+$313147.8   & 1/76        &  1   & 0.06 \\
SDSS J144318.56$+$472556.7   & 45/76       &  59  & 0.05 \\
SDSS J150506.47$+$032630.8   & 421/421     &  100 & 0.74 \\
SDSS J154817.92$+$351128.0   & 69/85       &  81  & 0.06 \\
SDSS J162901.30$+$400759.9   & 166/339     &  49  & 0.10 \\
SDSS J164442.53$+$261913.2   & 60/84       &  71  & 0.16  \\
SDSS J172206.03$+$565451.6   & 4/82        &  5   & 0.02  \\
\end{tabular}
\label{tab:ovrosources}
\end{table*}

\subsection{\textit{Planck} satellite}
\label{sec:planck}

\textit{Planck}{\footnote{http://www.esa.int/planck/}} satellite was operated by the European Space Agency (ESA) from 2009 to 2013. During its lifetime it mapped the whole sky every six months several times, 
the number depending on frequency. The Low Frequency Instrument{\footnote{http://sci.esa.int/planck/34730-instruments/}} (LFI) observed at frequencies 30, 44, and 70 GHz, and High Frequency Instrument (HFI) at 
frequencies 100, 143, 217, 353, 545, and 857 GHz. 

We have \textit{Planck} data for nine of the brightest sources, spanning from August 13, 2009 to October 3, 2013 for LFI (8 surveys) and from August 13, 2009 to January 14, 2012 for HFI (5 surveys). 
The observing times were calculated with the Planck On-Flight Forecaster tool \citep[{\tt POFF},][]{massardi10} which computes when the sources were visible at each of the satellite's frequencies. 
Given the scanning strategy of the satellite, for some sources and frequencies the \textit{Planck} flux densities are averages of several pointings over the source's visibility period in one survey.
The \textit{Planck} flux densities were extracted from the full mission maps from the 2015 data release using the Mexican Hat Wavelet~2 source detection and flux density estimation pipelines in the 
\textit{Planck} LFI and HFI Data Processing Centres. For LFI, data detection pipeline photometry (DETFLUX) was used; for HFI, aperture photometry (APERFLUX) was used. 
The calibration of \textit{Planck} is based on the dipole signal, and is consistent at approximately the 0.2\% level \citep{planck2014-a01}. The systematic uncertainties of the absolute flux density 
scale are under 1\% for the seven lowest frequencies and under 7\% for the two highest. 
See the Second \Planck\ Catalogue of Compact Sources  \citep{planck2014-a35} for further details of the data processing procedures. 

The nine sources for which we have \textit{Planck} data at our disposal are shown in Table~\ref{tab:otherdata}. We use S/N > 4 as a detection limit, which for 
these faint sources means that we have \textit{Planck} detections of only three sources, however, not necessarily at all of the frequencies.
Flux density curves are shown in Figs.~\ref{fig:lc1H0323}--\ref{fig:lcJ164442}.

\section{Mets\"{a}hovi observations}
\label{sec:metsahovi-obs}
The Mets\"{a}hovi observations of NLS1 sample 1 started in February 2012, and of the NLS1 sample 2 in November 2013. Three of the brighter sources 
(J084957.97$+$510829.0, also known as J0849$+$5108; J094857.31$+$002225.4, also known as J0948$+$0022; J150506.47$+$032630.8, also known as J1505$+$0326) have been observed already earlier because 
they have been targets of multifrequency campaigns; the longest light curve is for J0849$+$5108, 
observed since January 1986. Of these, J0849$+$5108 and J0948$+$0022 have been observed significantly more often than any other source, with 
138 and 132 observations, respectively. The typical number of observations for one source is on average 6.7 (median 5), and varies between 3 and 24 for both samples, excluding J0849$+$5108 and J0948$+$0022. 

The aim was to get at least three observations of each source separated by several months, first, to see if they are detectable at 37~GHz, and second, to examine their possible variability. 
Ultimately, we aimed to compile a sample of detected sources for follow-up observations. By the end of April 2015 the criterion was fulfilled for every source in samples 1 and 2. We were able to 
detect 15 out of 78 sources, the detection rate for the whole sample being 19.2\%. Seven of them were detected only once, the rest of the measurements being nondetections. This indicates that the 
sources are variable, and that we are not always able to detect them due to their faintness. The weather-dependent detection limit of the telescope undoubtedly also plays a role. The typical 
number of detections $N_{\text{det}}$ for one source is on average 4 (median 1), and varies between 1 and 18, excluding J0849$+$5108 and J0948$+$0022.

The more sensitive OVRO system can observe fainter sources than the Mets\"{a}hovi antenna. The detection limits of the telescopes are on the order of 
10~mJy and 200~mJy, respectively. Comparison of the flux density curves of the brightest sources
from both observatories show that the sources are detected at Mets\"{a}hovi at 37~GHZ when they are flaring in the OVRO 15~GHz curve (Fig.~\ref{fig:lcJ094857}). This trend can 
also been seen in the flux density curves of the fainter sources (Fig.~\ref{fig:lcJ164442}).
This confirms that the apparently sporadic detections at 37~GHz reflect genuine variability that would be observable with a more sensitive system. 
However, single 37~GHz detections with sharply inverted spectrum at the higher frequencies, such as in the case of J154817.92+351128.0 (Fig.~\ref{fig:lcJ154817}) that coincides 
with an uneventful period in the 15~GHz flux density curve, are somewhat suspicious. Another example is the source J103123.73+423439.3 (Fig.~\ref{fig:lcJ103123}) for which there 
unfortunately are no 15~GHz data. Even though the data reduction has been performed with care and 
healthy scepticism, false detections are possible. On the other hand, such sharply inverted spectrum, if authentic, would 
suggest that extreme processes are taking place in these sources.

Known radio sources are located in close proximity of five Sample 2 targets. These are indicated with asterisks in 
Table~\ref{tab:data2-stat}. Four of them have been detected at 37~GHz. The distances of the known radio sources from the targets 
vary between approximately 0.6 and 1 arcminutes. The flux densities vary between 60 and 300~mJy at 1.4~GHz, and 0 and 130~mJy at 8.6~GHz, i.e., they are most likely very faint at 37~GHz.
The spectral indices for possibly contaminating sources with radio observations in more than one band seem to be steep, further indicating their negligible impact on the 37~GHz observations.
We ran a set of simple tests to check how the possibility of a known radio source lying in the antenna beam (2.4 arcminutes at 37~GHz) together with the faint target might affect our observations, by observing them in 
pairs several times. In most cases there was no correspondence between the detections or non-detections of the NLS1 galaxies, and the known radio sources. Possible exception is the source J110542.72$+$020250.9 
that has a fairly bright, but, according to archival data, rather steep-spectrum AGN (SDSS~J110538.99$+$020257.3) close by (separation of 0.94 arcminutes). In this case the occurrence of 
detections in the test pairs suggests potential contamination. However, this is not a precise method and can be considered only as an approximation.

\section{Variability}

We calculated variability indices for Mets\"{a}hovi and OVRO data, for sources with at least two detections. This left us with 8 and 14 sources, 
respectively. We used equation \citep{1992aller1}

\begin{equation}
\Delta{S} = \frac{(S_{max} - \sigma_{S_{max}}) - (S_{min} + \sigma_{S_{min}})}{(S_{max} - \sigma_{S_{max}}) + (S_{min} + \sigma_{S_{min}})}
\end{equation}

that describes fractional variability; $S_{max}$ is the maximum observed flux density, $\sigma_{S_{max}}$ its error, and $S_{min}$ is the minimum
flux density and $\sigma_{S_{min}}$ its error. 
For a faint population, the number of detections affects the variability indices; sources that are observed more often are also more likely to 
be detected as they go through active and quiet periods. Their variability is therefore enhanced \citep{nieppola07}. Our sample also 
includes sources that are particularly bright and variable compared to others so we need to estimate the indices for them separately. For these 
reasons we have divided the sources into several groups, depending on the number of detections $N_{det}$.

For Mets\"{a}hovi data we have three groups:
\begin{itemize}
\item the four most frequently observed sources that are also the brightest (1H~0323$+$342, J0849$+$5108, J0948$+$0022, and J1505$+$0326). 
These are clearly different from the other sources with significantly higher variability indices. They all have $N_{\text{det}}$ > 10.
\item sources for which $N_{\text{det}}$ < 10.
\item all sources.
\end{itemize}

For OVRO data we have four groups:
\begin{itemize}
\item the four most frequently observed sources that are also the brightest (1H~0323$+$342, J0849$+$5108, J0948$+$0022, and J1505$+$0326), 
for which $N_{\text{det}}$ > 20. These are clearly different from the other sources with significantly higher variability indices. They all have $N_{det}$ > 300.
\item sources for which $N_{\text{det}}$ > 20 but excluding the four brightest sources.
\item sources for which $N_{\text{det}}$ < 20.
\item all sources. 
\end{itemize}

\begin{figure}
        \centering
        \includegraphics[width=0.5\textwidth]{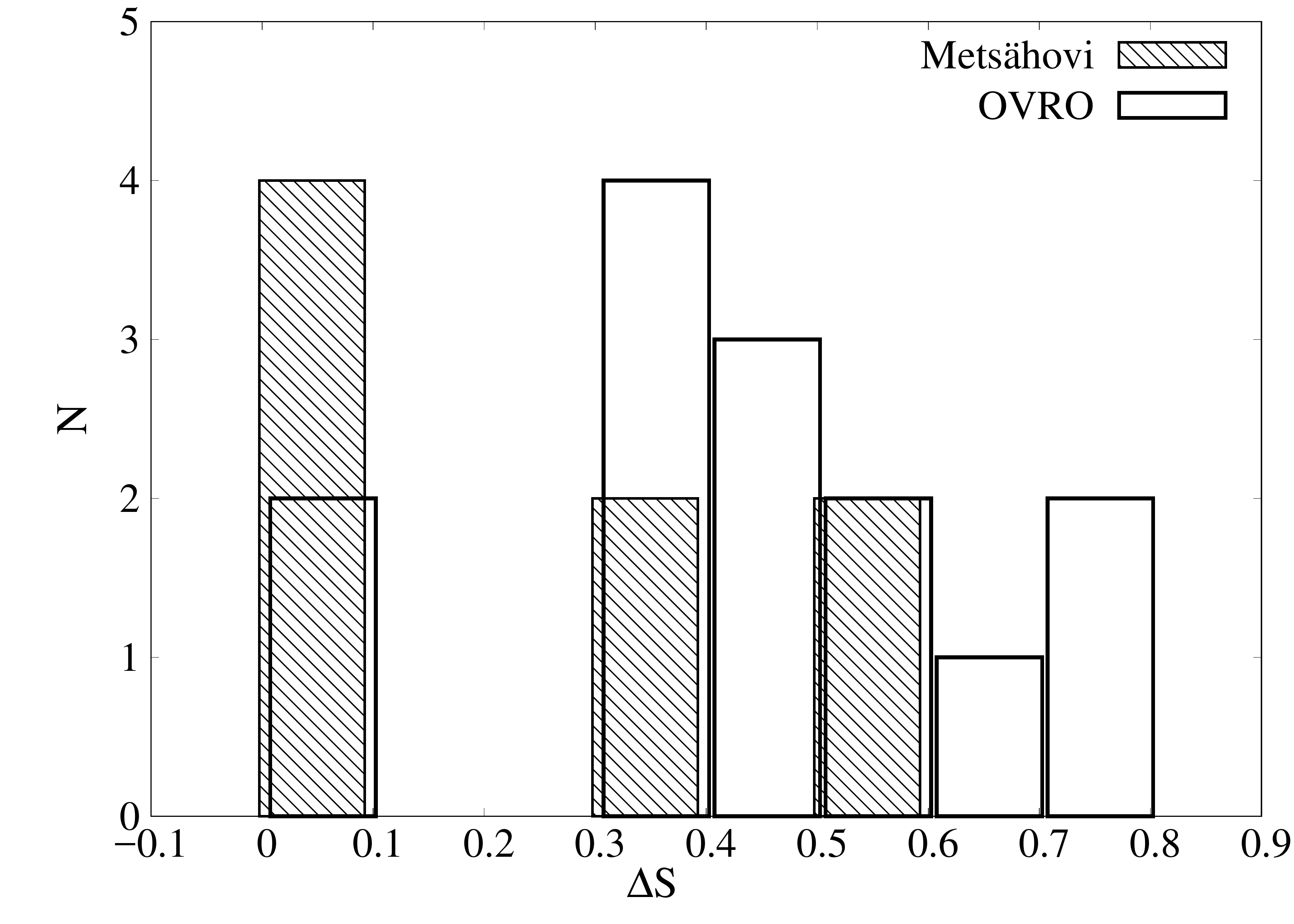}
        \caption{Distribution of variability indices at 15 and 37~GHz. Note that negative variability indices are depicted as zero.}
        \label{fig:histdeltas}
\end{figure}

The distributions of variability indices at 15 and 37~GHz are shown in Fig.~\ref{fig:histdeltas}, and the average and median indices and $N_{\text{det}}$ of each group 
are shown in Table~\ref{tab:varind}. A negative variability index means that the uncertainties in the flux densities are larger than the difference between the flux
densities and thus the source is not detectably variable. As expected, the variability index of the group that contains the four brightest sources is largest for both data sets. The 
dependence of the index on $N_{\text{det}}$ is evident. It should be noted that as we have not taken into account non-detections, true variability is in fact larger than the indices indicate.
\citet{2015richards} report that the nine brighter sources in the OVRO sample show variability of 13 -- 38~\% and the remaining sources 5 -- 20~\%. They conclude that this is similar 
compared to the OVRO blazar monitoring sample. In Mets\"{a}hovi data the difference between the four brightest and the other sources is also considerable. In addition to $N_{\text{det}}$, 
the rather high detection limit of the telescope causes the variability of the fainter sources to be diminished (see also discussion in Sect.~\ref{sec:metsahovi-obs}).

Statistical values of the flux densities at 37 and 15~GHz are presented in Table~\ref{tab:fluxstat}, where the minimum, maximum, average, and median 
flux densities, the standard deviations, and the total number of detections are listed for each of the above mentioned groups.

\begin{table*}[ht]
\caption[]{Variability indices for Mets\"{a}hovi and OVRO data.}
\centering
\begin{tabular}{l l l l l }
\hline\hline
                     & $\Delta{S}$         &          & $N_{\text{det}}$ &        \\
Mets\"{a}hovi        & Average    & Median   & Average    & Median \\ \hline
Brightest            & 0.44       & 0.43     & 48.5       & 42.5  \\
$N_{\text{det}}$<10  & -0.03      & -0.02    & 3.3        & 3.0   \\
All                  & 0.20       & 0.19     & 25.9       & 9.5   \\
                     &            &          &            &        \\
OVRO                 &            &          &            &        \\ \hline
Brightest            & 0.62      & 0.65    & 394.5      & 386.0 \\
$N_{\text{det}}$>20  & 0.38      & 0.40    & 190.6      & 71.0  \\
$N_{\text{det}}$<20  & 0.27      & 0.38    & 7.3        & 7.0    \\
All                  & 0.42      & 0.43    & 151.4      & 66.5  \\

\end{tabular}
\label{tab:varind}
\end{table*}

\begin{table*}[ht]
\caption[]{Flux density statistics of the detections for Mets\"{a}hovi and OVRO data.}
\centering
\begin{tabular}{l l l l l l l}
\hline\hline
Mets\"{a}hovi        & Min    & Max   & Average    & Median & SD & $N_{\text{det}}$ \\ \hline
Brightest            & 0.23      & 1.18    & 0.54       & 0.51 & 0.19 & 194  \\
$N_{\text{det}}$<10  & 0.21      & 0.55      & 0.34    & 0.34  & 0.07 & 20   \\
All                  & 0.21      & 1.18    & 0.52       & 0.48 & 0.19 & 214   \\
                     &            &          &            &    &    \\
OVRO                 &            &          &            &    &    \\ \hline
Brightest            & 0.10      & 0.94    & 0.40     &  0.38   & 0.14 & 1578 \\
$N_{\text{det}}$>20  & 0.01      & 0.23    & 0.07   &  0.04 & 0.05 & 519  \\
$N_{\text{det}}$<20  & 0.01      & 0.06    & 0.02     & 0.02  & 0.01 & 23    \\
All                  & 0.01      & 0.94    & 0.32   & 0.32   & 0.19 & 2120  \\

\end{tabular}
\label{tab:fluxstat}
\end{table*}

\section{Radio spectra}
\label{sec:spectra}

Radio spectra were compiled for those sources for which we have data at least at two frequencies (41 sources, 
Figs.~\ref{fig:spec1h0323}--\ref{fig:specrxj2314}).
Spectral indices were calculated for 20 sources for which we had quasi-simultaneous data. We consider data taken within one month quasi-simultaneous.
The amount of data points for each source and epoch varies from two to six. For nine sources we have only one spectral index value (i.e., one epoch).
The spectral index using quasi-simultaneous data was calculated either simply between two frequencies, or
between points chosen by eye to best represent the general shape of the spectrum. The latter was used for sources with numerous data points for which the simple
method would not produce a realistic value. 
In some cases there was an obvious break in the spectrum, and the indices were therefore determined for both slopes. The indices are listed in Table~\ref{tab:specind} 
where Col. 1 gives the source name, Cols. 2, 3, and 4 the start and end times of the epochs, and the frequency range used for calculating the indices, 
respectively, and the last column the spectral index. In the last column an asterisk denotes a 15 -- 37~GHz index that has been calculated separately, in 
case an index was originally calculated with more than the two data points for the same epoch.

For many sources there are several one-month epochs, and for some of those only two data points taken at OVRO and Mets\"{a}hovi. For sources observed 
frequently, this resulted in large numbers of indices calculated with only two data points at 15 and 37~GHz 
(45 for J0948$+$0022 and 26 for J0849$+$5108). 
Only statistical values are listed in these cases. In Table~\ref{tab:specmean} Col. 1 gives the source name, 
Cols. 2 and 3 the start and end times of the epochs used for calculating the indices, Col. 4 the number of spectral indices, Cols. 5, 6, 7, and 8 
the minimum, maximum, average, and median values of the indices, and Col. 9 the standard error of the mean (SEM).

\begin{figure}
        \centering
        \includegraphics[width=0.5\textwidth]{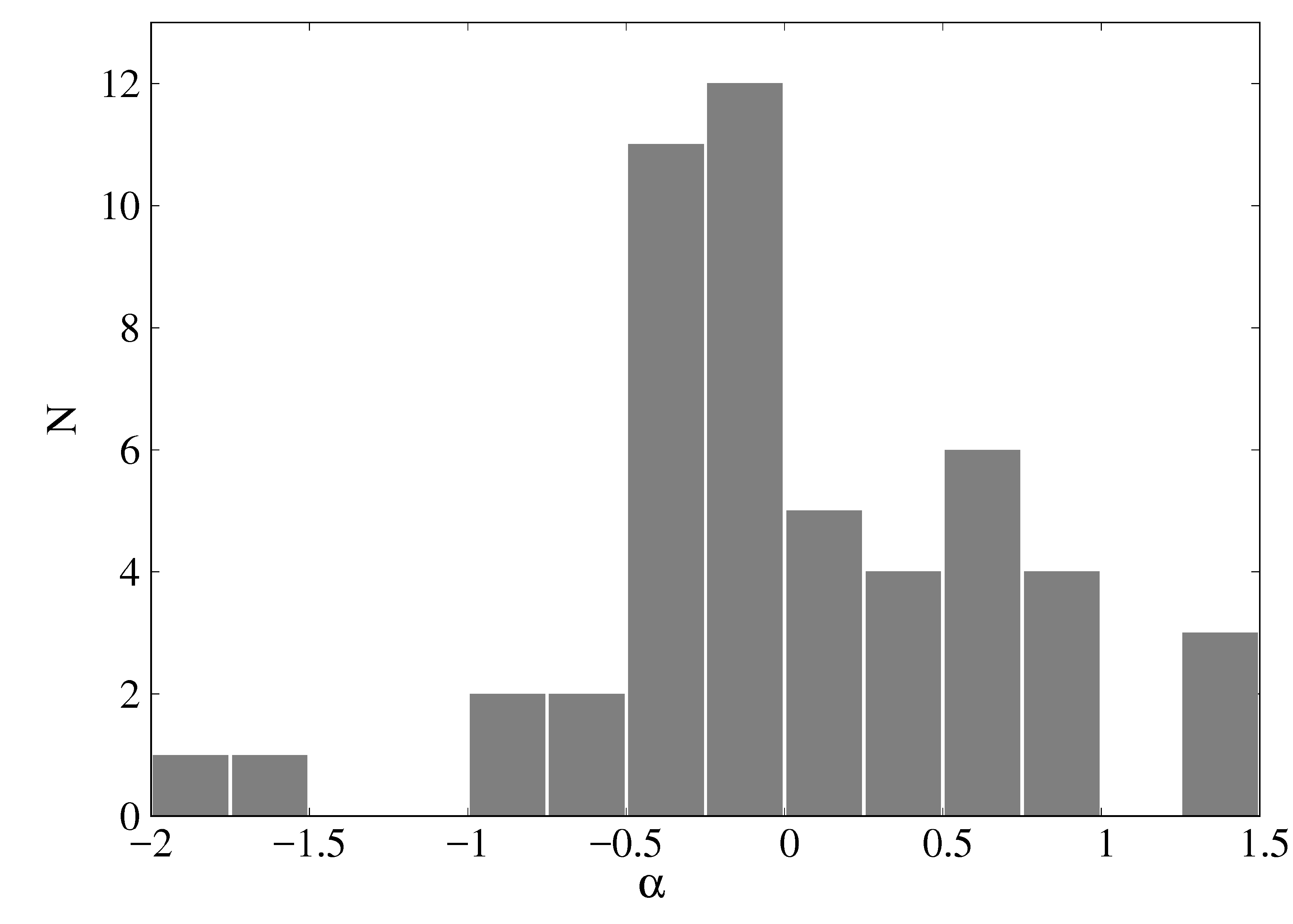}
        \caption{Distribution of spectral indices using all frequencies, except 15 -- 37~GHz.}
        \label{fig:alphaother}
\end{figure}

\begin{figure}
        \centering
        \includegraphics[width=0.5\textwidth]{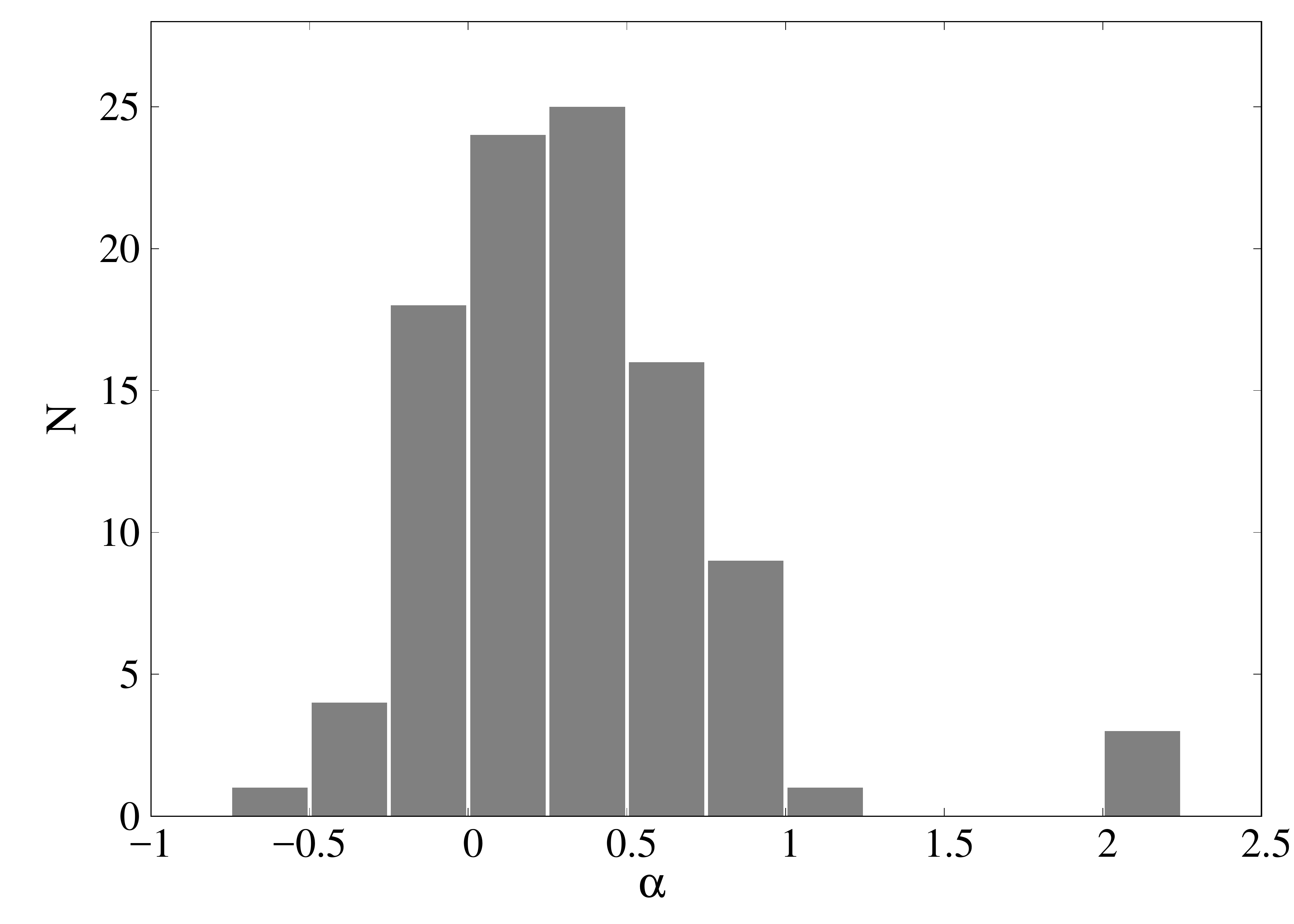}
        \caption{Distribution of 15 -- 37~GHz spectral indices.}
        \label{fig:alpha1537}
\end{figure}

The distribution of spectral indices for all frequencies, excluding indices between 15 and 37~GHz, is shown in Fig.~\ref{fig:alphaother}, and for frequencies 15 -- 37~GHz in Fig.~\ref{fig:alpha1537}. 
The distributions include all epochs, i.e., there can be several indices for one source. Sources observed more frequently than the others dominate the distribution of the 15 -- 37~GHz indices 
(see also Table~\ref{tab:specmean}). They are mostly flat, but can also be strongly inverted. The spectral indices determined using the other frequencies have a much wider distribution, particularly at the steeper end.

\begin{figure}
        \centering
        \includegraphics[width=0.5\textwidth]{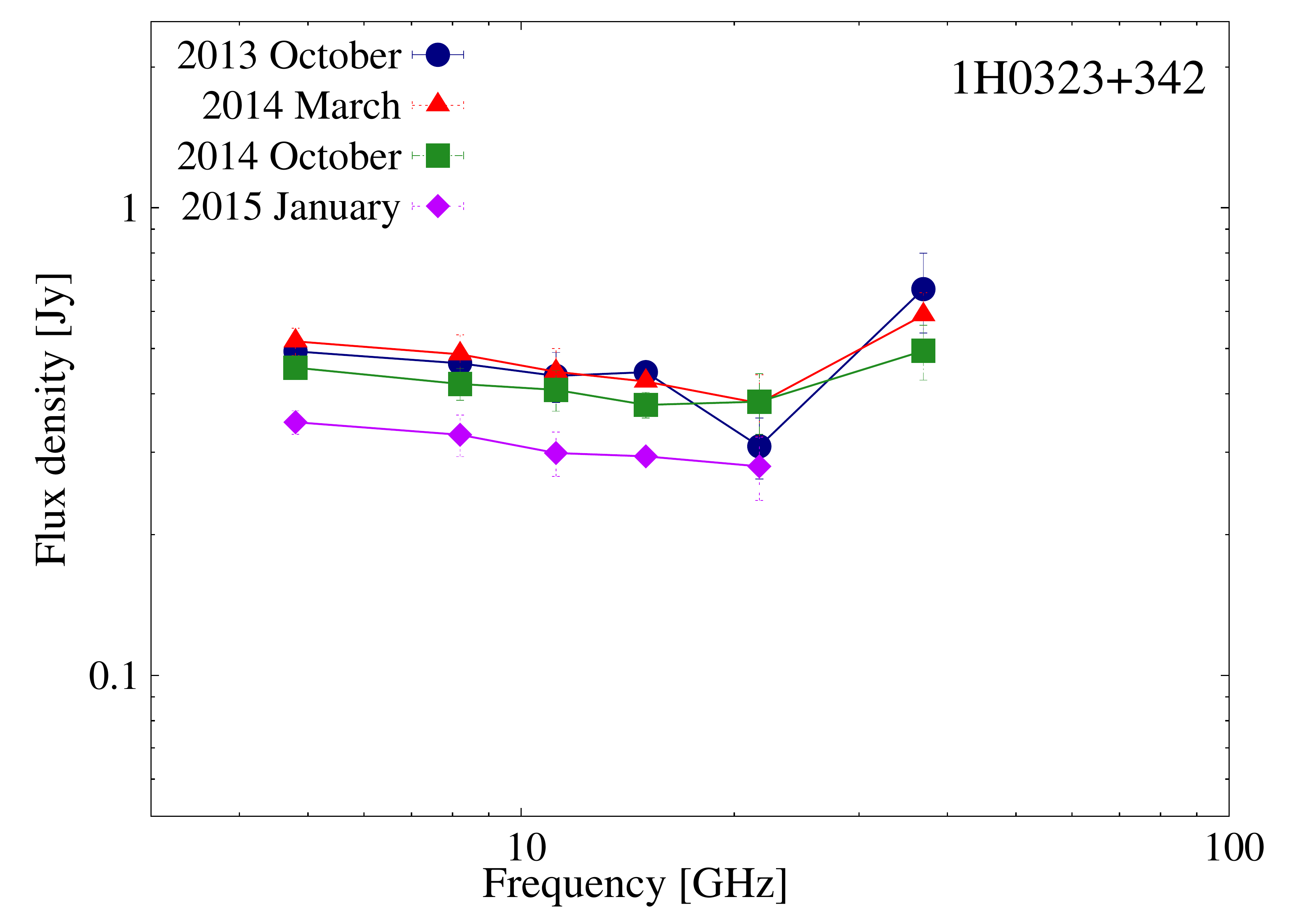}
        \caption{Simultaneous radio spectra of 1H~0323$+$342.}
        \label{fig:h0323epochs}
\end{figure}

\begin{figure}
        \centering
        \includegraphics[width=0.5\textwidth]{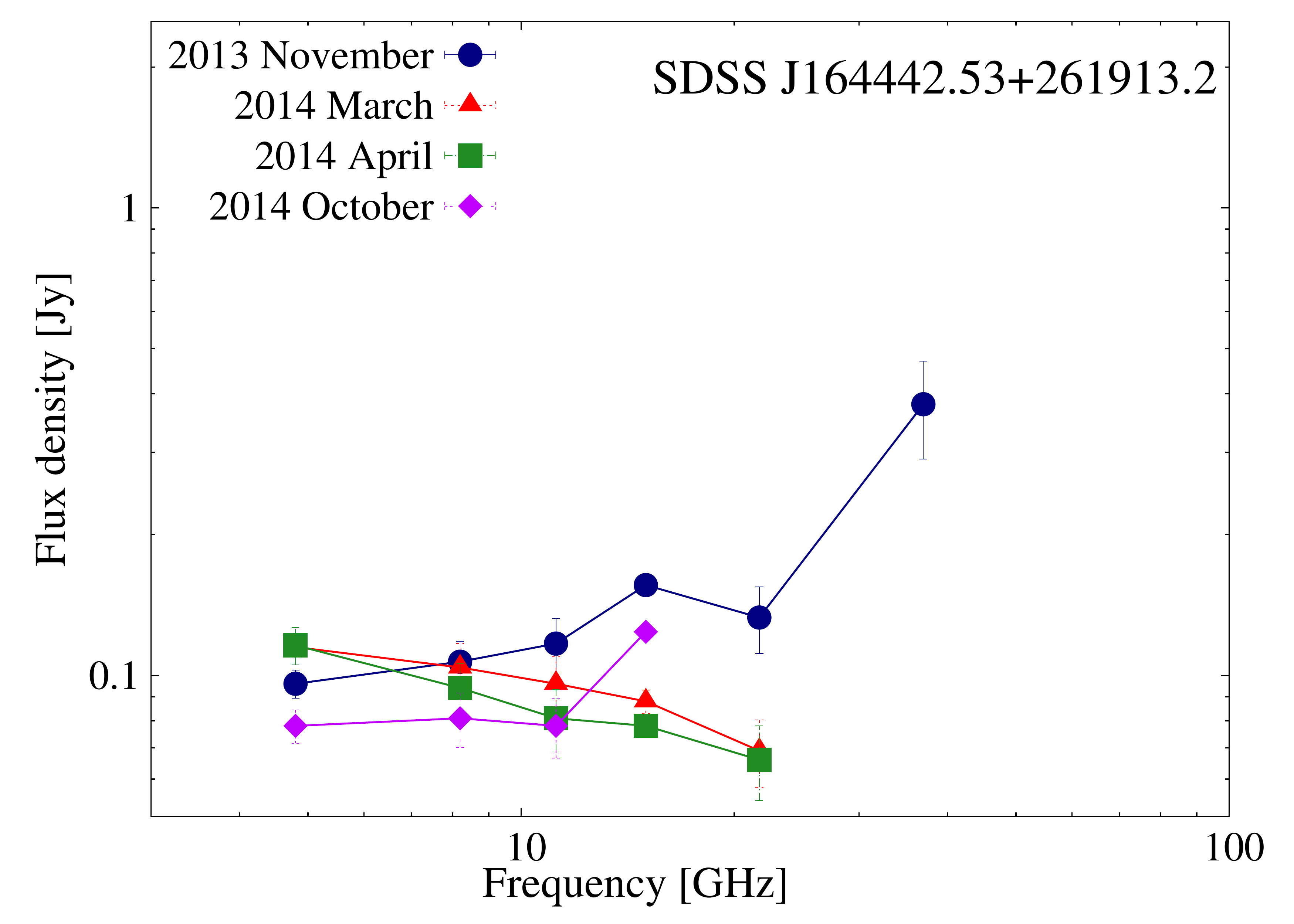}
        \caption{Simultaneous radio spectra of J164442.53$+$261913.2.}
        \label{fig:j1644epochs}
\end{figure}

In general the indices show significant variability between the various epochs in individual sources. The spectra are often inverted when 
37~GHz observations are available, resembling that of Gigahertz-Peaked Spectrum (GPS) sources.  Looking at the indices within one source in 
Table~\ref{tab:specind}, the spectra can be rather steep or inverted, and they only occasionally flatten. It is a clear indication of flaring that both steep and flat spectra can be seen in some sources, 
and even in those that constantly keep the steeper shape, the variability can be considerable. The slopes and the shapes of the spectra frequently change
from one epoch to another. This is well illustarated in the simultaneous spectra of, for example, 1H~0323$+$342 and J164442.53$+$261913.2 in Figs.~\ref{fig:h0323epochs} and 
\ref{fig:j1644epochs}, plotted using spectra with more than just two data points. See Sect.~\ref{sec:indiv} for details of the individidual cases.

\begin{figure}
        \centering
        \includegraphics[width=0.5\textwidth]{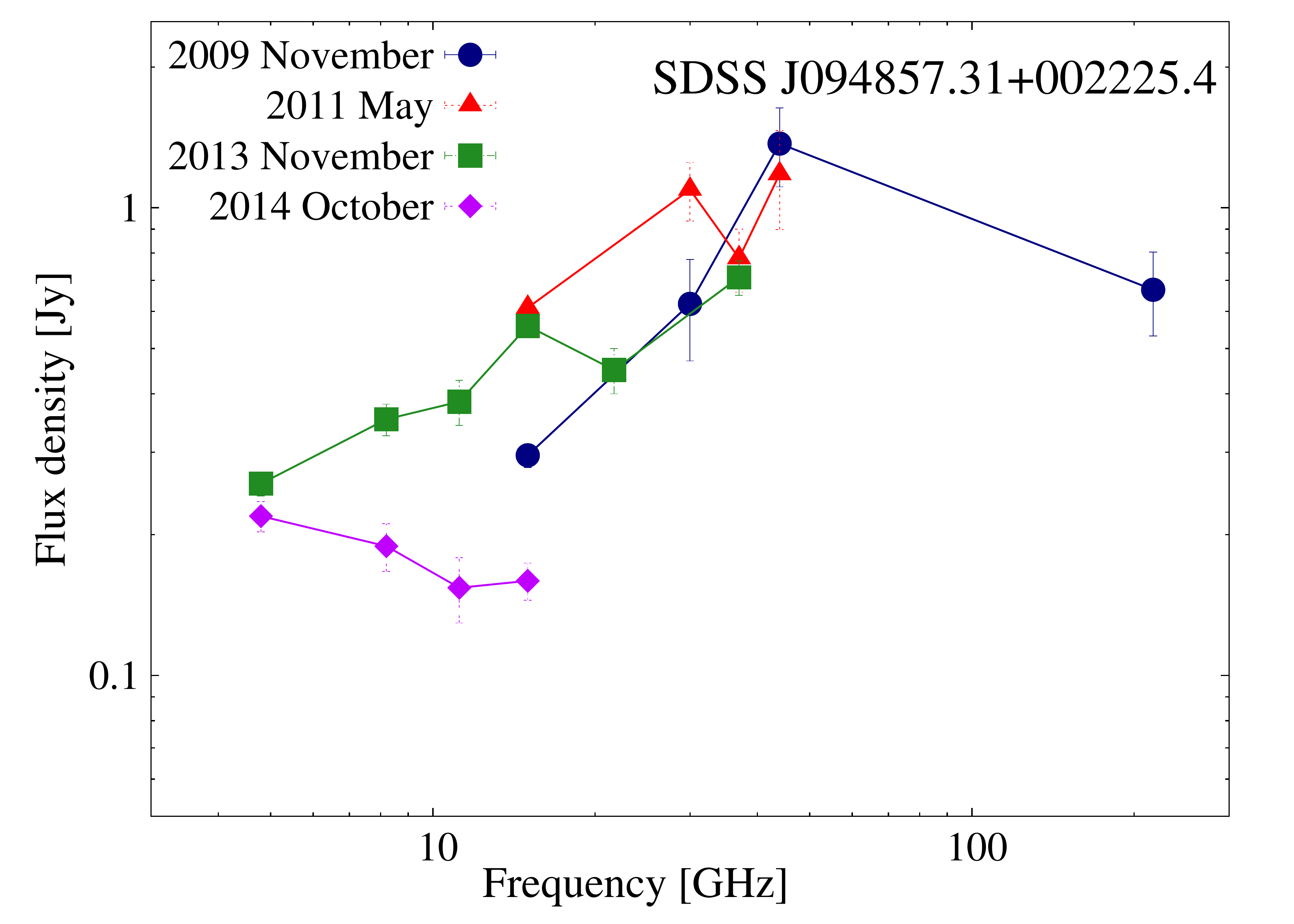}
        \caption{Simultaneous radio spectra of J0948$+$0022.}
        \label{fig:j0948epochs}
\end{figure}

\begin{figure}
        \centering
        \includegraphics[width=0.5\textwidth]{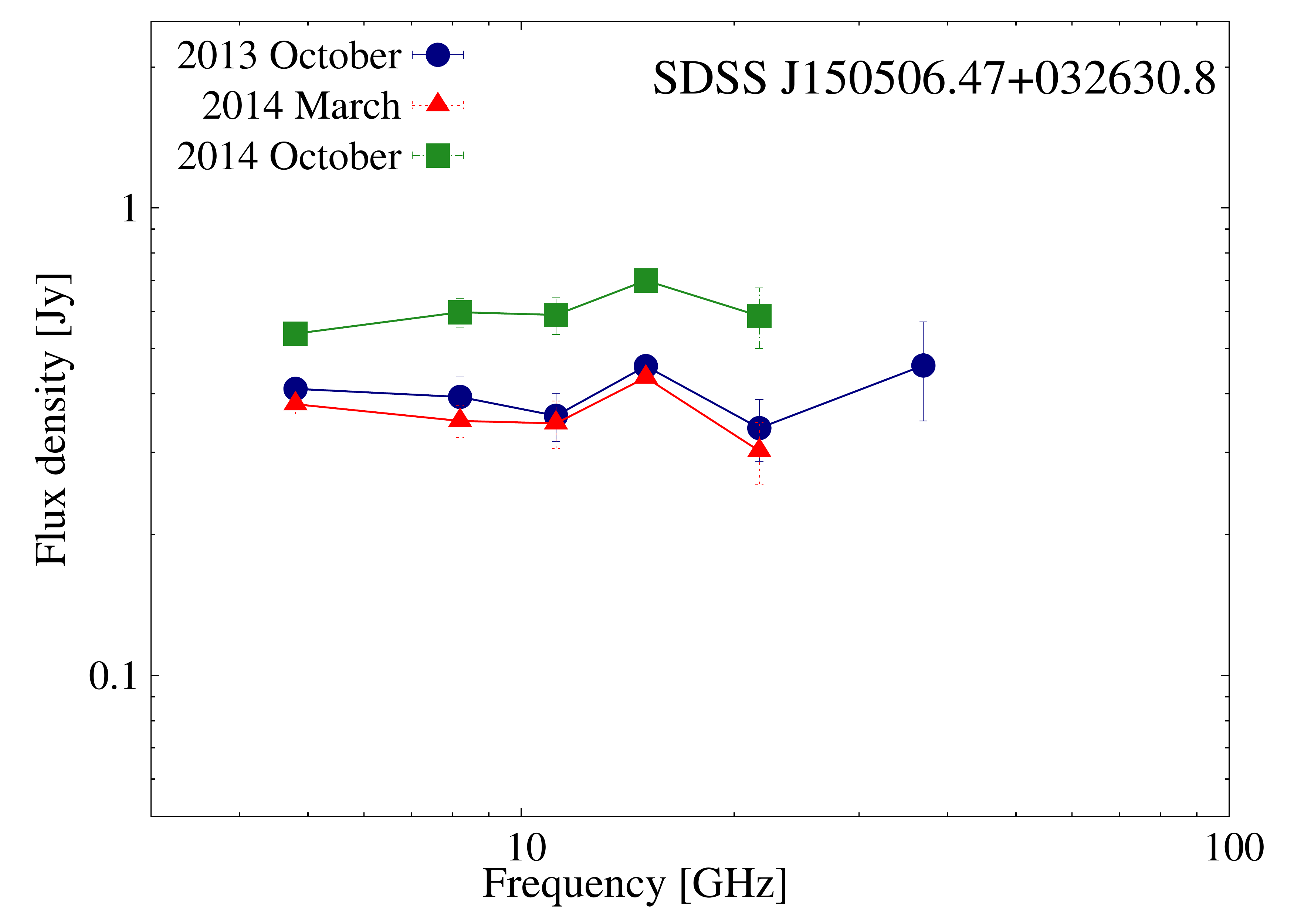}
        \caption{Simultaneous radio spectra of J1505$+$0326.}
        \label{fig:j1505epochs}
\end{figure}

\section{Synchrotron peak frequencies}
\label{sec:nupeak}

Spectral energy distributions (SEDs) were constructed for the 41 sources using the radio data and archival data collected using the ASI 
(Agenzia Spaziale Italiana) Science Data Center (ASDC) search tool\footnote{http://www.asdc.asi.it}. A third degree polynomial function was fitted to them in 
order to determine the synchrotron peak frequency $\nu_{\text{peak}}$. This was done in two ways: by using data only in a fixed frequency range of 8 to 14
($\log \nu_{\text{peak}}$) for each source, and by selecting an exclusive top frequency for each source. This allowed us in many cases to check
for a more sensible fit than what the fixed frequency range could produce, as in many cases the peak frequency was simply set at the highest value of 14
because the SED was still rising when the data range ran out. In the latter case when the disk component was clearly separate from the synchrotron component, it was left out from the fits. 
This caused a variance of 13.4 --- 15.4 in the individual top limits of the fitting range. The quality of the fits was checked by eye, and sources with not enough (or too much) data to yield 
an unambiguous fit were flagged as bad (7 for the first case, 11 for the latter). For the fixed frequency range fits the average $\nu_{\text{peak}}$ is 13.06, and for the individual
fits 13.20, excluding the bad ones. For all sources together, including the bad ones, $\nu_{\text{peak}}$ is 13.15 in both cases. \citet{planck2011-6.3a} found that the
average $\nu_{\text{peak}}$ of a large sample of the brightest AGN --mostly blazars-- is 13.2. The $\nu_{\text{peak}}$ of the fainter NLS1 sources
is almost exactly the same.

\begin{figure}
        \centering
        \includegraphics[width=0.5\textwidth]{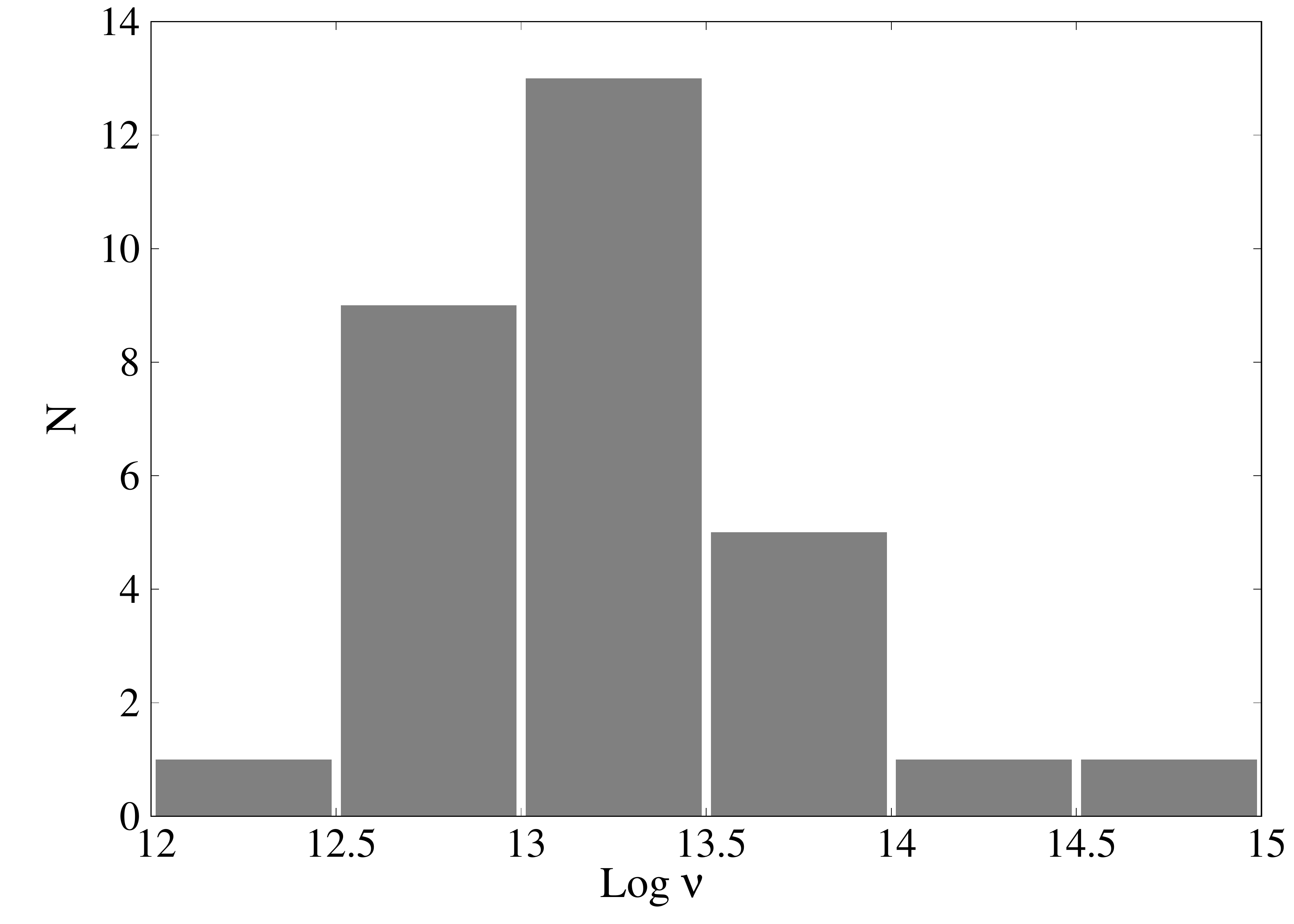}
        \caption{Distribution of the synchrotron peak frequencies.}
        \label{fig:nupeak}
\end{figure}

The distribution of the individually fitted peak frequencies is shown in Fig.~\ref{fig:nupeak}.

\section{Notes on individual sources}
\label{sec:indiv}

The sources detected at 37~GHz show an assortment of radio-loudness values from less than ten to over 4000, and their black hole masses are
all in the range from a few times 10$^6$ to almost 10$^8$ $M_{\sun}$ (Tables~\ref{tab:data1-stat} and \ref{tab:data2-stat}). A table
containing flux densities and their errors from Mets\"{a}hovi, RATAN-600, and \textit{Planck}, presented in this paper, are available at the CDS. An excerpt of the datafile can be found in Table~\ref{tab:data1h0323}. In the following, notes on individual sources are presented.

\textit{1H~0323$+$342}
This source has been observed at Mets\"{a}hovi since early 2012. Despite its relative brightness --up to around 1~Jy-- it is occasionally so faint
that it is not detected at all. This is in fact typical for all of the brightest sources in this sample. The OVRO flux density curve shows several well-defined
flares that coincide with the 37~GHz flares well. This is the only source that has averaged \emph{Planck} detections; the two 143~GHz detections are averages of 
observations done on 2009-Sep-01 and 2010-Feb-13 (2009.8932), and on 2010-Sep-01 and 2011-Feb-13 (2010.8932). These are marked with asterisks in Table~\ref{tab:data1h0323}.
The source was detected at gamma \citep{2009abdo3} and both parsec and kiloparsec-scale 
structures have been observed in it \citep{2012doi1,2014wajima1}. \citet{2016lister} found superluminal features in 
the jet with speeds of 9.0$c$. \citet{2014leontavares1} studied the host galaxy which exhibits spiral structure, or more likely, a ring-like structure 
caused by a merger.

The simultaneous radio spectra of this source (Fig.~\ref{fig:h0323epochs}) is inverted, and at its highest in October 2013 -- March 2014, showing some features 
between 11 and 37~GHz. In October 2014 it is already declining, and is at its lowest in January 2015 and the spectrum is fairly flat, at least without
the 37~GHz data point. The nearest 37~GHz measurement in February 2015 is not considered simultaneous by our criteria and is therefore omitted from the last spectra, however, it is on the order of 0.5~Jy. 
Because the source is known to exhibit superluminal motion, it can be argued that the variability in the spectrum may be
due to components moving in the jet according to the Marscher \& Gear shock-in-jet model \citep{marscher85}. Our results comply with those found in
a comprehensive spectral analysis by \citet{angelakis2015}. 

\textit{J0849$+$5108}
The source was first observed at 37~GHz in 1986. At its brightest in 1989 the flux density was about 1.1~Jy, however, for the past 15 years the source
has been in a fainter state of around 0.5~Jy or slightly less. The OVRO flux density curve exhibits similar, moderate variability. The source has been 
detected at gamma-rays and shows parsec-scale core-jet structure \citep{2012dammando1}. A superluminal speed of 5.8$c$ was measured by \citet{2016lister}.

For this source we have data only between 15 and 37~GHz, with one additional \textit{Planck} 30~GHz data point in April 2012. The simultaneous radio spectra is convex;
it is inverted between 15--30~GHz, and very steep between 30--37~GHz, however, the 15 -- 37~GHz spectral index calculated without the \textit{Planck} data point is flat.

\textit{J0948$+$0022} 
One of the brightest and most studied NLS1 galaxies \citep[{see, e.g.},][]{2012foschini1,2015foschini1}, this source has been observed at Mets\"{a}hovi 
since 2009. It emits in gamma-rays \citep{2009abdo2}, and shows parsec and kiloparsec-scale jet structures \citep{2016lister,2012doi1} with a superluminal
speed of 11.5$c$. It also varies in the optical regime, and the jet has been suggested as the probable cause \citep{2013maune1}. 
15 and 37~GHz flux density curves show large flares, up to 1.1~Jy at the higher frequency.

The simultaneous radio spectra are shown in Fig.~\ref{fig:j0948epochs} and are in general inverted. The spectra in November 2009 and 2011 May show a high state, 
and subsequent spectra in November 2013 a decreasing trend and October 2014 an apparently low state. Unfortunately the 37~GHz observations in November 2009 and 
October 2014 are both non-detections made in relatively poor weather. The features between 11 and 37~GHz again hint at a possible (superluminal) component moving 
in the jet according to the shock-in-jet model.

\textit{J1505$+$0326}
Observed in Mets\"{a}hovi since 2005 and, despite having been detected at gamma-rays \citep{2009abdo3} and in general being one of the most studied
NLS1 galaxies, this source does not exhibit particularly strong variability at 37~GHz. The flux density has never exceeded 0.7~Jy. Admittedly the number of observations is
relatively low. In contrast, the 15~GHz flux density curve mainly shows moderate variability, incresing in 2014 and ending in a large flare and
a substantial drop in brightness. The drop is also seen in the 37~GHz data following the period presented in this paper. The ground level of the 15~GHz
flux densities is higher than for any of the other sources (around 0.4~Jy until 2013, and around 0.3~Jy after that), with the exception
of the brightening of J0849$+$5108 towards the end of 2013. Parsec-scale features have been observed in this source \citep{2013dammando2}, 
and \citet{2016lister} reports sub-luminal jet features moving at a speed of 1.1$c$.

The simultaneous radio spectra of this source (Fig.~\ref{fig:j1505epochs}) show achromatic variablity: the spectra retain more or less the same shape 
but move up and down exhibiting both a high state (October 2014) and a low state (October 2013 and March 2014). There are no 37~GHz observations in October
2014, but a detection of around 0.5~Jy in April 2014 suggests that the shape remains exactly similar.

\textit{J161259.83$+$421940.3} This source was detected five times. Its radio-loudness is only 24  \citep{2011foschini1}, one of the lowest in the
sample, yet it is one of the brightest detections in our ``non-bright'' category.

\textit{J164442.53$+$261913.2}
Both parsec and kiloparsec-scale jets have been observed in this source \citep{2011doi1,2012doi1} and it has also been detected in gamma-rays 
\citep{2015dammando1}. Three of four 37~GHz detections coincide nicely with the 15~GHz flares (Fig.~\ref{fig:lcJ164442}). For the first 37~GHz 
detection there exists no simultaneous 15~GHz data.
 
The simultaneous radio spectra of this source (Fig.~\ref{fig:j1644epochs}) shows similar variability as that of 1H~0323$+$342. In November 2013 it is at its highest
with an inverted shape due to the 37~GHz detection. In March and April 2014 the spectrum is low, steep, and straight, and the source is not detected at 
37~GHz. In October 2014 it again starts to rise towards the higher frequencies. Unfortunately there are no simultaneous 37~GHz observations for October 
2014, but a detection in September is on the order of 0.3~Jy. The two simple 15 to 37~GHz spectral indices both indicate an inverted 
spectrum, particulary that occurring in September 2014 but not considered simultaneous according to our criteria of one month. It does, nevertheless, 
coincide nicely with the rising multifrequency spectrum in October. There is some structure between 11 and 37~GHz, possibly indicating a component
moving in the jet. However, there are no measurements or superluminal motion in this source so far.

\textit{Other sources}
There are two detections of J110542.72$+$020250.9 and J145041.93$+$591936.9. A kiloparsec-scale jet has been observed in the latter 
\citep{2012doi1}. There is a rather bright radio source close to J110542.72$+$020250.9 which may contribute to the 37~GHz flux 
(see Sect.~\ref{sec:metsahovi-obs} for details); caution is advisable.

Sources
J080535.17$+$302201.7, J103123.73$+$423439.3, J125635.89$+$500852.4, J133345.47$+$414127.7, J150832.91$+$583422.5, J154817.92$+$351128.0, and 
J162901.30$+$400759.9 have all been detected once at 37~GHz.

The flux density of the source J104732.68$+$472532.0 at 15~GHz is consistently around 0.2 Jy, however, it has not been detected at 37~GHz. In
addition to the three observations reported in this paper (until end of April 2015), there are four further 37~GHz observations, all non-detections. Most
other sources detected at 15~GHz but not at 37~GHz have fairly uneventful flux density curves, with the exception of J144318.56$+$472556.7 that 
shows activity towards the end of 2014 and early 2015. Nevertheless, the source was not detected at 37~GHz during this period.

\section{Discussion}
\label{sec:discussion}

Recurrent detections and variability imply that the origin of the radio emission in NLS1 galaxies at 37~GHz is a relativistic jet rather than star formation processes. It is unlikely that star formation activity alone could generate variable radio emission at a minimum level of 200 mJy, observed in distant galaxies at such high radio frequencies, considering that the spectrum of a radio supernova typically turns over at low frequencies and the peak amplitude is of the order of 100~mJy or less.
The radio spectra of the radio-brightest NLS1 galaxies also show similar variability as blazars, i.e., components moving in the jet or achromatic variability
\citep[{see, e.g.},][]{angelakis12,angelakis2015,planck2016-XLV}. It is therefore
expected that jets will be found in all sources detected at 37~GHz. Most of the detected sources in this paper are radio-loud, however, a couple are on
the verge of being radio-quiet. In samples 3 and 4, of which work is in progress and will be published in subsequent papers, we have detections of sources classified
as radio-quiet or even radio-silent; evidently the division of NLS1 galaxies into categories based on one-epoch low frequency radio observations is 
seriously misguided. This has been earlier found true also for other fainter AGN classes such as BLOs and GPS 
sources \citep{nieppola06,nieppola07,nieppola09,tornikoski01_gps,torniainen05,2009tornikoski1}. BLOs  as a class resemble NLS1 galaxies, i.e., they were 
previously assumed to be very faint, particularly those classified as high-energy peaked BLOs (HBLs). However, \citet{nieppola07} studied a large sample 
of BLOs, and it turned out that 34\% of all BLOs and 15\% of HBLs were detected at 37~GHz. The result is similar to ours.

\begin{figure}

        \centering
        \includegraphics[width=0.5\textwidth]{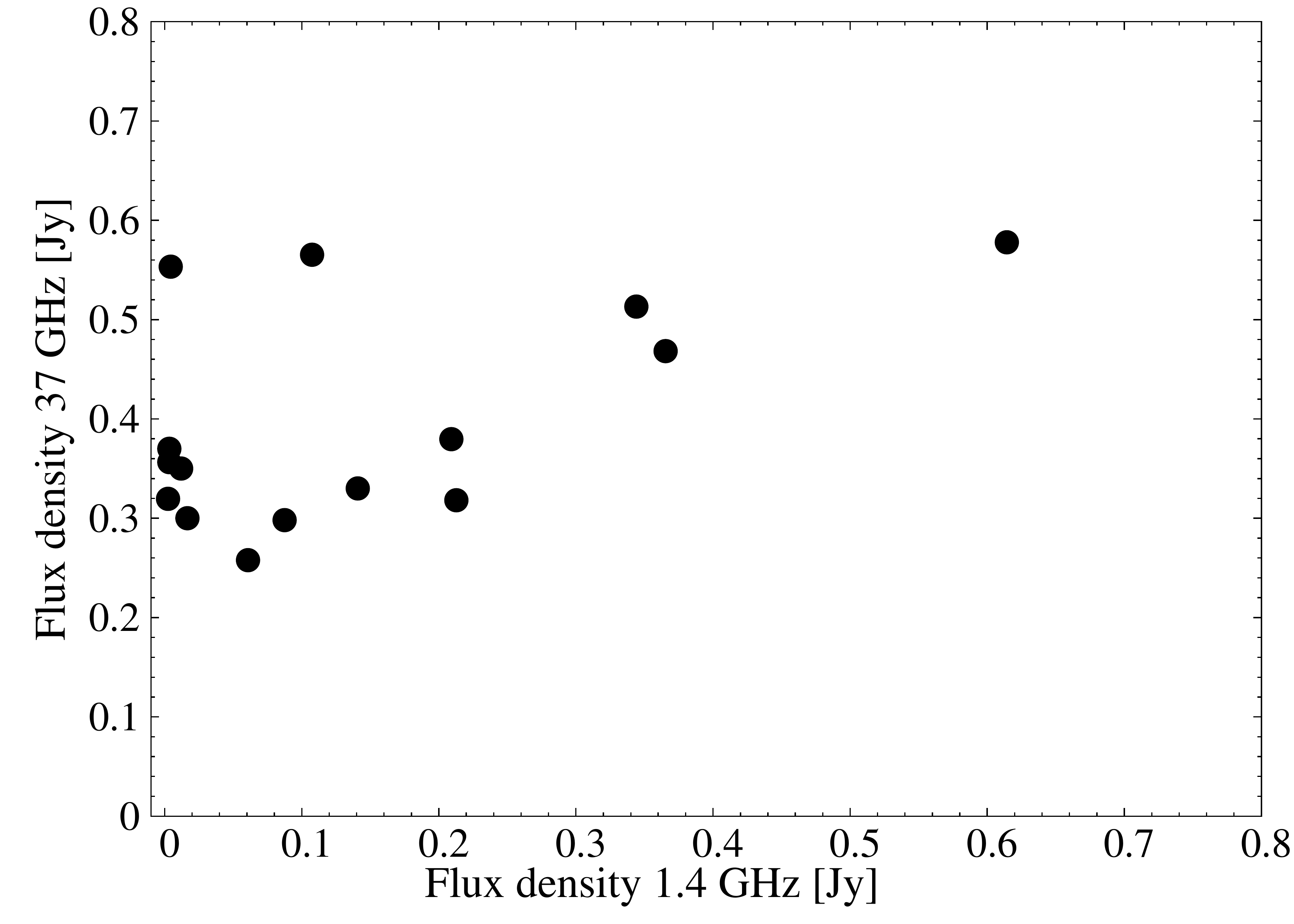}
        \caption{Historical flux density at 1.4~GHz plotted against the average flux density at 37~GHz from this campaign. The values are not simultaneous.}
        \label{fig:radcorr}
\end{figure}

Certainly there may exist a NLS1 type where star formation
activity is the main source of radio emission, particularly at lower frequencies, or jets may not exist at all, or any combination of these at
variable contribution levels \citep[{see, e.g.},][]{2015caccianiga1}. However, we conclude that there are relativistic jets in a much larger
number of NLS1 sources than previously assumed, and that their contribution in many cases is considerable. Whether lower frequency, one-epoch
observations, such as the VLA FIRST or NVSS observations, or radio-loudness estimated based on them, can be used for predicting how bright the
sources are at higher frequencies is dubious. Pearson's product-moment and Spearman's rank correlation coefficients show that the 1.4 and 37~GHz flux
densities for the 15 detected sources are only weakly correlated: $r_{\text{Pearson}}$=0.54, $P_{\text{Pearson}}$=0.04; $r_{\text{Spearman}}$=0.34,
$P_{\text{Spearman}}$=0.22 (the correlation is significant and the null hypothesis rejected if $P$<0.05).
The correlation displayed in Fig.~\ref{fig:radcorr} shows bimodal structure. In some sources the flux
densities appear nicely correlated, for some sources faint at 1.4~GHz the 37~GHz flux density is relatively high. It is the latter that provide
the unanticipated results.

The existence of fully-developed relativistic jets in spirals or in young, evolving galaxies has implications on, for example, models of AGN evolution and jet formation. 
However, some NLS1 source are clearly harboured in interacting and/or irregular systems \citep{2007ohta1, 2014leontavares1}. Usually this division between the 
different types of host galaxies has been made between radio-quiet and radio-loud sources. If we now observe substantial amounts of high-frequency radio emission 
from radio-quiet NLS1 galaxies, the assumptions of the hosts ---or the models--- must be re-examined. Without extensive host galaxy studies of all NLS1 types the 
question will remain open.

Even if NLS1 galaxies challenge our current knowledge of AGN and relativistic jet systems, at the same time they present an opportunity to study them starting 
from an alternative set of initial properties compared to blazars. NLS1 galaxies may, for example, aid us to figure out what kind of 
properties are needed to trigger and maintain AGN activity. It is also clear that the evolutionary lines and unfication scenarios of the various AGN 
populations need closer inspection to achieve a coherent picture.

\section{Conclusions}

We have presented multi-epoch observations of a sample of 78 NLS1 galaxies at 37~GHz, with additional quasi-simultaneous radio data at other frequencies. 
Our main conclusions are the following.
\begin{enumerate}

\item The detection rate at 37~GHz is around 19\%. This includes both radio-loud and radio-quiet sources.
\item The high-frequency radio variability of NLS1 galaxies is substantial. Frequent observations reveal this more effectively. Variability is therefore
expected to increase when more data become available.
\item The radio spectra show variablity from one epoch to another, in some cases resembling that of blazars (i.e., shocks moving in the jet or
  achromatic variability).
\item The average synchrotron peak frequencies of NLS1 galaxies and bright blazars are almost exactly the same.
\item The source of the high-frequency radio emission in sources detected at 37~GHz is likely to be relativistic jets rather than star formation, 
and the number of jets found in NLS1 galaxies is expected to increase significantly as more observations become available.
\end{enumerate}

We will publish the results for the next NLS1 samples at 37~GHz, and all samples at 22~GHz in the near future. In addition to NLS1 galaxies, we are
monitoring a sample of low-luminosity AGN (LLAGN) at 22 and 37~GHz. Follow-up observations of the current sample at various wavelengths are also in progess.

\begin{acknowledgements}
The OVRO 40-m monitoring programme is supported in part by NASA grants NNX08AW31G, NNX11A043G and NNX14AQ89G, and NSF grants AST-0808050 and AST-1109911. 
TH was supported by the Academy of Finland project number 267324.
MGM acknowledges support through the Russian Government Program of Competitive Growth of Kazan Federal University.
This research has made use of the NASA/IPAC Extragalactic Database (NED) which is operated by the Jet Propulsion Laboratory, 
California Institute of Technology, under contract with the National Aeronautics and Space Administration. 

Based on observations obtained with Planck (http://www.esa.int/Planck), an ESA science mission with instruments and contributions 
directly funded by ESA Member States, NASA, and Canada.

\end{acknowledgements}

\bibliographystyle{aa}
\bibliography{artikkeli.bib,RATAN600_nls.bib,Planck_bib.bib,enbib_141016.bib}

\begin{table*}[ht]
\caption[]{Example of the datafile, available at the CDS, containing Mets\"{a}hovi, RATAN-600, and \emph{Planck} observations. Non-detections are shown only for Mets\"{a}hovi data.}
\centering
\begin{tabular}{l l l l l}
\hline\hline
Source name & Frequency  & Time        & Flux density & Error \\
            & (GHz)       & (decimal year) & (Jy)   & (Jy)     \\ \hline
1H0323+342  & 4.8 & 2013.810959 & 0.49 & 0.03 \\
1H0323+342  & 4.8 & 2014.197260 & 0.52 & 0.03 \\
1H0323+342  & 4.8 & 2014.780822 & 0.46 & 0.02 \\
1H0323+342  & 4.8 & 2015.010959 & 0.35 & 0.02 \\
1H0323+342  & 8.2 & 2013.810959 & 0.47 & 0.03 \\
1H0323+342  & 8.2 & 2014.197260 & 0.49 & 0.05 \\
1H0323+342  & 8.2 & 2014.780822 & 0.42 & 0.03 \\
1H0323+342  & 8.2 & 2015.010959 & 0.33 & 0.03 \\
1H0323+342  & 11.2 & 2013.810959 & 0.44 & 0.05 \\ 
1H0323+342  & 11.2 & 2014.197260 & 0.45 & 0.05 \\
1H0323+342  & 11.2 & 2014.780822 & 0.41 & 0.04 \\
1H0323+342  & 11.2 & 2015.010959 & 0.30 & 0.03 \\
1H0323+342  & 21.7 & 2013.810959 & 0.31 & 0.05 \\
1H0323+342  & 21.7 & 2014.197260 & 0.38 & 0.06 \\
1H0323+342  & 21.7 & 2014.780822 & 0.39 & 0.06 \\
1H0323+342  & 21.7 & 2015.010959 & 0.28 & 0.04 \\
1H0323+342  & 37 & 2012.103123 & nd & \\
1H0323+342  & 37 & 2012.103250 & nd & \\
1H0323+342  & 37 & 2012.108219 & nd & \\
1H0323+342  & 37 & 2013.228524 & 0.64 & 0.08 \\
1H0323+342  & 37 & 2013.234888 & 0.46 & 0.07 \\
1H0323+342  & 37 & 2013.515495 & 0.46 & 0.07 \\
1H0323+342  & 37 & 2013.526617 & 0.60 & 0.06 \\
1H0323+342  & 37 & 2013.647215 & 0.40 & 0.06 \\
1H0323+342  & 37 & 2013.781315 & 0.67 & 0.13 \\
1H0323+342  & 37 & 2013.783215 & nd & \\
1H0323+342  & 37 & 2013.837960 & nd & \\
1H0323+342  & 37 & 2013.849637 & 1.05 & 0.15 \\
1H0323+342  & 37 & 2013.871182 & 0.56 & 0.08 \\
1H0323+342  & 37 & 2014.078259 & 0.80 & 0.07 \\
1H0323+342  & 37 & 2014.207397 & 0.59 & 0.07 \\
1H0323+342  & 37 & 2014.592094 & nd & \\
1H0323+342  & 37 & 2014.695785 & 0.39 & 0.07 \\
1H0323+342  & 37 & 2014.739778 & 0.50 & 0.07 \\
1H0323+342  & 37 & 2014.887168 & nd  & \\
1H0323+342  & 37 & 2015.125504 & 0.53 & 0.06 \\
1H0323+342  & 37 & 2015.160275 & nd  & \\
1H0323+342  & 37 & 2015.199265 & 0.45 & 0.06 \\
1H0323+342  & 70 & 2010.654795 & 1.19 & 0.26 \\
1H0323+342  & 70 & 2011.128767 & 1.12 & 0.28 \\
1H0323+342  & 143 & 2009.893200* & 0.75 & 0.15 \\
1H0323+342  & 143 & 2010.893200* & 0.82 & 0.18 \\
1H0323+342  & 217 & 2009.660274 & 0.46 & 0.10 \\
1H0323+342  & 217 & 2010.660274 & 0.70 & 0.13 \\

\end{tabular}
\tablefoot{Column 1: source name; Col. 2: frequency; Col. 3: date. For RATAN-600 this is the average date of the observations stacked together. This also applies to those \emph{Planck} observations marked with an asterisk. Col. 4: flux density. RATAN-600 flux densities are averages of multiple observations. This also applies to those \emph{Planck} observations marked with an asterisk in Col. 3. Col. 5: error in the flux density.}
\label{tab:data1h0323}
\end{table*}

\begin{appendix}

\section{Additional tables and figures}

\begin{table*}[ht]
\caption[]{Sources with additional data from RATAN-600, OVRO, and \textit{Planck}. Detections (det) at any frequency and non-detections (nd) are marked separately.}
\centering
\begin{tabular}{l l l l l}
\hline\hline
Source                      & Mets\"{a}hovi & RATAN-600  & OVRO       & \textit{Planck} \\ \hline
1H 0323$+$342               & det & det & det        & det \\
FBQS J0713$+$3820           & nd  & det & \dots      & \dots  \\
SDSS J075800.05$+$392029.0  & nd & det & \dots      & \dots  \\
SDSS J080409.23$+$385348.8  & nd & det & \dots   &   \dots  \\
SDSS J081432.11$+$560956.6  & nd & nd  & det        & \dots \\
SDSS J084957.97$+$510829.0  & det & \dots      & det        & det \\
SDSS J085001.17$+$462600.5  & nd & det & \dots      & \dots  \\
SDSS J090227.16$+$044309.5  & nd & det & det        & \dots \\
IRAS 09426$+$1929           & nd & det & \dots      & \dots \\
SDSS J094857.31$+$002225.4  & det & det & det        & det \\
SDSS J095317.09$+$283601.5  & nd & det & det        & \dots \\
SDSS J103123.73$+$423439.3  & det & det & \dots      & nd \\
SDSS J103727.45$+$003635.6  & nd & det & \dots      & \dots  \\
SDSS J104732.68$+$472532.0  & nd & det & det        & \dots  \\
SDSS J104816.58$+$222239.0  & nd & det & \dots      & \dots \\
SDSS J111005.03$+$365336.3  & nd & det & \dots      & \dots  \\
SDSS J111438.89$+$324133.4  & nd & det & \dots      & \dots \\
SDSS J113824.54$+$365327.1  & nd & det & \dots      & \dots  \\
SDSS J114654.28$+$323652.3  & nd & det & \dots      & \dots  \\
SDSS J115917.32$+$283814.5  & nd & nd  & \dots      & \dots  \\
SDSS J122749.14$+$321458.9  & nd & det & \dots      & \dots  \\
SDSS J123852.12$+$394227.8  & nd & det & \dots      & \dots  \\
SDSS J124634.65$+$023809.0  & nd & det & det        & \dots \\
SDSS J130522.75$+$511640.3  & nd & det & \dots      & \dots \\
SDSS J142114.05$+$282452.8  & nd & det & \dots      & \dots  \\
SDSS J143509.49$+$313147.8  & nd & det & det        & \dots \\
SDSS J144318.56$+$472556.7  & nd & det & det        & \dots \\
SDSS J150506.47$+$032630.8  & det & det & det        & nd \\
SDSS J154817.92$+$351128.0  & det & det & det        & nd \\
SDSS J161259.83$+$421940.3  & det & det & \dots      & nd \\
SDSS J162901.30$+$400759.9  & det & det & det        & nd \\
SDSS J163323.58$+$471858.9  & nd & det & \dots      & \dots  \\
SDSS J164442.53$+$261913.2  & det & det & det        & nd \\
SDSS J170330.38$+$454047.1  & nd & det & \dots      & \dots  \\
SDSS J172206.03$+$565451.6  & nd & \dots      & det        & \dots \\
RX J2314.9$+$2243           & nd & det & \dots      & \dots \\
\end{tabular}
\label{tab:otherdata}
\end{table*}

\begin{table*}[ht]
\caption[]{Spectral indices.}
\centering
\begin{tabular}{l l l l l}
\hline\hline
Source                   & Start time & End time   & Frequency range (GHz) & $\alpha$    \\ \hline
1H0323+342               & 2009-08-16 & 2009-09-12 & 15-217                & -0.2048 $\pm$ 0.1946     \\    
                         & 2010-08-19 & 2010-09-10 & 15-70                 &  0.6535 $\pm$ 0.3257   \\
                         & 2010-08-19 & 2010-09-10 & 70-217                & -0.4690 $\pm$ 0.5788   \\                                                    
                         & 2011-02-12 & 2011-03-02 & 15-70                 & 0.6672 $\pm$ 0.3716    \\                                                    
                         & 2013-03-13 & 2013-03-25 & 15-37                 & 0.3103 $\pm$ 0.3476    \\                                                    
                         & 2013-03-27 & 2013-04-07 & 15-37                 & -0.1916 $\pm$ 0.4008   \\                                                    
                         & 2013-06-28 & 2013-07-08 & 15-37                 & 0.5884 $\pm$ 0.3940    \\                                                    
                         & 2013-07-12 & 2013-07-19 & 15-37                 & 0.8388 $\pm$ 0.2673    \\                                                    
                         & 2013-08-12 & 2013-09-06 & 15-37                 & 0.1461 $\pm$ 0.3951    \\                                                    
                         & 2013-10-02 & 2013-10-27 & 4.8-21.7              & -0.3097 $\pm$ 0.2431    \\                                                    
                         & 2013-10-02 & 2013-10-27 & 21.7-37               & 1.4504 $\pm$ 1.0548    \\
                         & 2013-10-02 & 2013-10-27 & 15-37                 & 0.4523* $\pm$ 0.4985    \\                                                    
                         & 2013-10-30 & 2013-11-07 & 15-37                 & 0.9398 $\pm$ 0.4018    \\                                                    
                         & 2013-11-12 & 2013-11-18 & 15-37                 & 0.1751 $\pm$ 0.3712    \\                                                    
                         & 2014-01-15 & 2014-02-06 & 15-37                 & 0.4988 $\pm$ 0.2362    \\                                                    
                         & 2014-03-13 & 2014-03-20 & 4.8-21.7              & -0.2019 $\pm$ 0.2515   \\                                                    
                         & 2014-03-13 & 2014-03-20 & 21.7-37               & 0.8121 $\pm$ 0.8254   \\
                         & 2014-03-13 & 2014-03-20 & 15-37                 & 0.3616* $\pm$ 0.3054   \\                                                    
                         & 2014-09-06 & 2014-09-16 & 15-37                 & -0.0856 $\pm$ 0.4642   \\                                                    
                         & 2014-09-28 & 2014-10-13 & 4.8-21.7              & -0.1107 $\pm$ 0.2382   \\                                                              
                         & 2014-09-28 & 2014-10-13 & 21.7-37               & 0.4691 $\pm$ 0.8635   \\
                         & 2014-09-28 & 2014-10-13 & 15-37                 & 0.2933* $\pm$ 0.3798    \\                                                              
                         & 2014-12-30 & 2015-01-15 & 4.8-21.7              & -0.1441 $\pm$ 0.2516   \\                                                    
                         & 2015-02-11 & 2015-02-26 & 15-37                 & 0.5362 $\pm$ 0.3007    \\                                                    
                         & 2015-03-05 & 2015-03-27 & 15-37                 & 0.4199 $\pm$ 0.3615    \\                                                    
                         &            &            &                       &             \\                                                    
SDSS J084957.97+510829.0 & 2012-04-17 & 2012-05-07 & 15-30                 & 0.6589 $\pm$  0.7835   \\                                                    
                         & 2012-04-17 & 2012-05-07 & 30-37                 & -1.8716 $\pm$ 3.2229   \\
                         & 2012-04-17 & 2012-05-07 & 15-37                 & 0.0711* $\pm$ 0.4469    \\                                                    
                         &            &            &                       &             \\                                                    
SDSS J085001.17+462600.5 & 2015-01-19 & 2015-01-19 & 4.8-11.2              & -0.1027 $\pm$ 0.7015   \\                                                    
                         &            &            &                       &             \\                                                    
SDSS J090227.16+044309.5 & 2013-10-22 & 2013-11-17 & 8.2-15                & 0.2343 $\pm$ 0.4565    \\                                                    
                         & 2013-10-22 & 2013-11-17 & 15-21                 & -0.9772 $\pm$ 1.0440   \\                                                    
                         & 2014-04-16 & 2014-05-02 & 4.8-15                & -0.2117 $\pm$ 0.3746   \\                                                    
                         &            &            &                       &             \\
SDSS J094857.31+002225.4 & 2009-11-12 & 2009-12-01 & 15-44                 & 1.4266 $\pm$ 0.4285    \\
                         & 2009-11-12 & 2009-12-01 & 44-217                & -0.4509 $\pm$ 0.4030   \\
                         & 2010-05-10 & 2010-05-24 & 15-70                 & 0.8229 $\pm$ 0.3000    \\
                         & 2011-05-04 & 2011-05-21 & 15-44                 & 0.6112 $\pm$ 0.5137    \\
                         & 2011-05-04 & 2011-05-21 & 15-37                 & 0.2696* $\pm$ 0.3955   \\
                         & 2013-10-19 & 2013-11-11 & 4.8-37                & 0.4976 $\pm$ 0.1161    \\
                         & 2014-10-07 & 2014-10-11 & 4.8-15                & -0.2793 $\pm$ 0.2385   \\
                         &            &            &                       &             \\
SDSS 095317.09+283601.5  & 2014-03-19 & 2014-03-22 & 4.8-15                & -0.0901 $\pm$ 0.4305    \\
                         & 2015-01-23 & 2015-01-29 & 4.8-15                & 0.7792 $\pm$ 0.4622    \\
                         &            &            &                       &             \\
SDSS J103727.45+003635.6 & 2014-03-19 & 2014-03-19 & 4.8-8.2               & -0.7571 $\pm$ 1.3106   \\
                         &            &            &                       &             \\
SDSS J104732.68+472532.0 & 2014-04-02 & 2014-04-24 & 4.8-21.7              & -0.4315 $\pm$ 0.2417   \\
                         &            &            &                       &             \\
SDSS J111438.89+324133.4 & 2015-01-24 & 2015-01-24 & 4.8-11.2              & -0.2368 $\pm$ 0.6845   \\
                         &            &            &                       &             \\
SDSS J114654.28+323652.3 & 2014-04-21 & 2014-04-21 & 4.8-11.2              & 0.9254 $\pm$ 0.6038    \\
                         &            &            &                       &             \\
SDSS J123852.12+394227.8 & 2014-04-25 & 2014-04-25 & 4.8-8.2               & 0.2610 $\pm$ 1.0188    \\                   
                         &            &            &                       &             \\
\end{tabular}
\tablefoot{In the last column an asterisk denotes a 15 -- 37~GHz index that has been calculated separately, in
case an index was originally calculated with more than the two data points for the same epoch.}
\label{tab:specind}
\end{table*}

\begin{table*}[ht]
\centering
\begin{tabular}{l l l l l}
\hline\hline
Source                   & Start time & End time   & Frequency range (GHz) & $\alpha$    \\ \hline
SDSS J124634.65+023809.0 & 2014-03-19 & 2014-03-22 & 4.8-15                & -0.3077 $\pm$ 0.4880   \\
                         &            &            &                       &             \\
SDSS J142114.05+282452.8 & 2013-10-30 & 2013-10-30 & 8.2-11.2              & -0.6530 $\pm$ 1.3842   \\
                         & 2014-03-17 & 2014-03-17 & 4.8-11.2              & -0.4785 $\pm$ 0.5824   \\
                         &            &            &                       &             \\      
SDSS J144318.56+472556.7 & 2014-04-08 & 2014-05-02 & 4.8-21.7              & -0.3664 $\pm$ 0.4545   \\
                         &            &            &                       &             \\
SDSS J150506.47+032630.8 & 2009-04-20 & 2009-05-05 & 15-37                 & 0.0308 $\pm$ 0.2764    \\
                         & 2012-03-03 & 2012-03-27 & 15-37                 & -0.1528 $\pm$ 0.3856   \\
                         & 2012-04-17 & 2012-05-14 & 15-37                 & -0.5908 $\pm$ 0.3670   \\
                         & 2013-03-25 & 2013-04-06 & 15-37                 & -0.3488 $\pm$ 0.4568   \\
                         & 2013-06-04 & 2013-06-29 & 15-37                 & -0.3342 $\pm$ 0.5830   \\
                         & 2013-08-16 & 2013-08-30 & 15-37                 & -0.0378 $\pm$ 0.4888   \\
                         & 2013-10-17 & 2013-11-13 & 4.8-21.7              & -0.1280 $\pm$ 0.2457   \\
                         & 2013-10-17 & 2013-11-13 & 21.7-37               & 0.5775 $\pm$ 1.2192    \\
                         & 2013-10-17 & 2013-11-13 & 15-37                 & 0.0036* $\pm$ 0.6115    \\
                         & 2014-03-10 & 2014-03-25 & 4.8-21.7              & -0.1523 $\pm$ 0.2410   \\
                         & 2014-04-07 & 2014-05-04 & 15-37                 & -0.0798 $\pm$ 0.2698   \\
                         & 2014-10-03 & 2014-10-18 & 4.8-21.7              & 0.0578 $\pm$ 0.2387   \\
                         &            &            &                       &             \\
SDSS J154817.92+351128.0 & 2013-10-20 & 2013-11-16 & 8.2-37                & 1.2934 $\pm$ 0.4226    \\
                         & 2013-10-20 & 2013-11-16 & 15-37                 & 2.2020* $\pm$ 0.5493   \\
                         & 2014-03-12 & 2014-03-25 & 4.8-21.7              & 0.0108 $\pm$ 0.3080    \\
                         &            &            &                       &             \\
SDSS J162901.30+400759.9 & 2012-03-18 & 2012-03-27 & 15-37                 & 2.4994 $\pm$ 0.4790    \\
                         & 2013-10-20 & 2013-11-11 & 4.8-21.7              & -0.4370 $\pm$ 0.3391   \\
                         & 2014-04-13 & 2014-04-20 & 4.8-15                & 0.0163 $\pm$ 0.2178    \\
                         & 2014-10-09 & 2014-10-09 & 4.8-15                & 0.2004 $\pm$ 0.4268    \\
                         &            &            &                       &             \\
SDSS J163323.58+471858.9 & 2014-03-20 & 2014-03-20 & 4.8-8.2               & -0.0612 $\pm$ 0.6844   \\
                         &            &            &                       &             \\
SDSS J164442.53+261913.2 & 2013-09-28 & 2013-10-19 & 15-37                 & 0.9453 $\pm$ 0.4489    \\
                         & 2013-10-27 & 2013-11-09 & 4.8-37                & 0.6737 $\pm$ 0.2782    \\
                         & 2013-10-27 & 2013-11-09 & 15-37                 & 0.9880* $\pm$ 0.6073   \\
                         & 2014-03-19 & 2014-03-25 & 4.8-21.7              & -0.3386 $\pm$ 0.2635   \\
                         & 2014-04-08 & 2014-04-27 & 4.8-21.7              & -0.3738 $\pm$ 0.3101   \\
                         & 2014-08-28 & 2014-09-10 & 15-37                 & 2.0471 $\pm$ 0.5583     \\
                         & 2014-09-28 & 2014-10-10 & 4.8-15                & 0.4075 $\pm$ 0.1791    \\
                         &            &            &                       &             \\
SDSS J170330.38+454047.1 & 2013-10-30 & 2013-10-30 & 4.8-11.2              & -1.5987 $\pm$ 0.6576   \\
                         & 2014-03-20 & 2014-03-20 & 4.8-11.2              & -0.0387 $\pm$ 0.4932   \\
                         & 2014-04-06 & 2014-04-06 & 4.8-8.2               & -0.5551 $\pm$ 0.8613   \\
\end{tabular}
\end{table*}

\begin{table*}[ht]
\caption[]{Spectral indices in the range of 15--37~GHz for sources with a large number of values.}
\centering
\begin{tabular}{l l l l l l l l l}
\hline\hline
Source                   & Start time & End time   & N  & Min     & Max    & Mean   & Median & SEM    \\ \hline
SDSS J084957.97+510829.0 & 2008-12-20 & 2015-03-24 & 26 & -0.1242 & 1.1019 & 0.4337 & 0.4322 & 0.0549 \\
SDSS J094857.31+002225.4 & 2009-04-02 & 2015-04-15 & 45 & -0.3486 & 0.9264 & 0.1961 & 0.1088 & 0.0476 \\
\end{tabular}
\label{tab:specmean}
\end{table*}


\begin{figure*}[ht!]
\centering
\begin{minipage}{0.47\textwidth}
\centering
\includegraphics[width=1\textwidth]{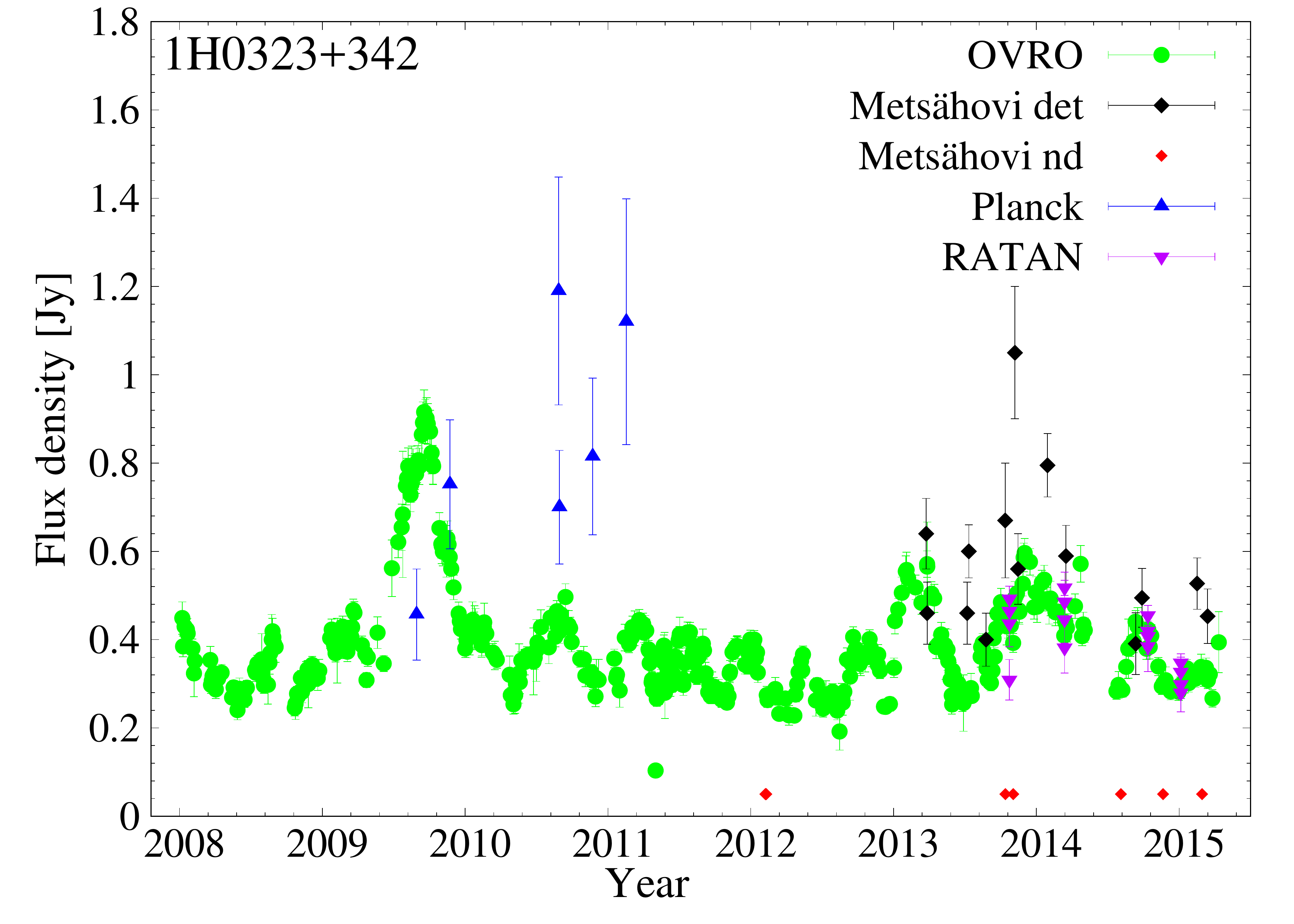}
\caption{Flux density curve of 1H 0323+342.} \label{fig:lc1H0323}
\end{minipage}\hfill
\begin{minipage}{0.47\textwidth}
\centering
\includegraphics[width=1\textwidth]{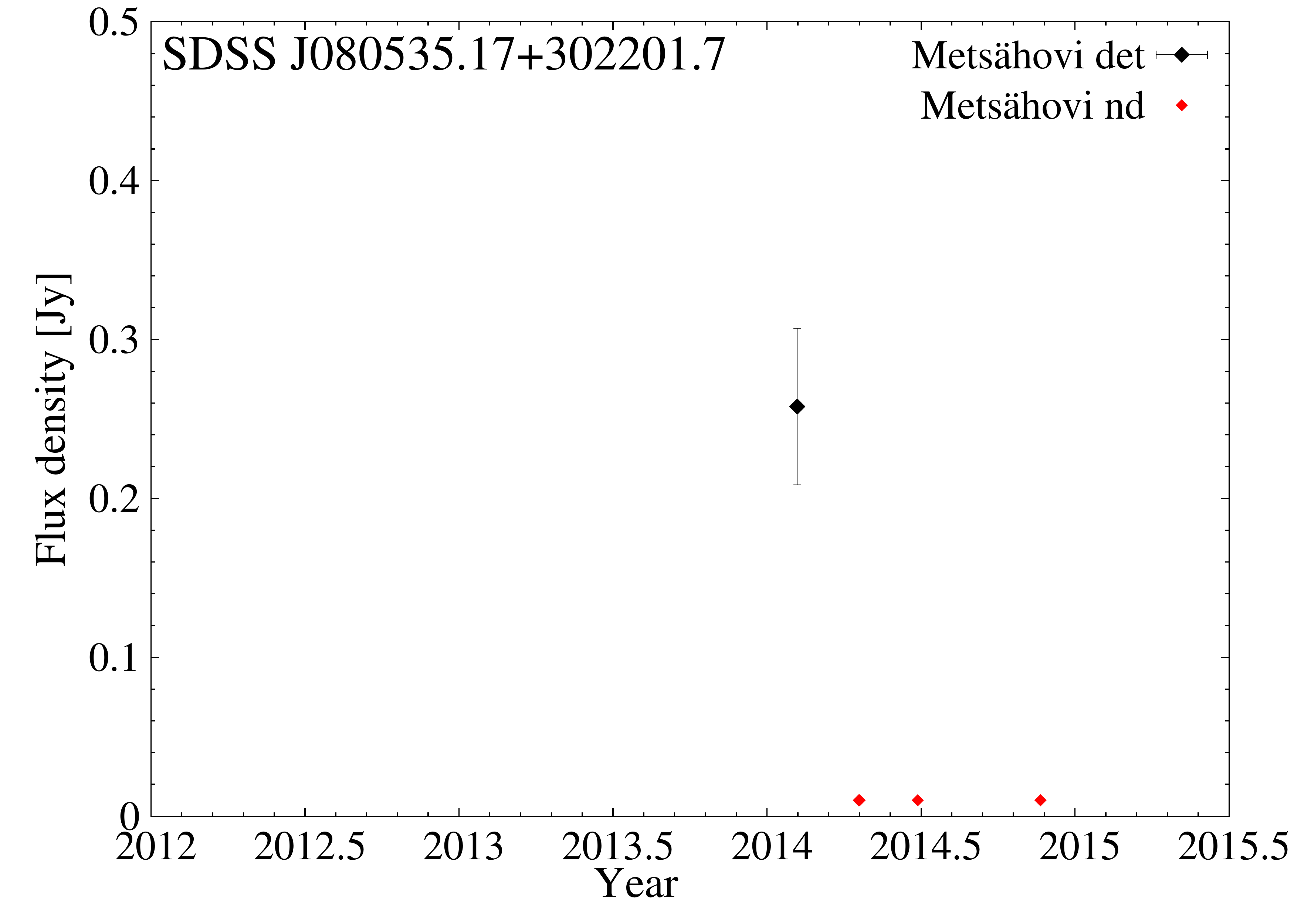}
\caption{Flux density curve of SDSS J080535.17+302201.7.} \label{fig:lcJ080535}
\end{minipage}
\end{figure*}

\begin{figure*}[ht!]
\centering
\begin{minipage}{0.47\textwidth}
\centering
\includegraphics[width=1\textwidth]{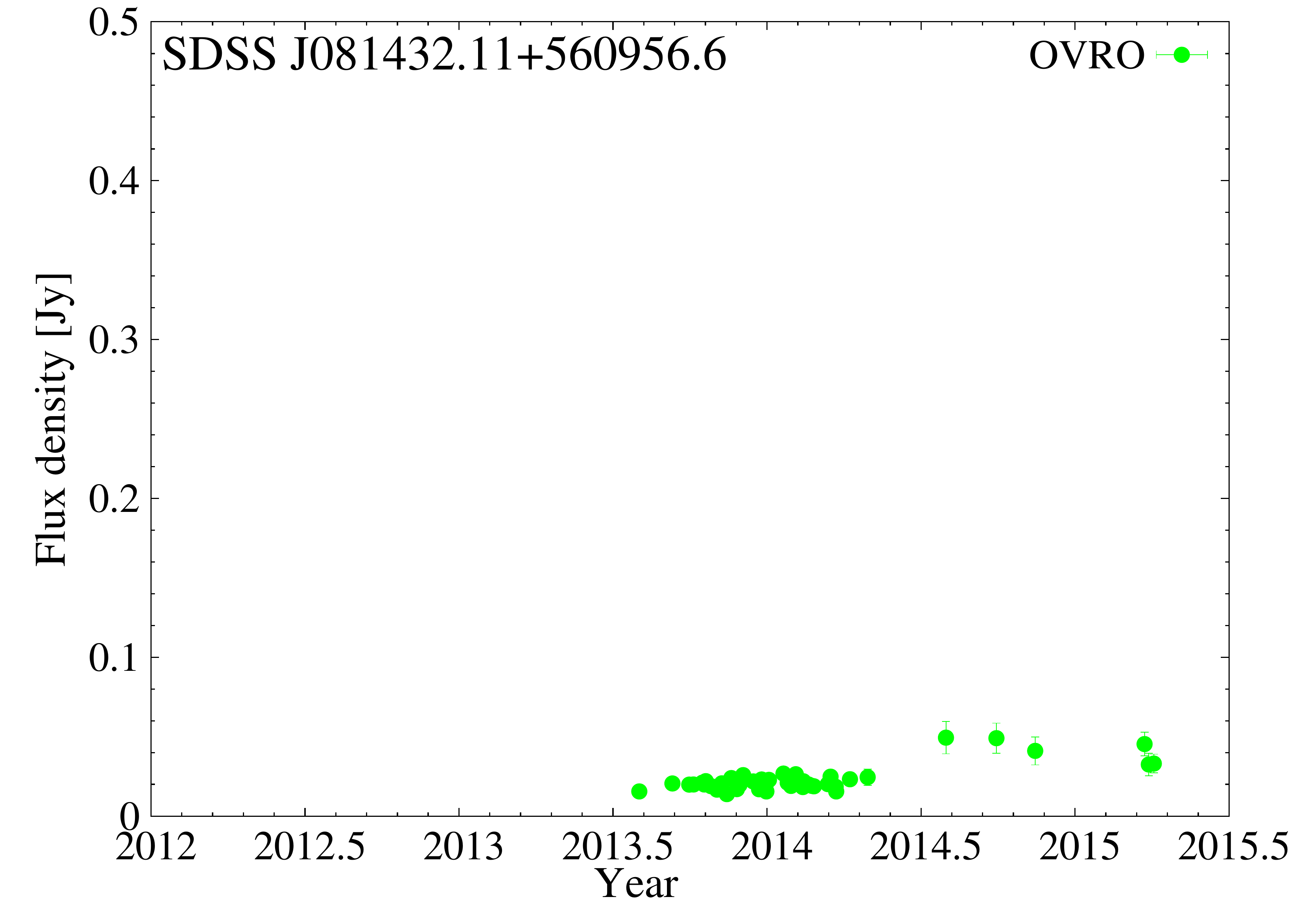}
\caption{Flux density curve of SDSS J081432.11+560956.6.} \label{fig:lcJ081432}
\end{minipage}\hfill
\begin{minipage}{0.47\textwidth}
\centering
\includegraphics[width=1\textwidth]{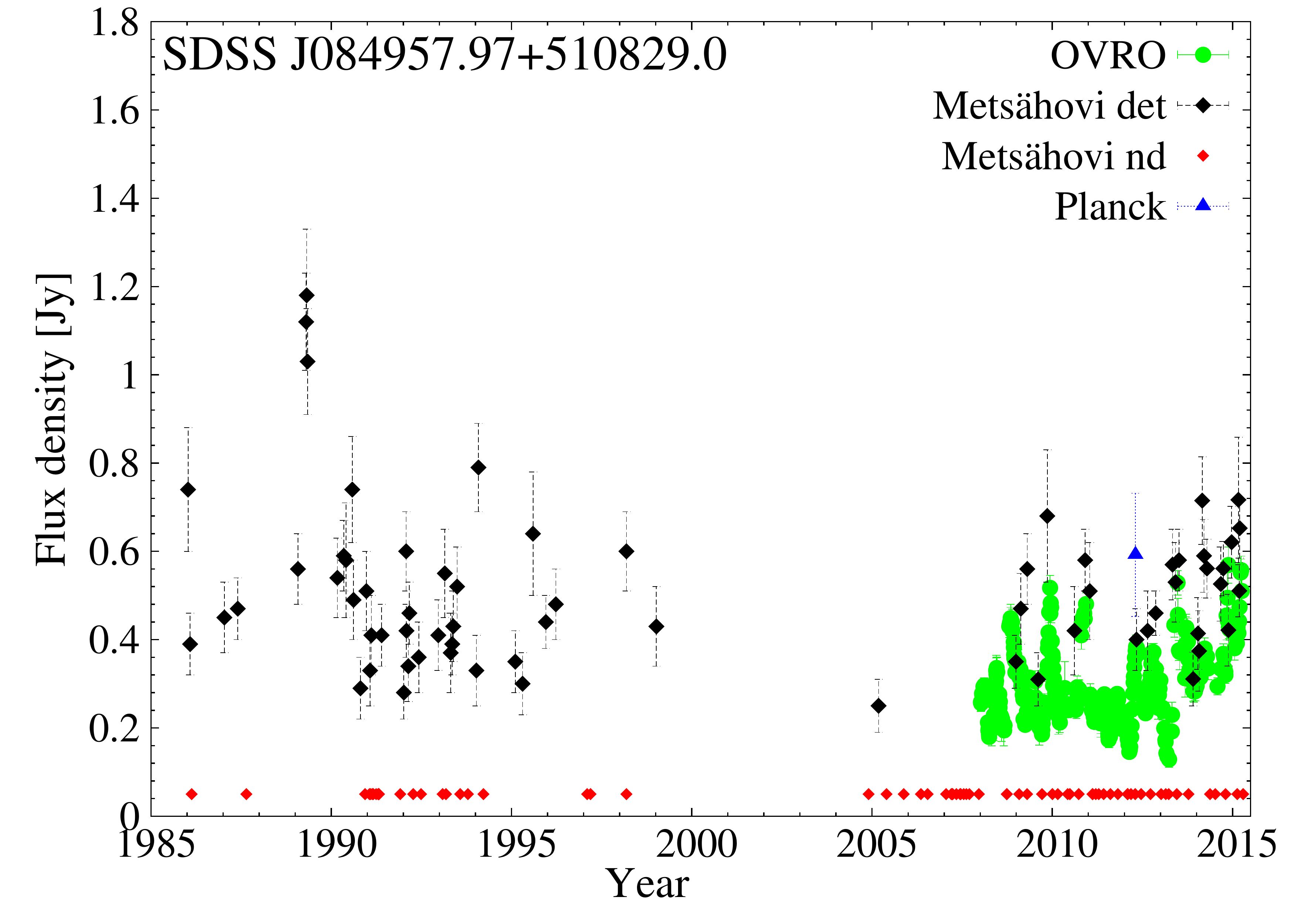}
\caption{Flux density curve of SDSS J084957.97+510829.0.} \label{fig:lcJ084957v1}
\end{minipage}
\end{figure*}

\begin{figure*}[ht!]
\centering
\begin{minipage}{0.47\textwidth}
\centering
\includegraphics[width=1\textwidth]{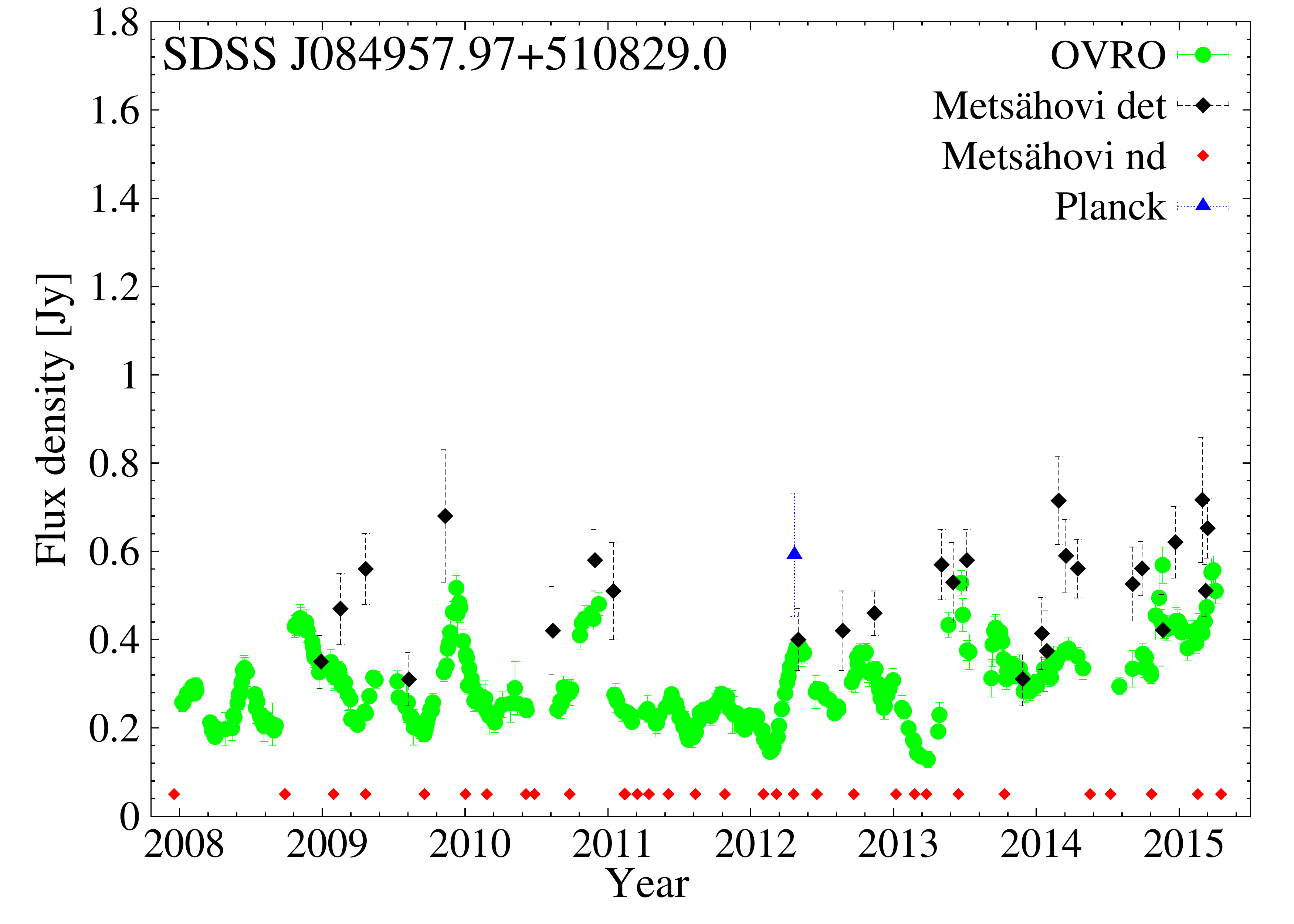}
\caption{Zoom-in to the flux density curve of SDSS J084957.97+510829.0.} \label{fig:lcJ084957v2}
\end{minipage}\hfill
\begin{minipage}{0.47\textwidth}
\centering
\includegraphics[width=1\textwidth]{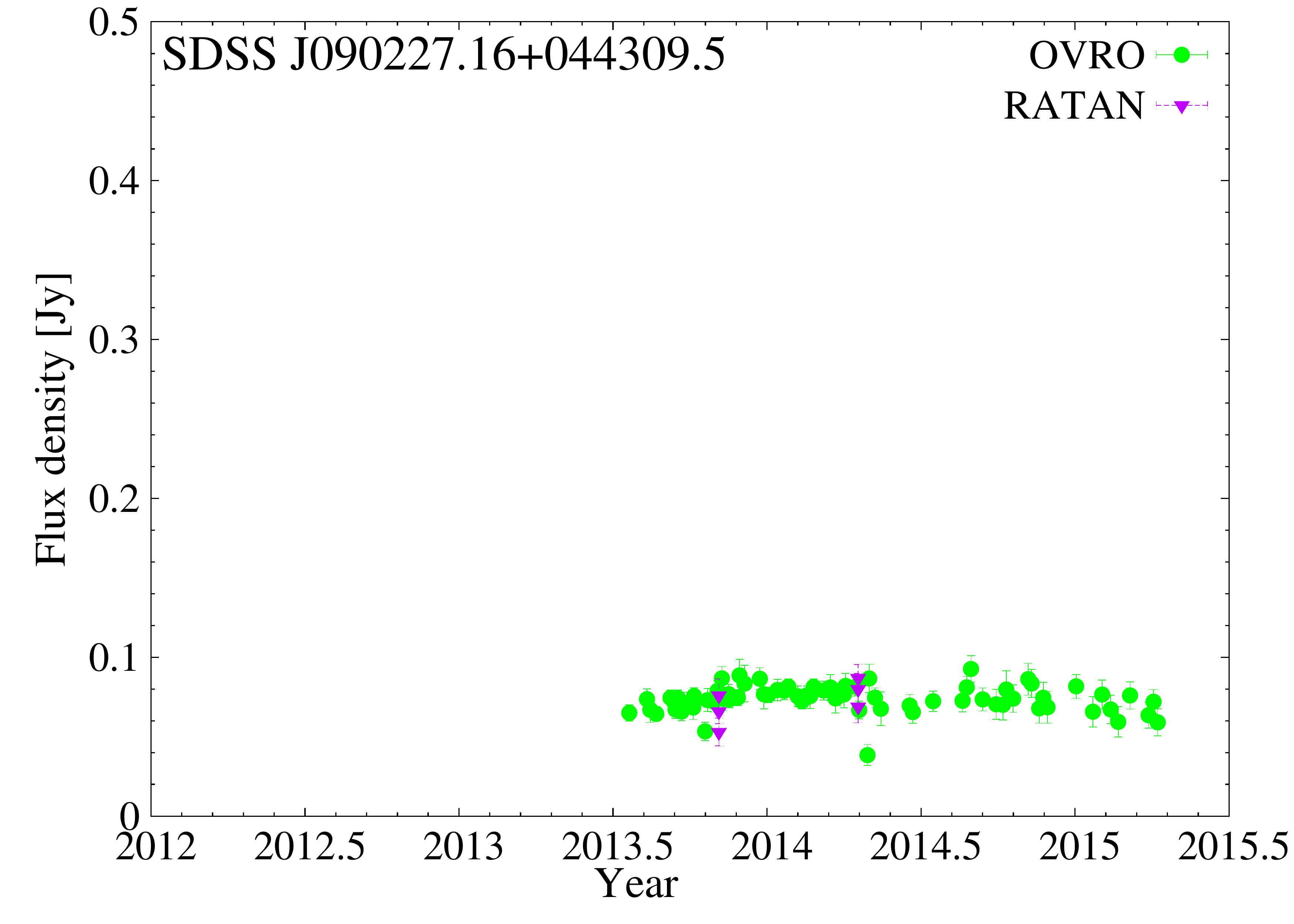}
\caption{Flux density curve of SDSS J090227.16+044309.5. } \label{fig:lcJ090227}
\end{minipage}
\end{figure*}

\clearpage

\begin{figure*}[ht!]
\centering
\begin{minipage}{0.47\textwidth}
\centering
\includegraphics[width=1\textwidth]{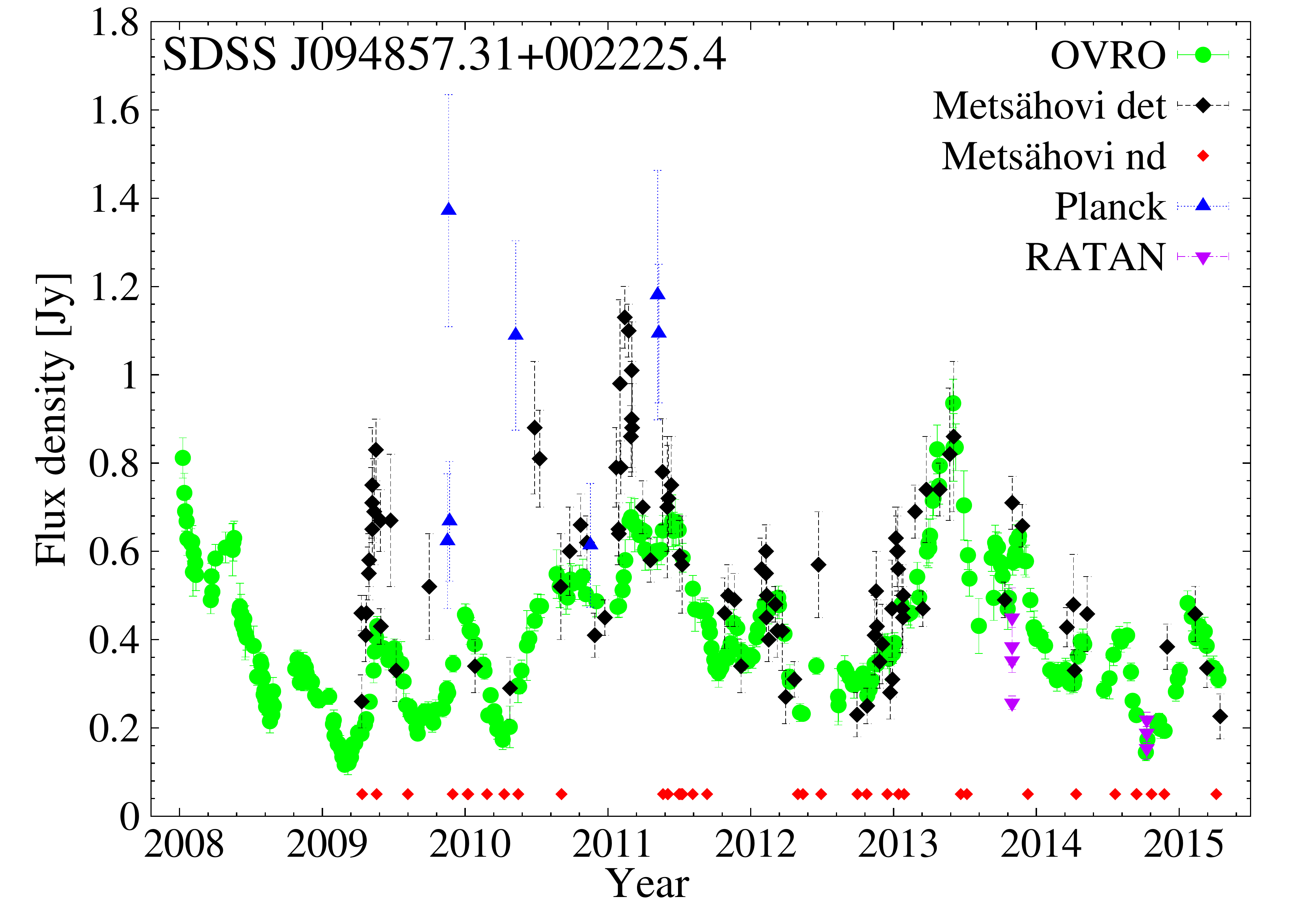}
\caption{Flux density curve of SDSS J094857.31+002225.4. } \label{fig:lcJ094857}
\end{minipage}\hfill
\begin{minipage}{0.47\textwidth}
\centering
\includegraphics[width=1\textwidth]{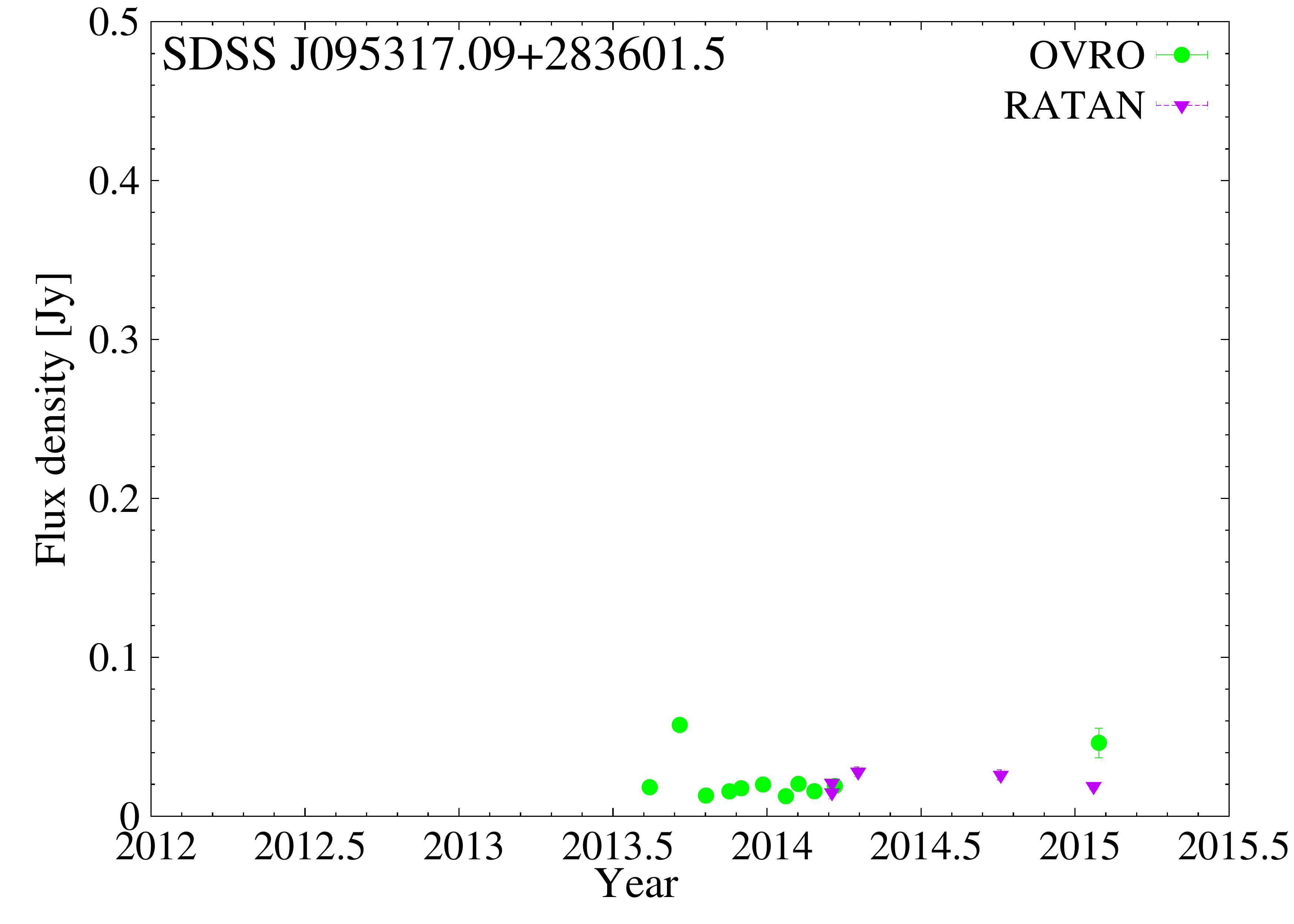}
\caption{Flux density curve of SDSS J095317.09+283601.5. } \label{fig:lcJ095317}
\end{minipage}
\end{figure*}

\begin{figure*}[ht!]
\centering
\begin{minipage}{0.47\textwidth}
\centering
\includegraphics[width=1\textwidth]{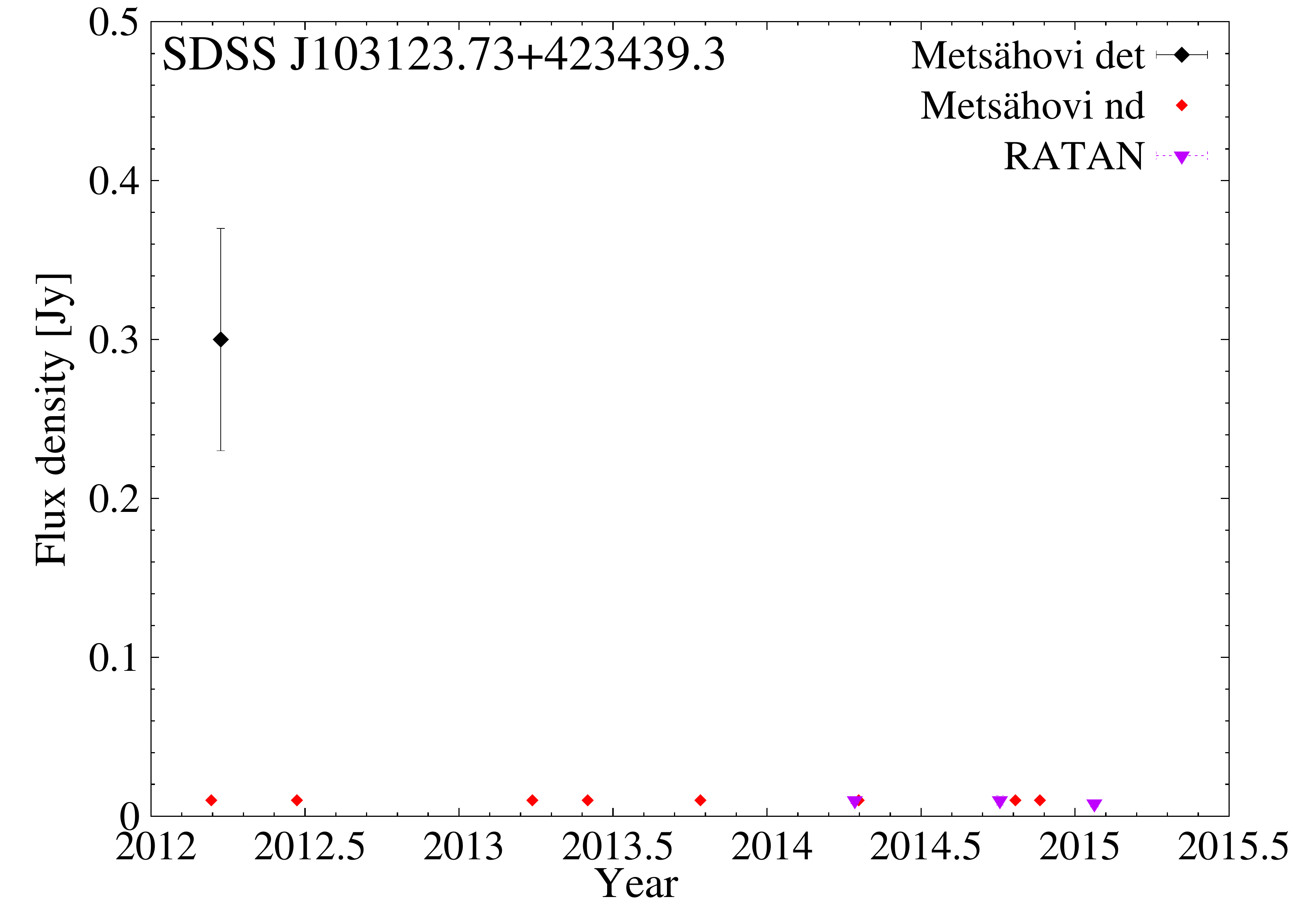}
\caption{Flux density curve of SDSS J103123.73+423439.3. } \label{fig:lcJ103123}
\end{minipage}\hfill
\begin{minipage}{0.47\textwidth}
\centering
\includegraphics[width=1\textwidth]{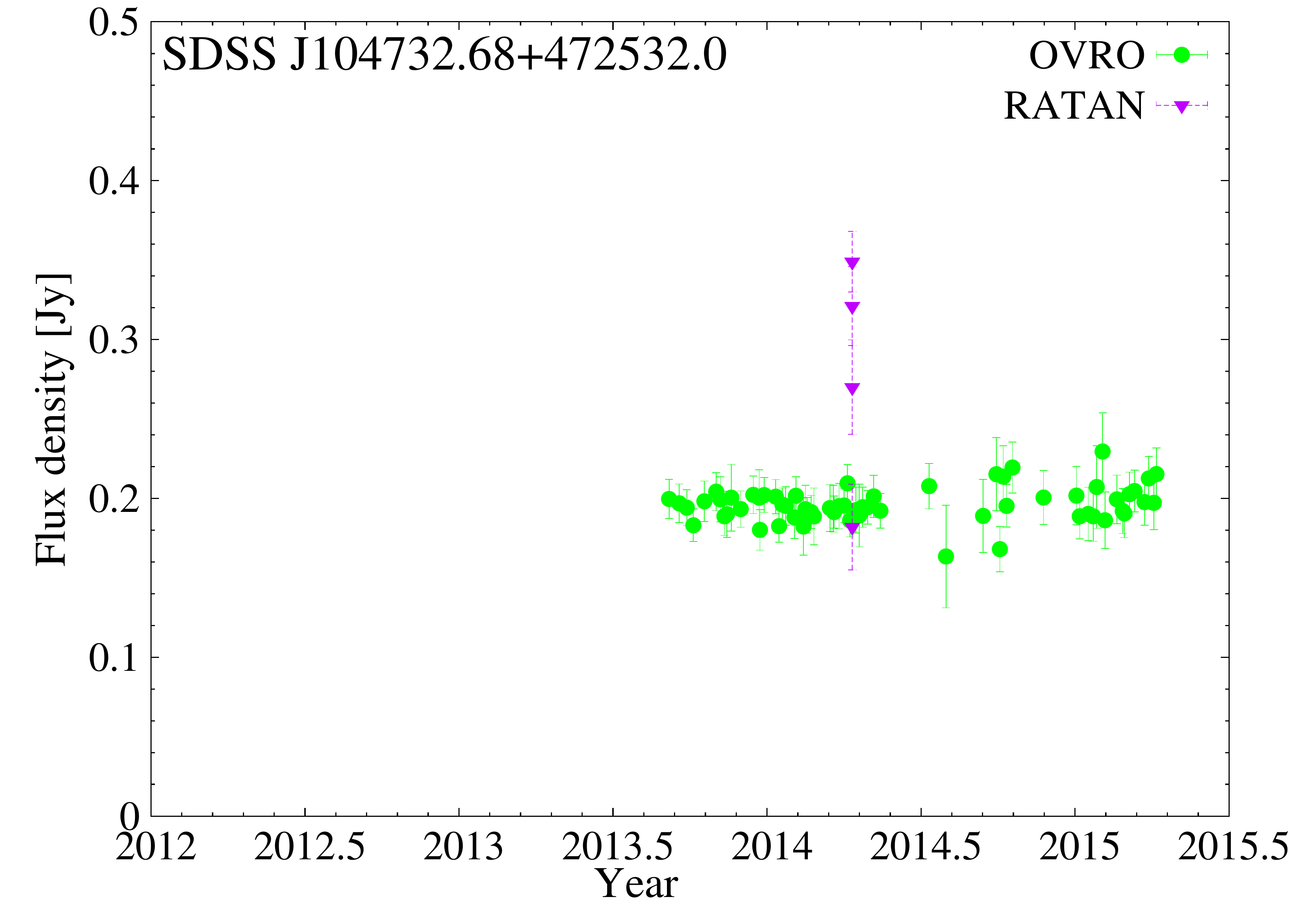}
\caption{Flux density curve of SDSS J104732.68+472532.0. } \label{fig:lcJ104732}
\end{minipage}
\end{figure*}

\begin{figure*}[ht!]
\centering
\begin{minipage}{0.47\textwidth}
\centering
\includegraphics[width=1\textwidth]{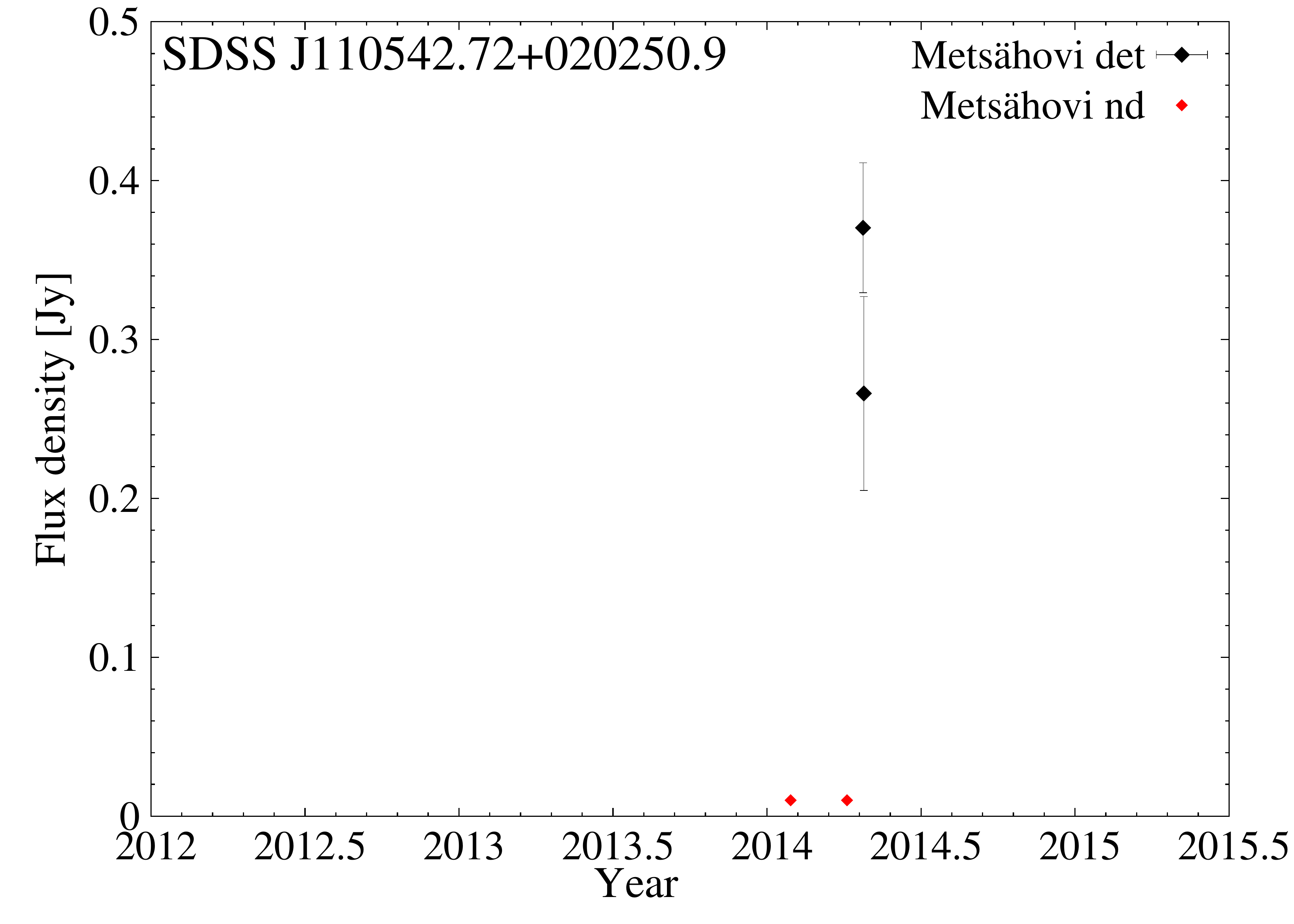}
\caption{Flux density curve of SDSS J110542.72+020250.9. } \label{fig:lcJ110542}
\end{minipage}\hfill
\begin{minipage}{0.47\textwidth}
\centering
\includegraphics[width=1\textwidth]{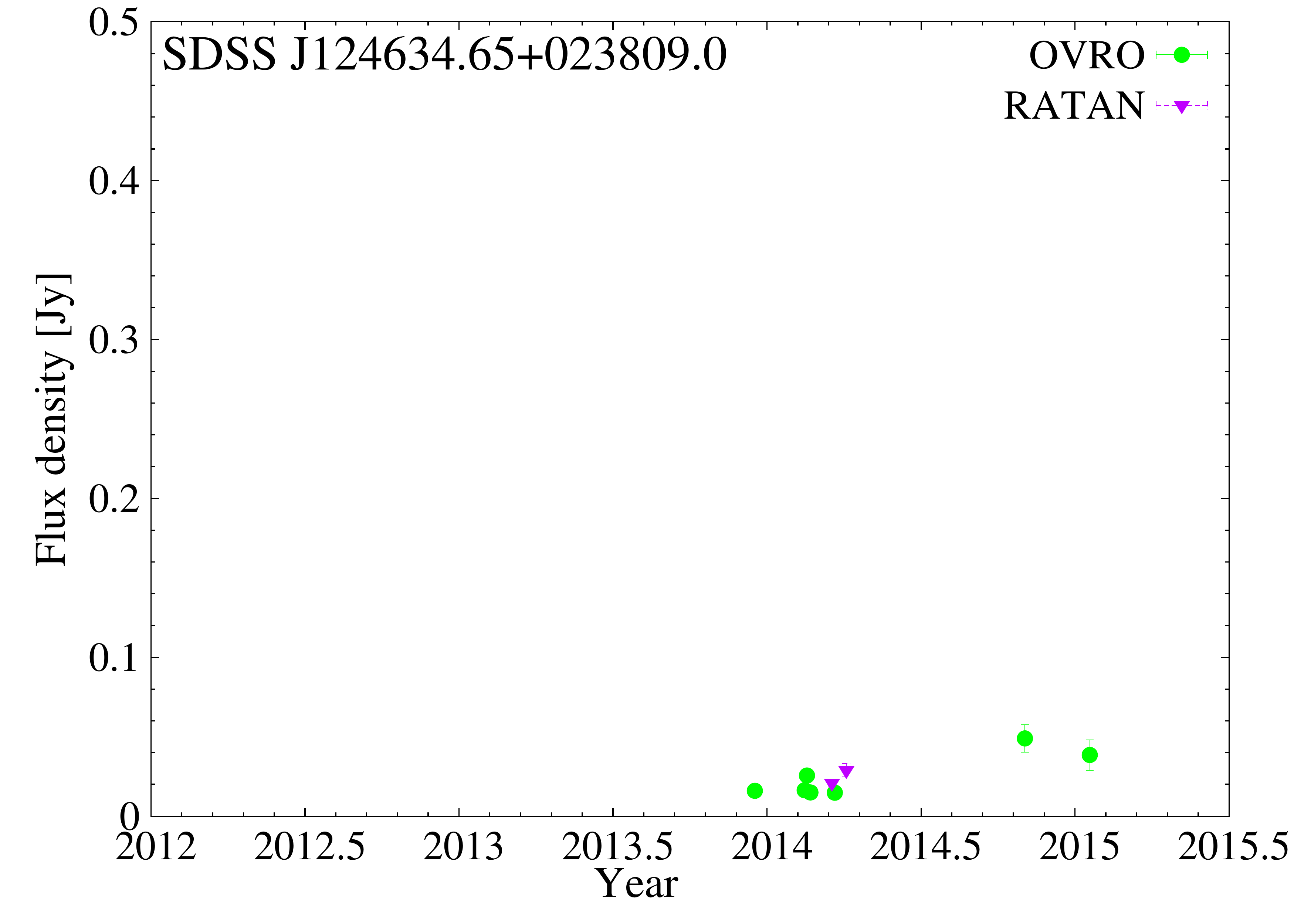}
\caption{Flux density curve of SDSS J124634.65+023809.0. } \label{fig:lcJ124634}
\end{minipage}
\end{figure*}

\clearpage

\begin{figure*}[ht!]
\centering
\begin{minipage}{0.47\textwidth}
\centering
\includegraphics[width=1\textwidth]{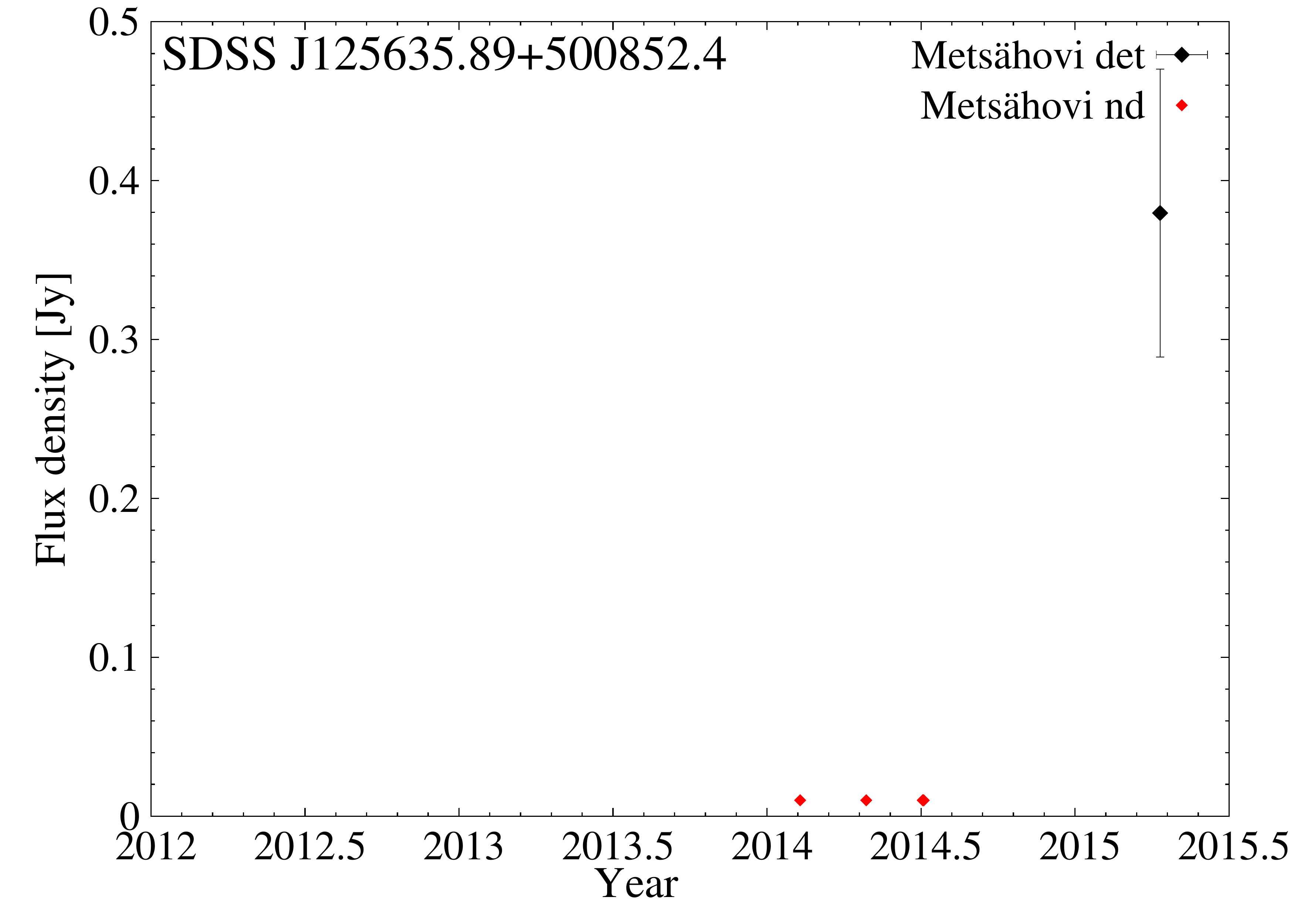}
\caption{Flux density curve of SDSS J125635.89+500852.4. } \label{fig:lcJ125635}
\end{minipage}\hfill
\begin{minipage}{0.47\textwidth}
\centering
\includegraphics[width=1\textwidth]{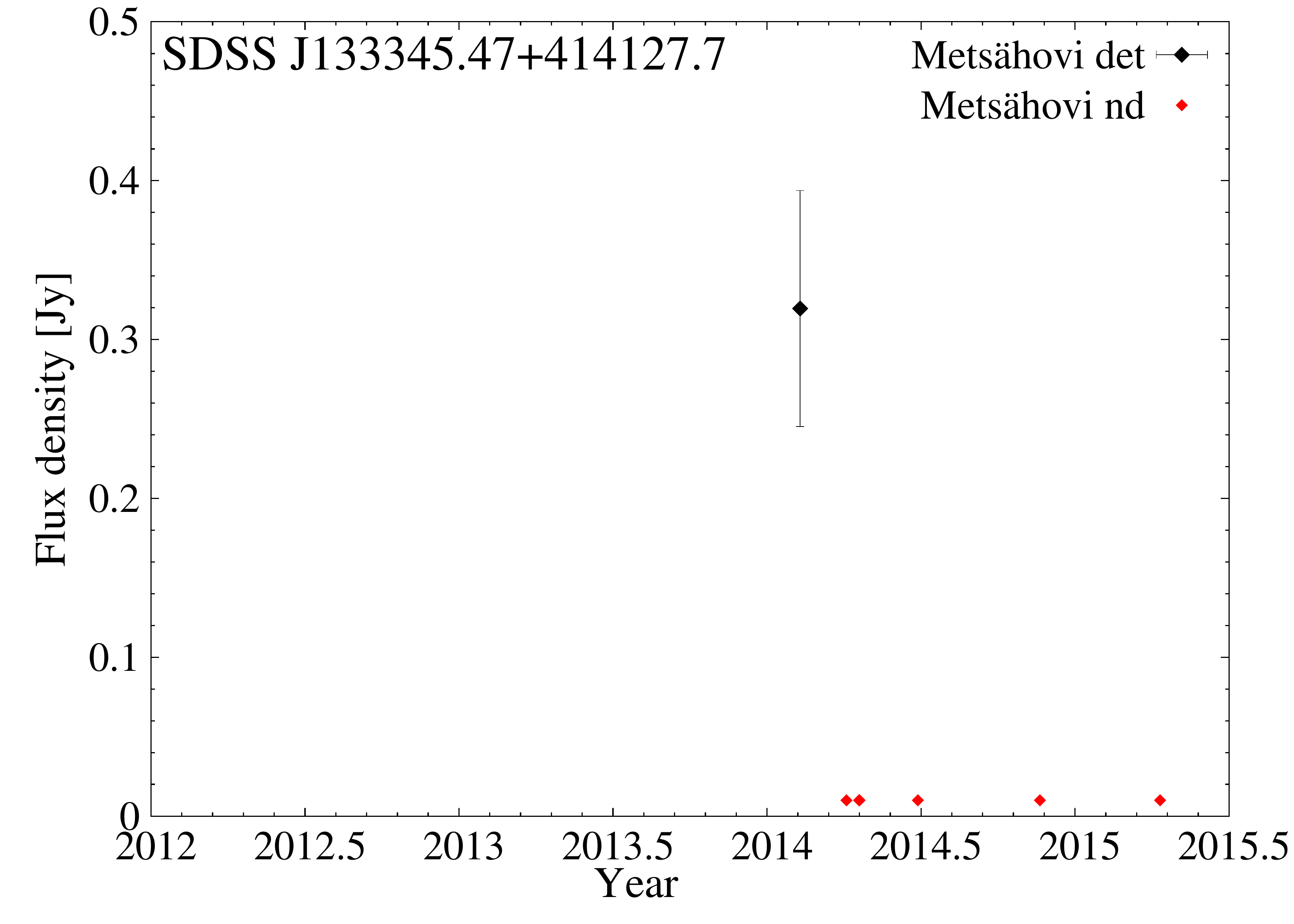}
\caption{Flux density curve of SDSS J133345.47+414127.7. } \label{fig:lcJ133345}
\end{minipage}
\end{figure*}

\begin{figure*}[ht!]
\centering
\begin{minipage}{0.47\textwidth}
\centering
\includegraphics[width=1\textwidth]{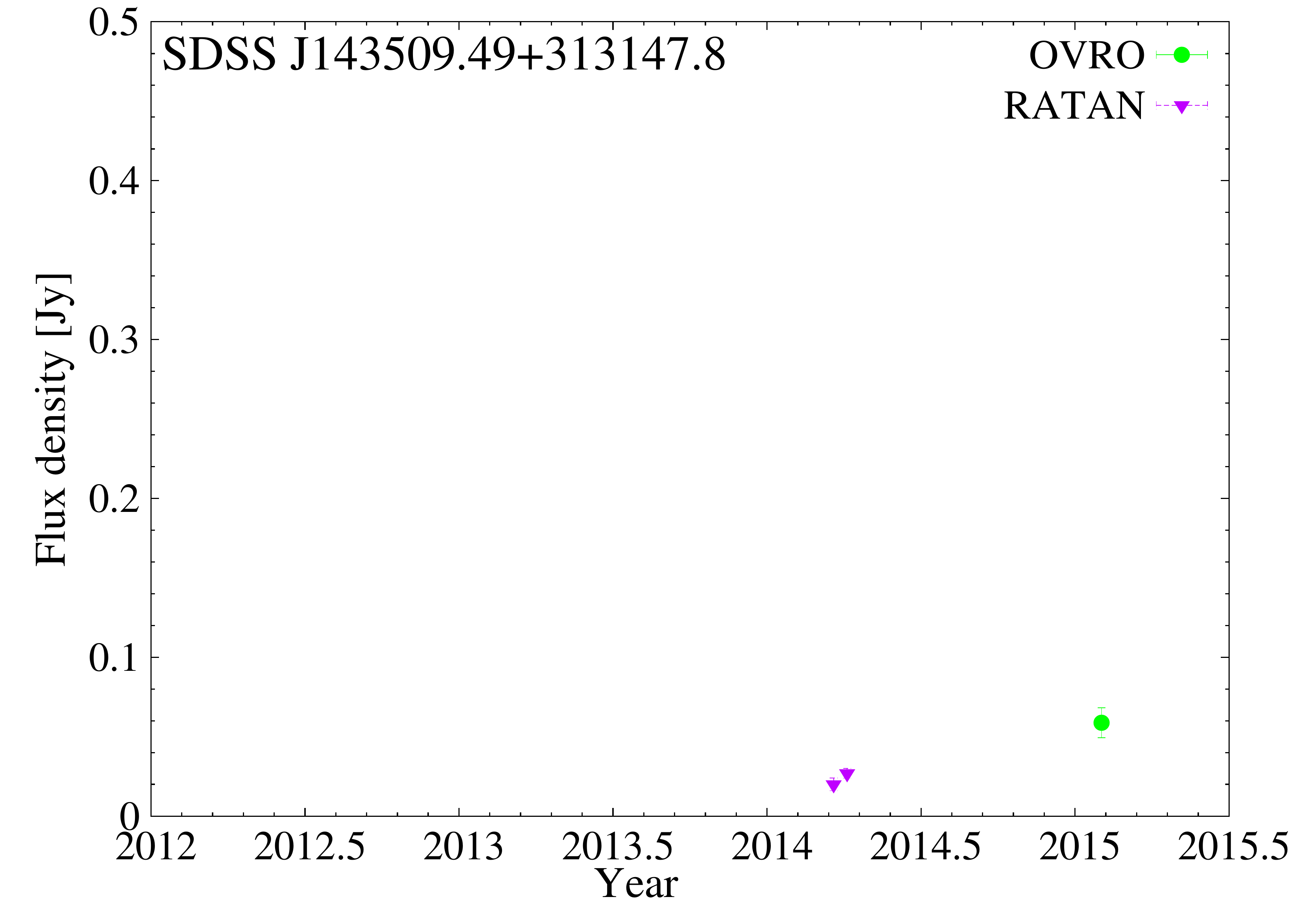}
\caption{Flux density curve of SDSS J143509.49+313147.8. } \label{fig:lcJ143509}
\end{minipage}
\begin{minipage}{0.47\textwidth}
\centering
\includegraphics[width=1\textwidth]{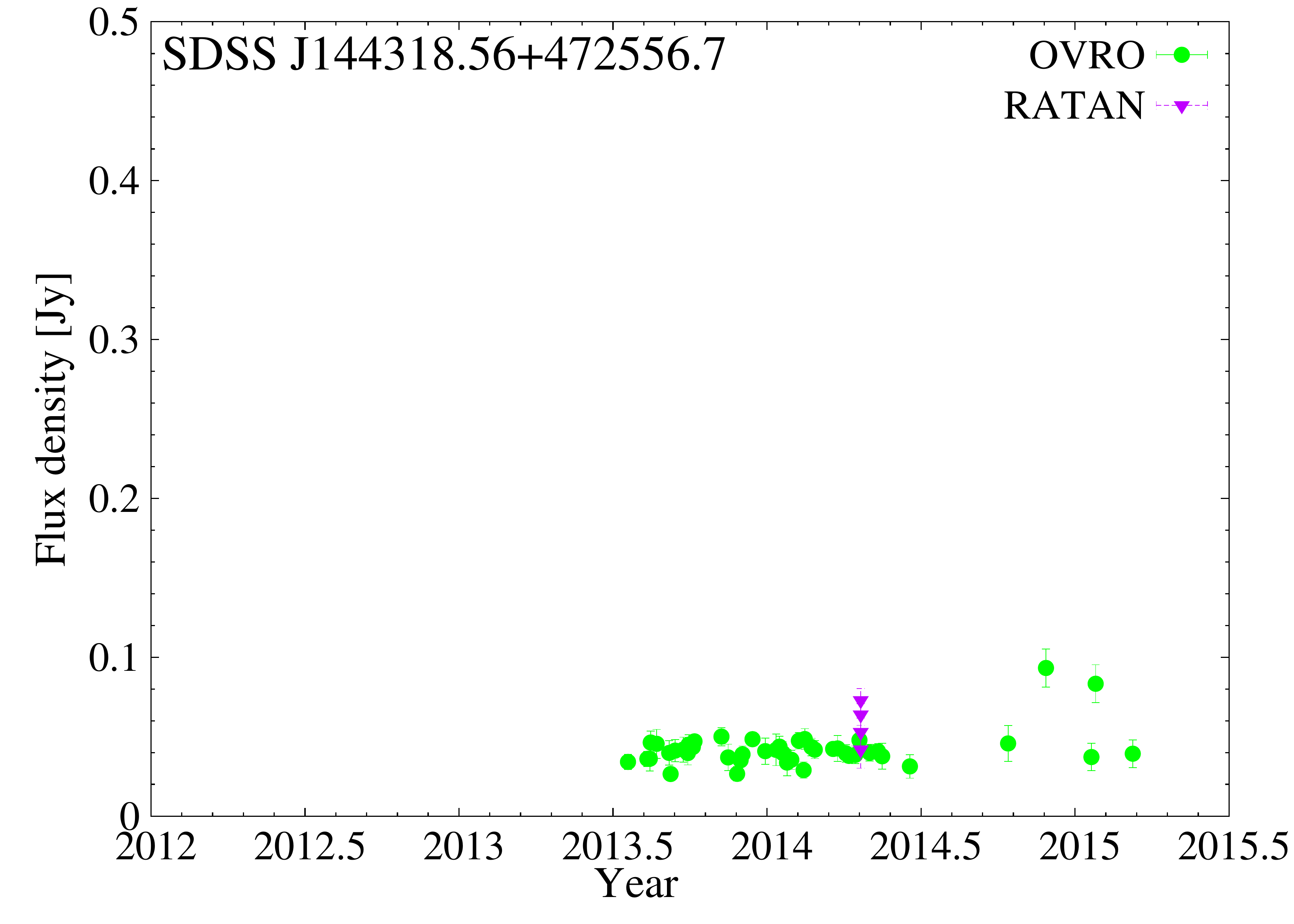}
\caption{Flux density curve of SDSS J144318.56+472556.7. } \label{fig:lcJ144318}
\end{minipage}\hfill
\end{figure*}

\begin{figure*}[ht!]
\centering
\begin{minipage}{0.47\textwidth}
\centering
\includegraphics[width=1\textwidth]{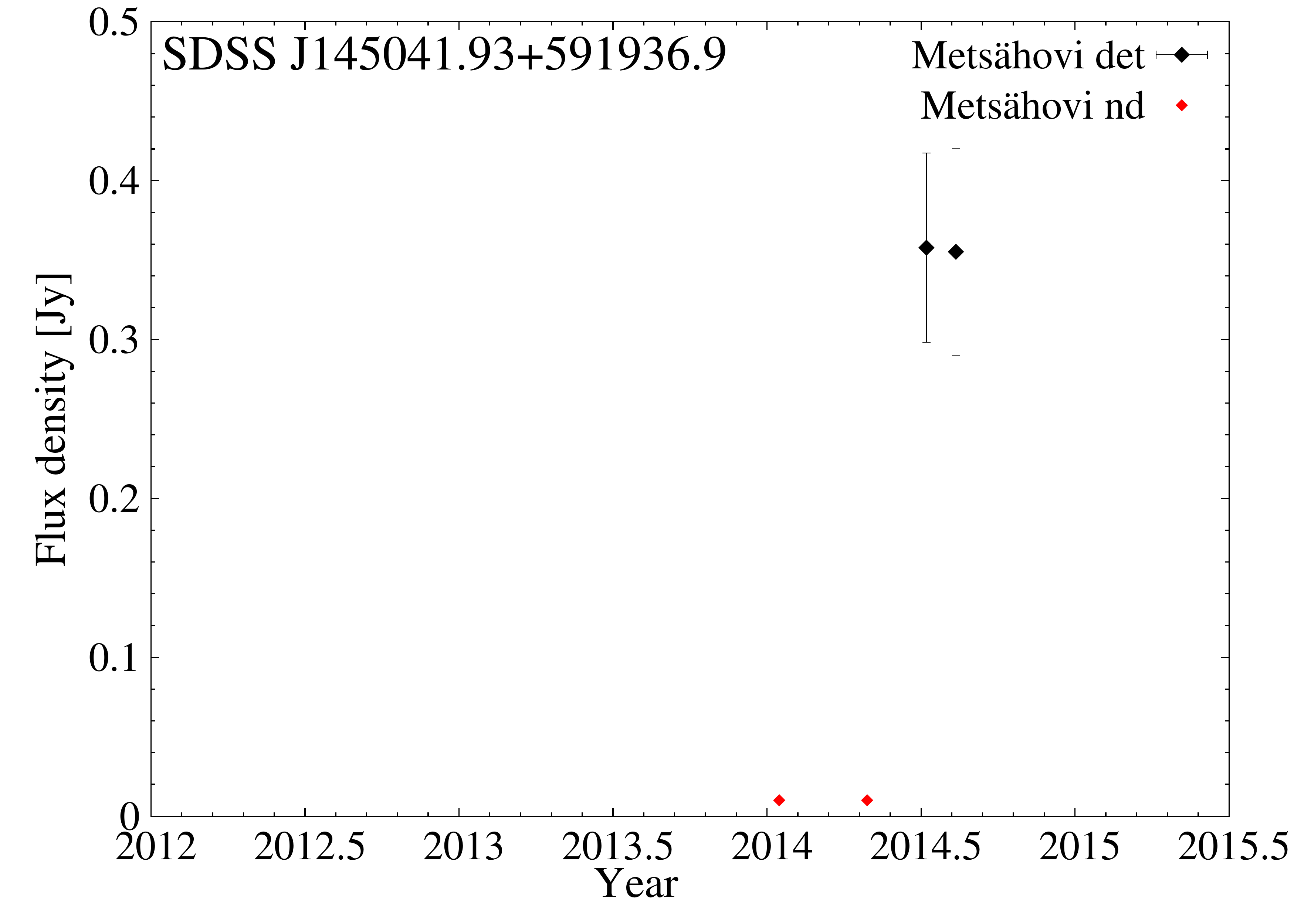}
\caption{Flux density curve of SDSS J145041.93+591936.9. } \label{fig:lcJ145041}
\end{minipage}
\begin{minipage}{0.47\textwidth}
\centering
\includegraphics[width=1\textwidth]{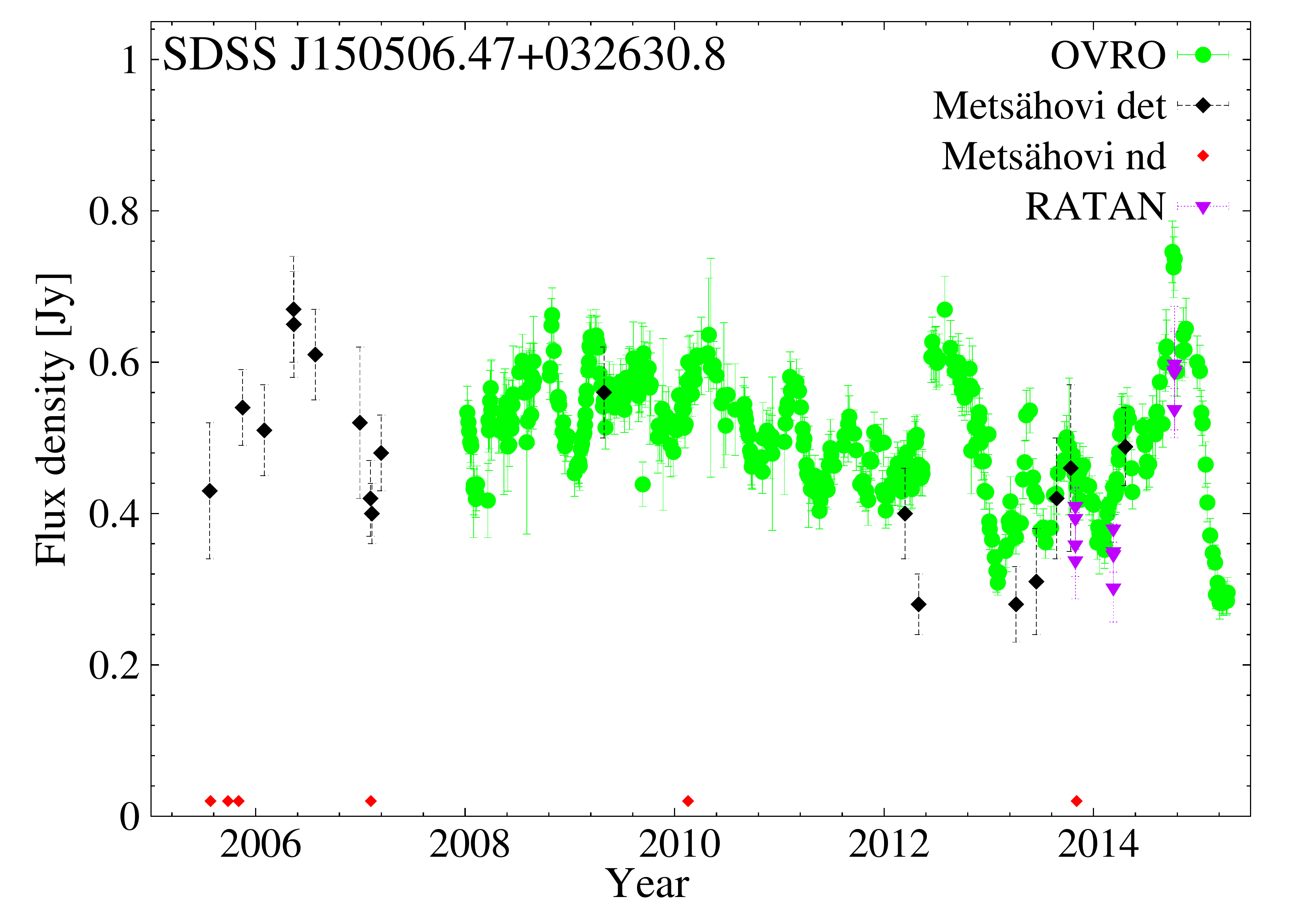}
\caption{Flux density curve of SDSS J150506.47+032630.8. } \label{fig:lcJ150506}
\end{minipage}\hfill
\end{figure*}

\clearpage

\begin{figure*}[ht!]
\centering
\begin{minipage}{0.47\textwidth}
\centering
\includegraphics[width=1\textwidth]{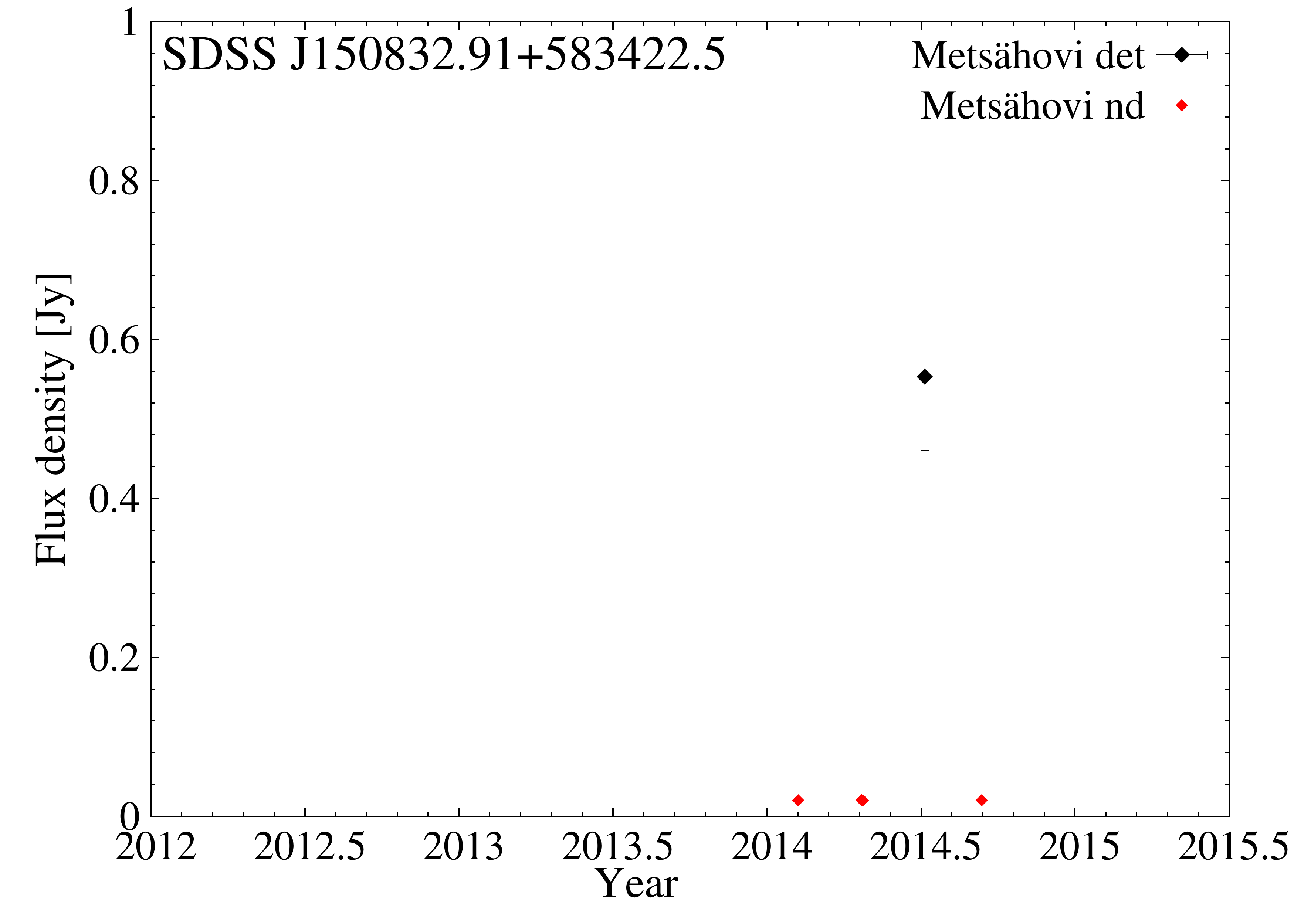}
\caption{Flux density curve of SDSS J150832.91+583422.5. } \label{fig:lcJ150832}
\end{minipage}
\begin{minipage}{0.47\textwidth}
\centering
\includegraphics[width=1\textwidth]{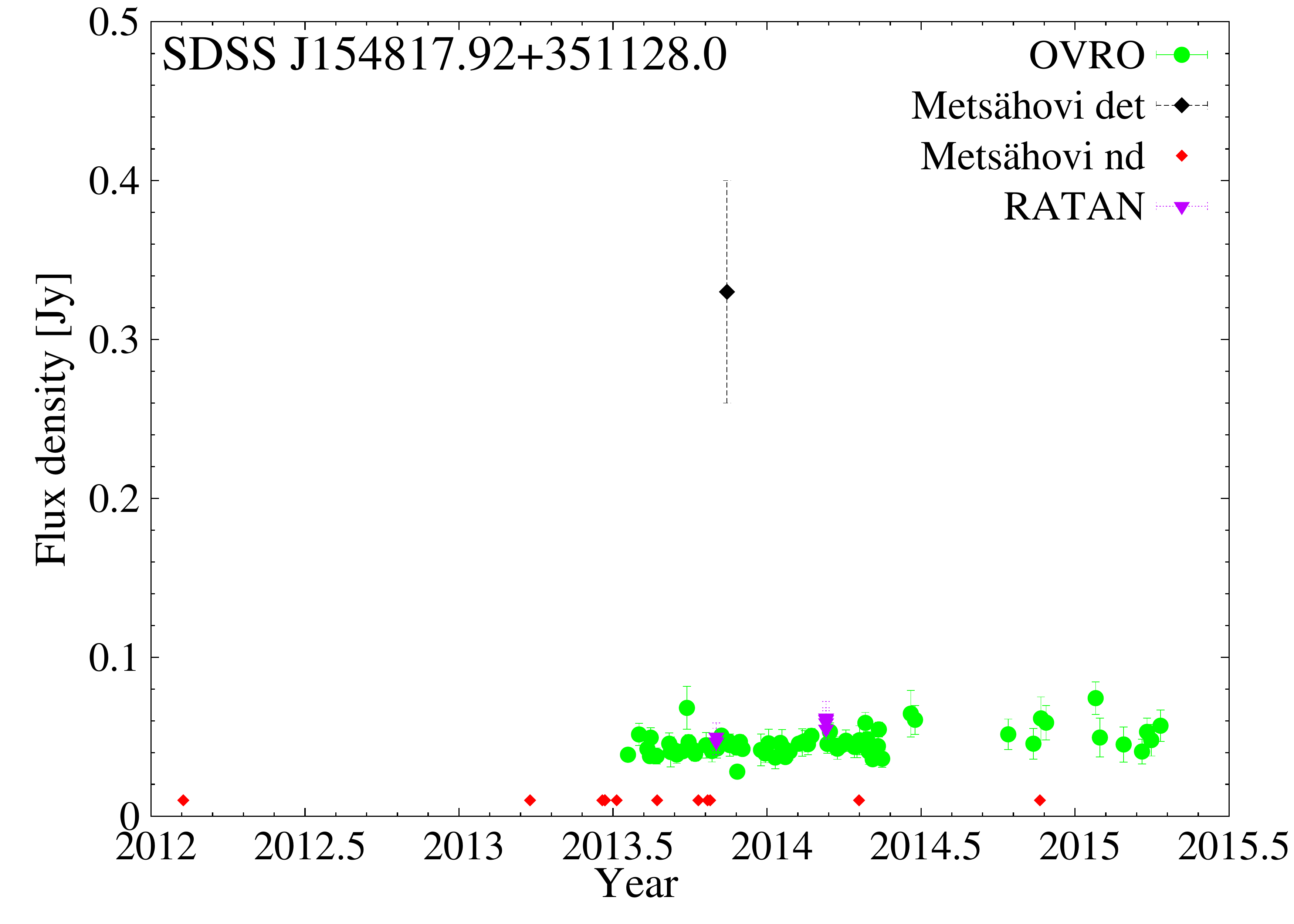}
\caption{Flux density curve of SDSS J154817.92+351128.0. } \label{fig:lcJ154817}
\end{minipage}\hfill
\end{figure*}

\begin{figure*}[ht!]
\centering
\begin{minipage}{0.47\textwidth}
\centering
\includegraphics[width=1\textwidth]{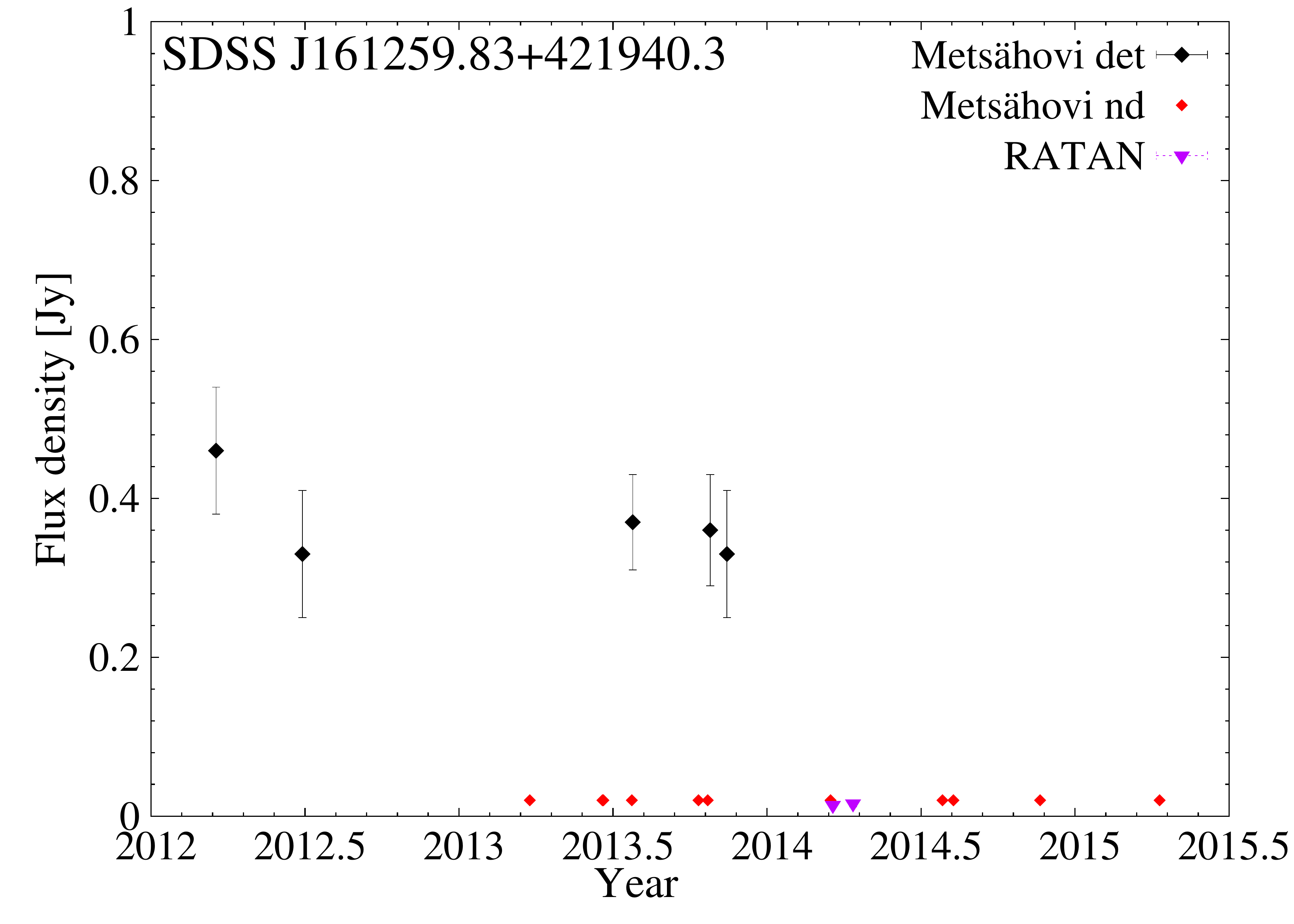}
\caption{Flux density curve of SDSS J161259.83+421940.3. } \label{fig:lcJ161259}
\end{minipage}
\begin{minipage}{0.47\textwidth}
\centering
\includegraphics[width=1\textwidth]{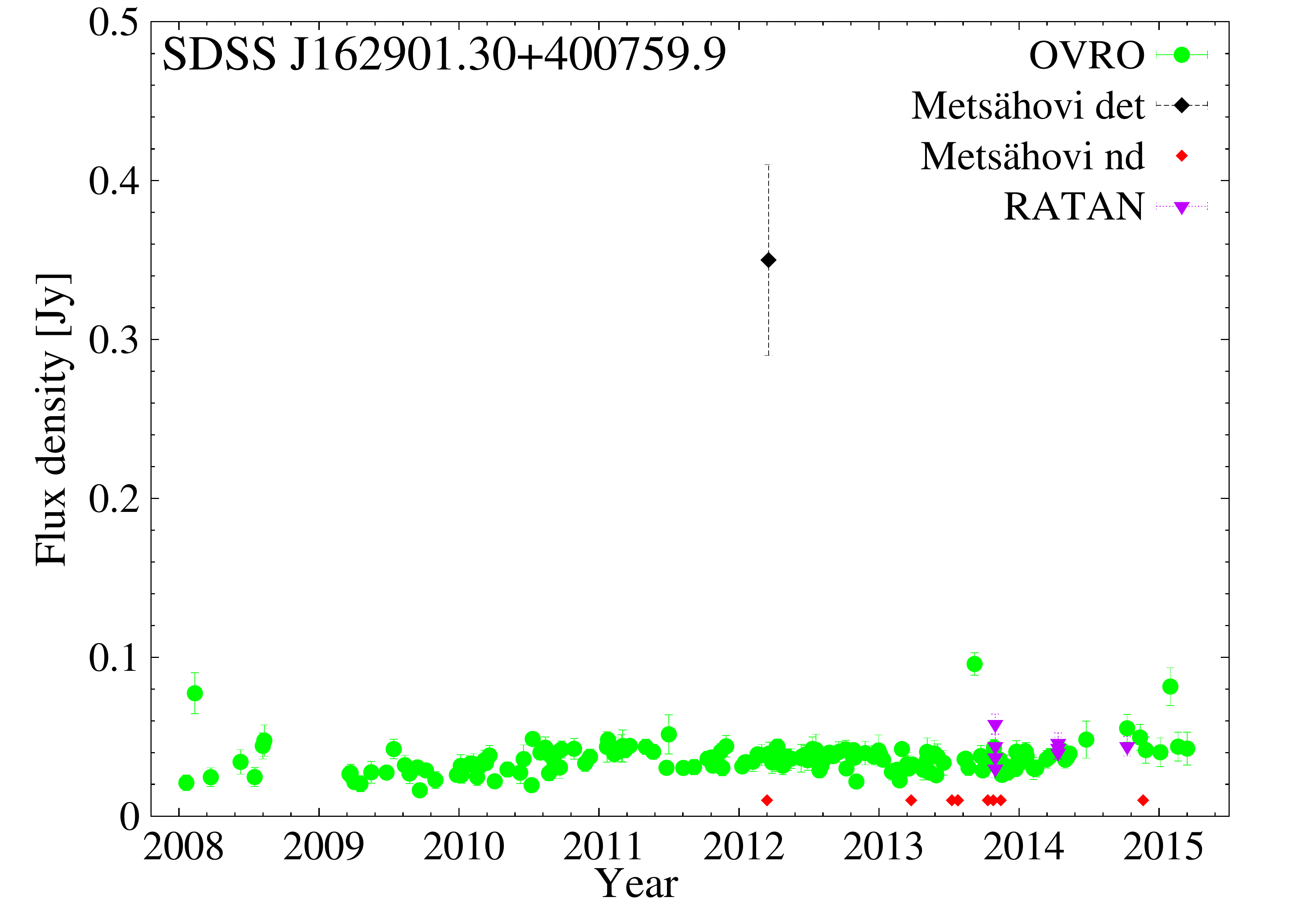}
\caption{Flux density curve of SDSS J162901.30+400759.9. } \label{fig:lcJ162901}
\end{minipage}\hfill
\end{figure*}

\begin{figure*}[ht!]
\centering
\begin{minipage}{0.47\textwidth}
\centering
\includegraphics[width=1\textwidth]{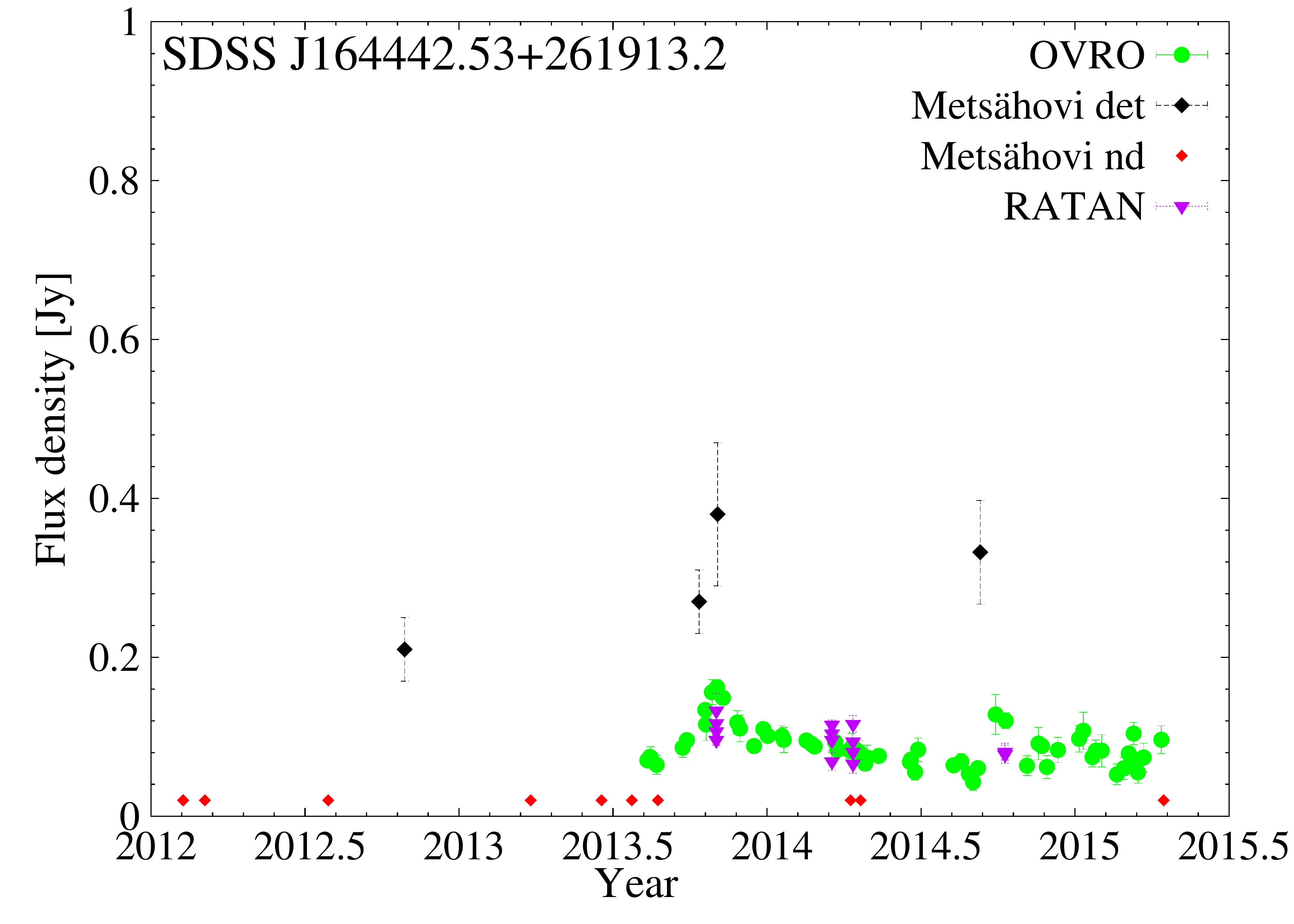}
\caption{Flux density curve of SDSS J164442.53+261913.2. } \label{fig:lcJ164442}
\end{minipage}
\begin{minipage}{0.47\textwidth}
\centering
\includegraphics[width=1\textwidth]{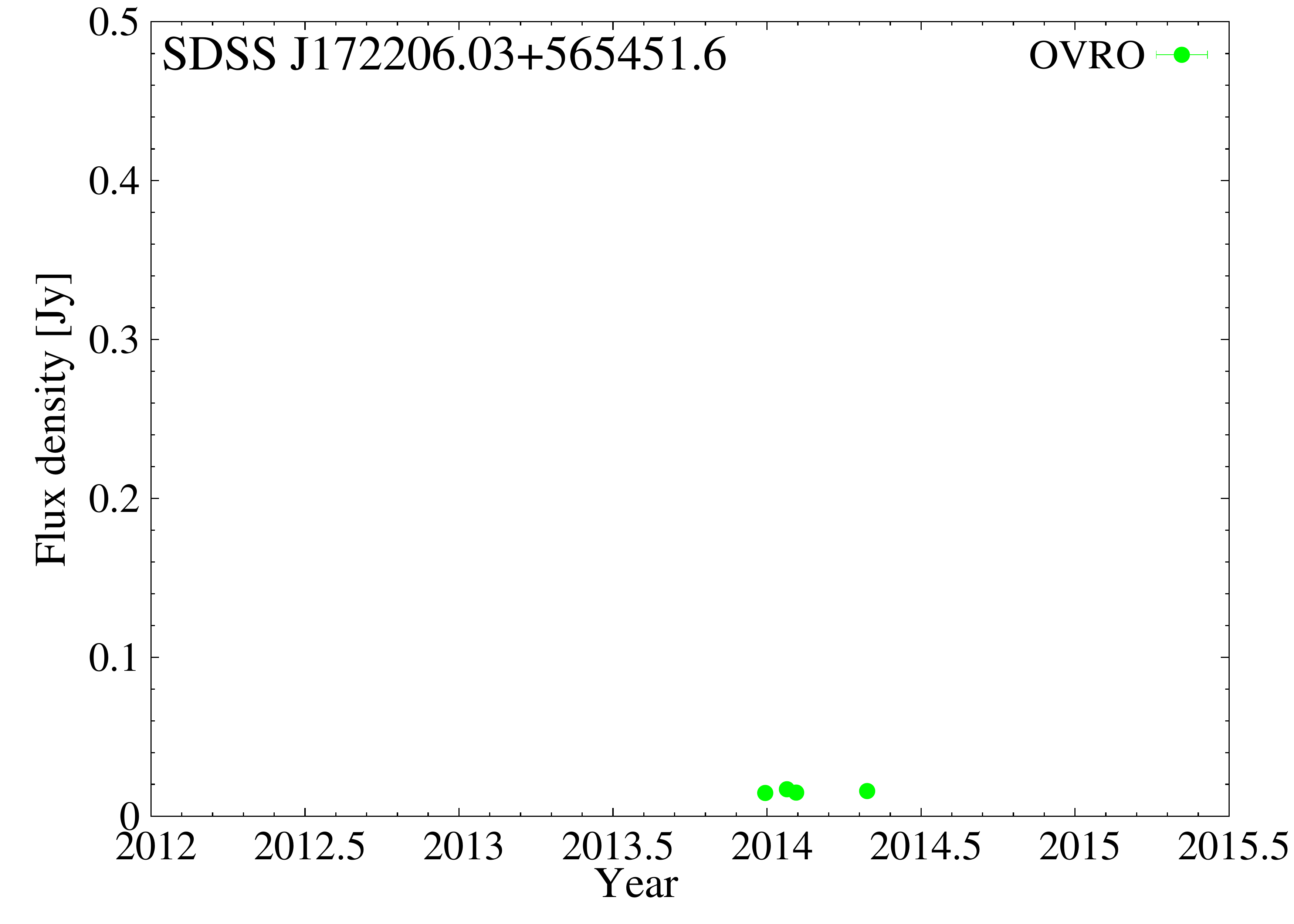}
\caption{Flux density curve of SDSS J172206.03+565451.6. } \label{fig:lcJ172206}
\end{minipage}\hfill
\end{figure*}

\clearpage

\begin{figure*}[ht!]
\centering
\begin{minipage}{0.47\textwidth}
\centering
\includegraphics[width=1\textwidth]{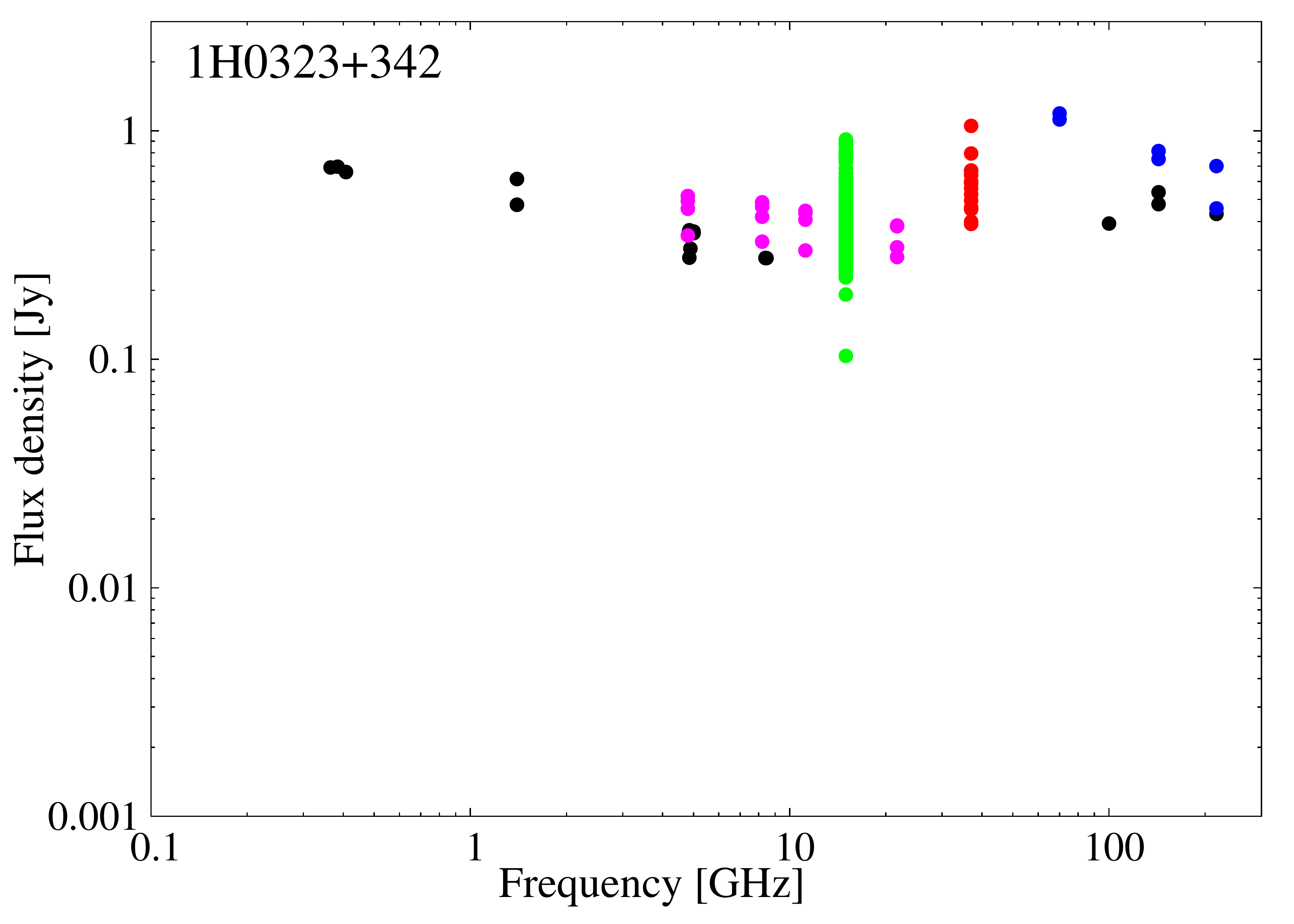}
\caption{Radio spectrum of 1H0323+342. Black: archival data, magenta: RATAN-600, green: OVRO, blue: \emph{Planck}, red: Mets\"{a}hovi. Only detections.} \label{fig:spec1h0323}
\end{minipage}
\begin{minipage}{0.47\textwidth}
\centering
\includegraphics[width=1\textwidth]{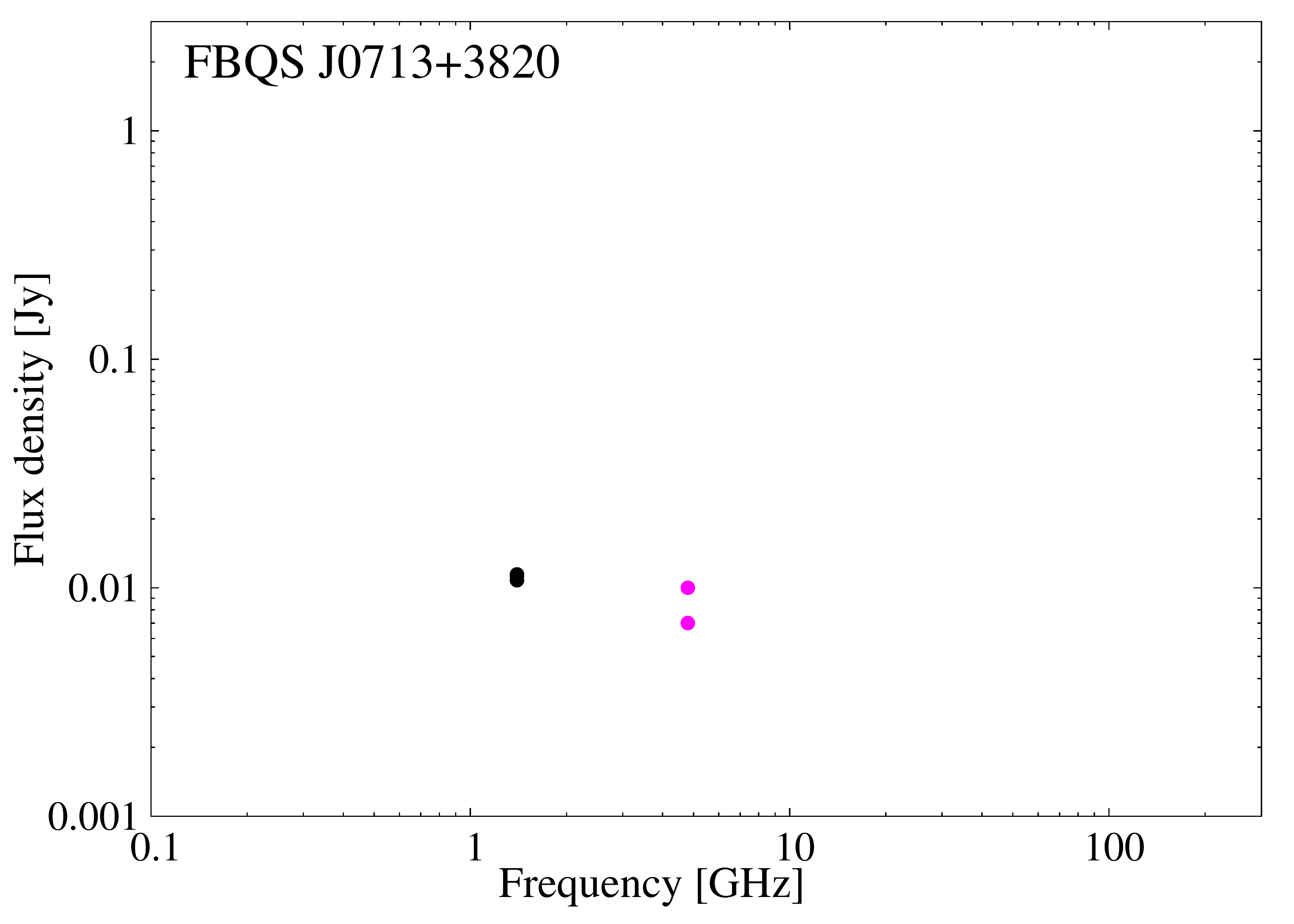}
\caption{Radio spectrum of FBQSJ0713+3820. Only detections. Colours as in Fig.~\ref{fig:spec1h0323}.} \label{fig:specfbqsj0713}
\end{minipage}\hfill
\end{figure*}

\begin{figure*}[ht!]
\centering
\begin{minipage}{0.47\textwidth}
\centering
\includegraphics[width=1\textwidth]{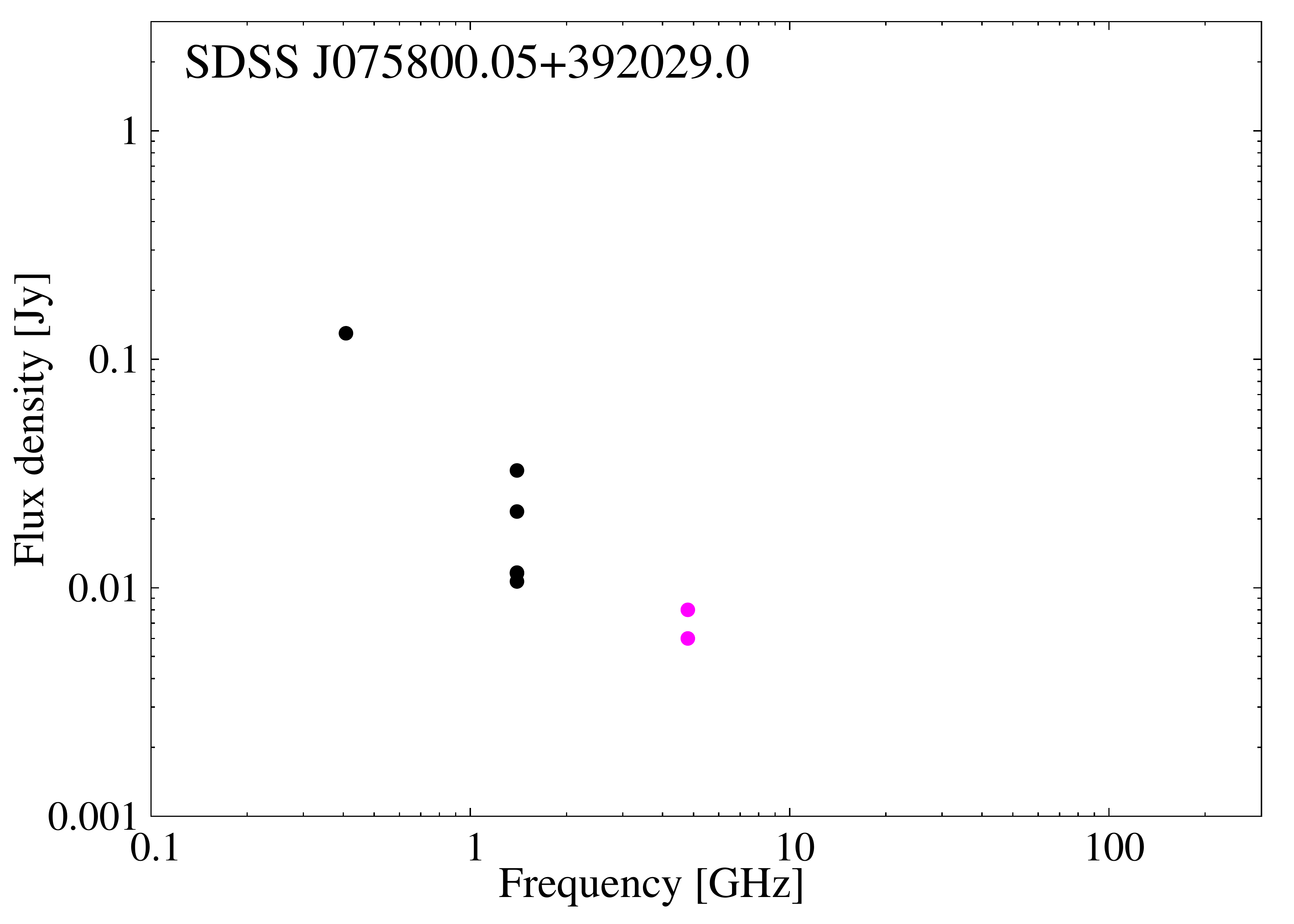}
\caption{Radio spectrum of J075800.05+392. Only detections. Colours as in Fig.~\ref{fig:spec1h0323}.} \label{fig:specj075800}
\end{minipage}
\begin{minipage}{0.47\textwidth}
\centering
\includegraphics[width=1\textwidth]{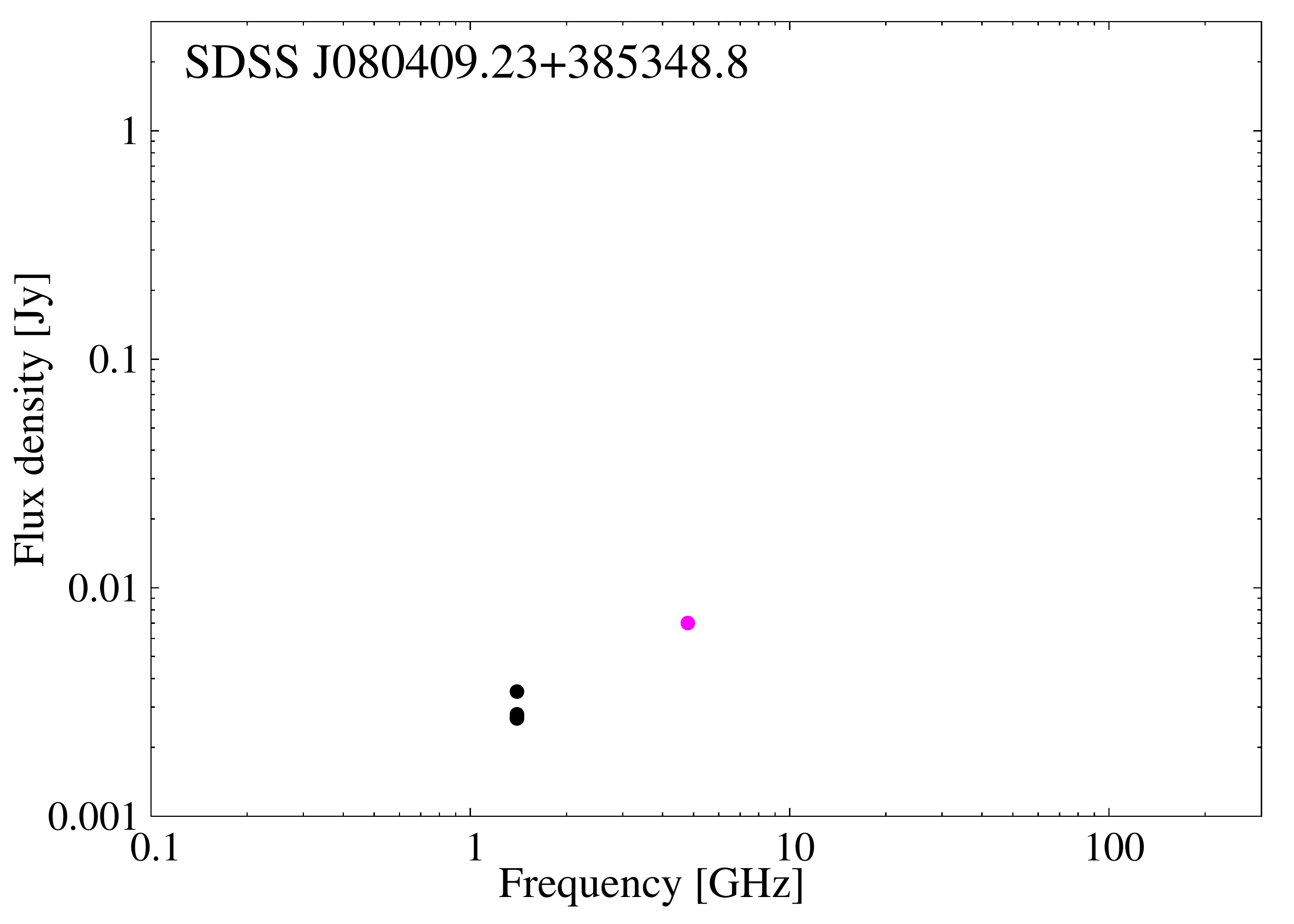}
\caption{Radio spectrum of J080409.23+385. Only detections. Colours as in Fig.~\ref{fig:spec1h0323}.} \label{fig:specj080409}
\end{minipage}\hfill
\end{figure*}

\begin{figure*}[ht!]
\centering
\begin{minipage}{0.47\textwidth}
\centering
\includegraphics[width=1\textwidth]{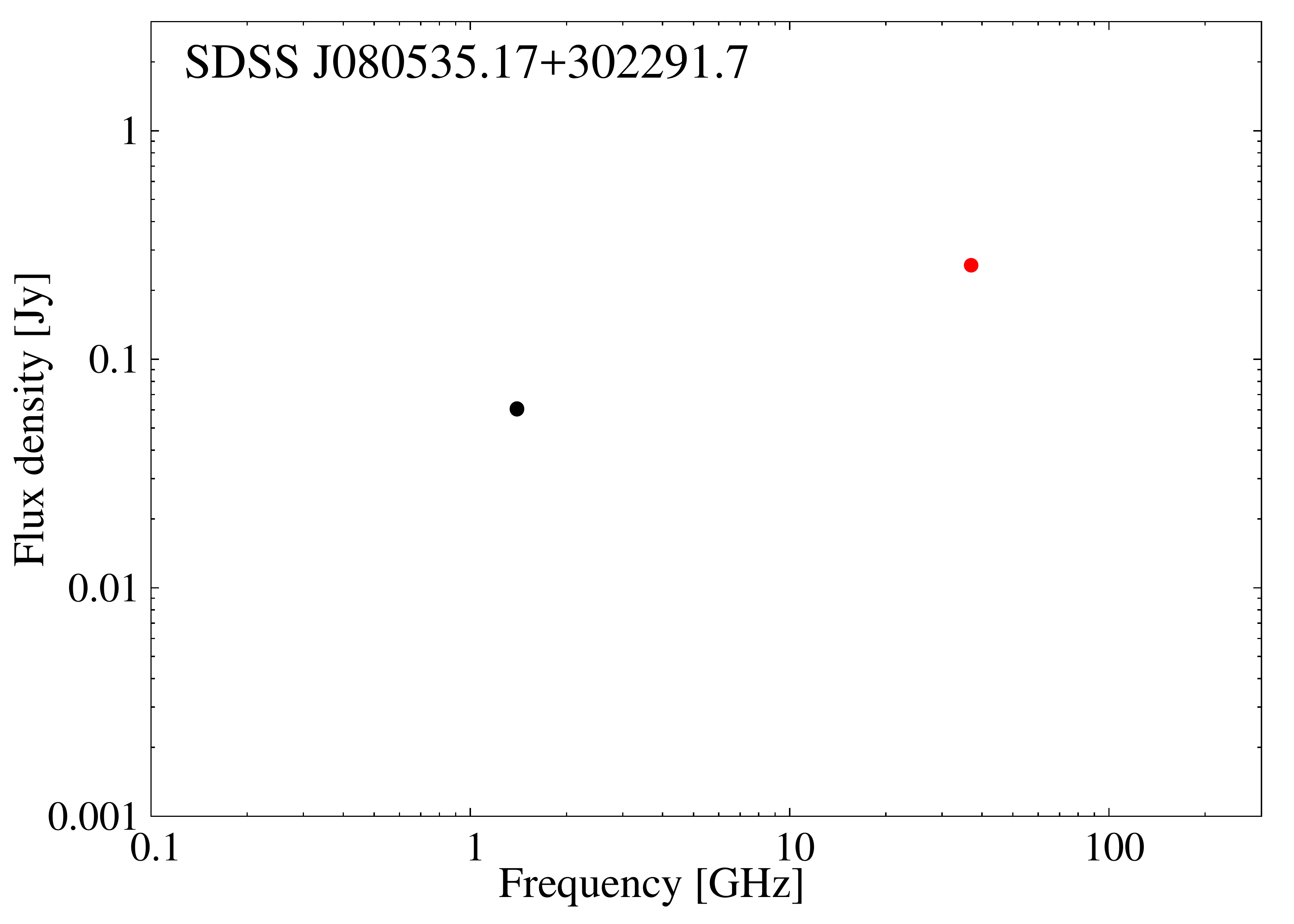}
\caption{Radio spectrum of J080535.17+302. Only detections. Colours as in Fig.~\ref{fig:spec1h0323}.} \label{fig:specj080535}
\end{minipage}
\begin{minipage}{0.47\textwidth}
\centering
\includegraphics[width=1\textwidth]{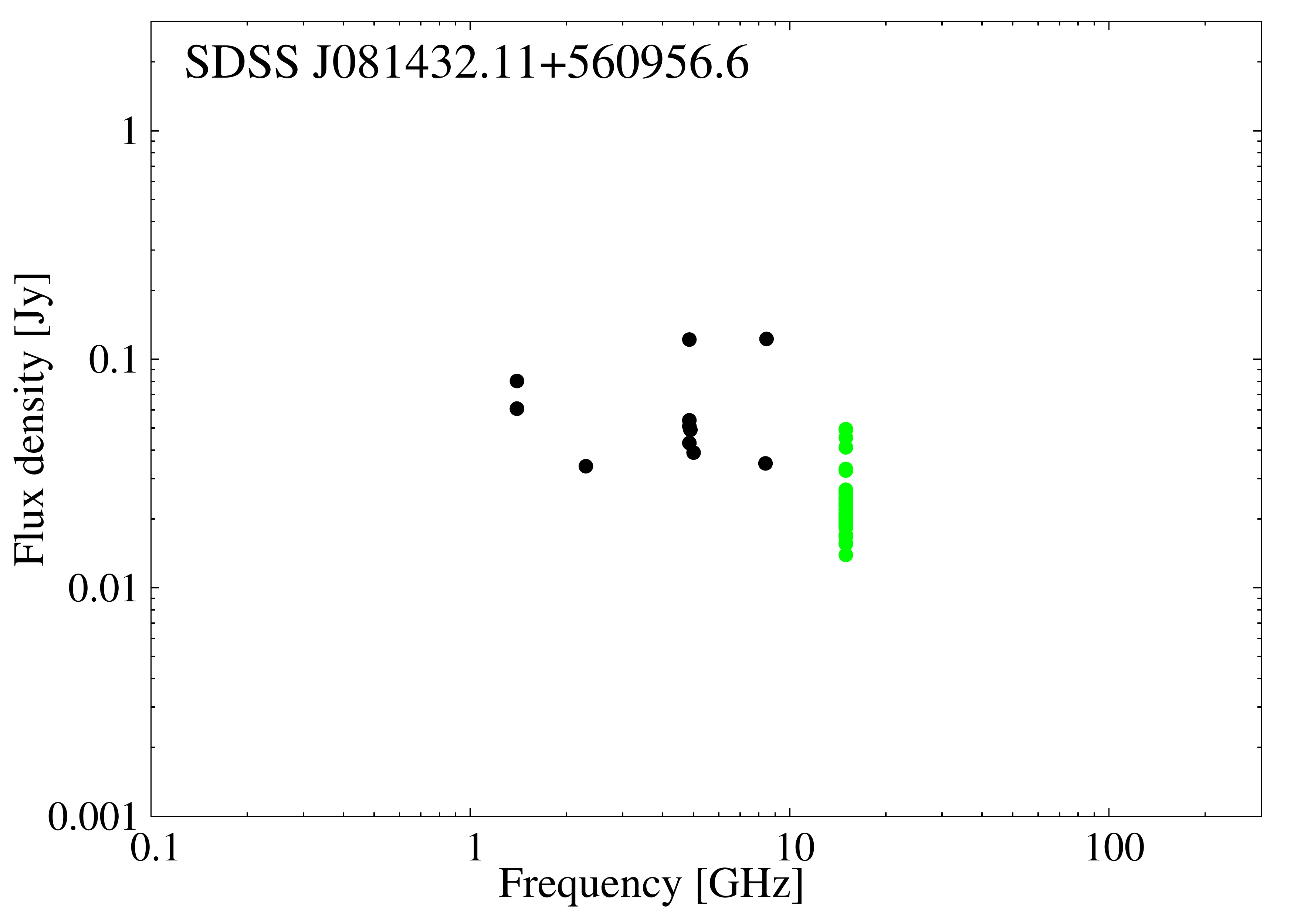}
\caption{Radio spectrum of J081432.11+560. Only detections. Colours as in Fig.~\ref{fig:spec1h0323}.} \label{fig:specj081432}
\end{minipage}\hfill
\end{figure*}

\clearpage

\begin{figure*}[ht!]
\centering
\begin{minipage}{0.47\textwidth}
\centering
\includegraphics[width=1\textwidth]{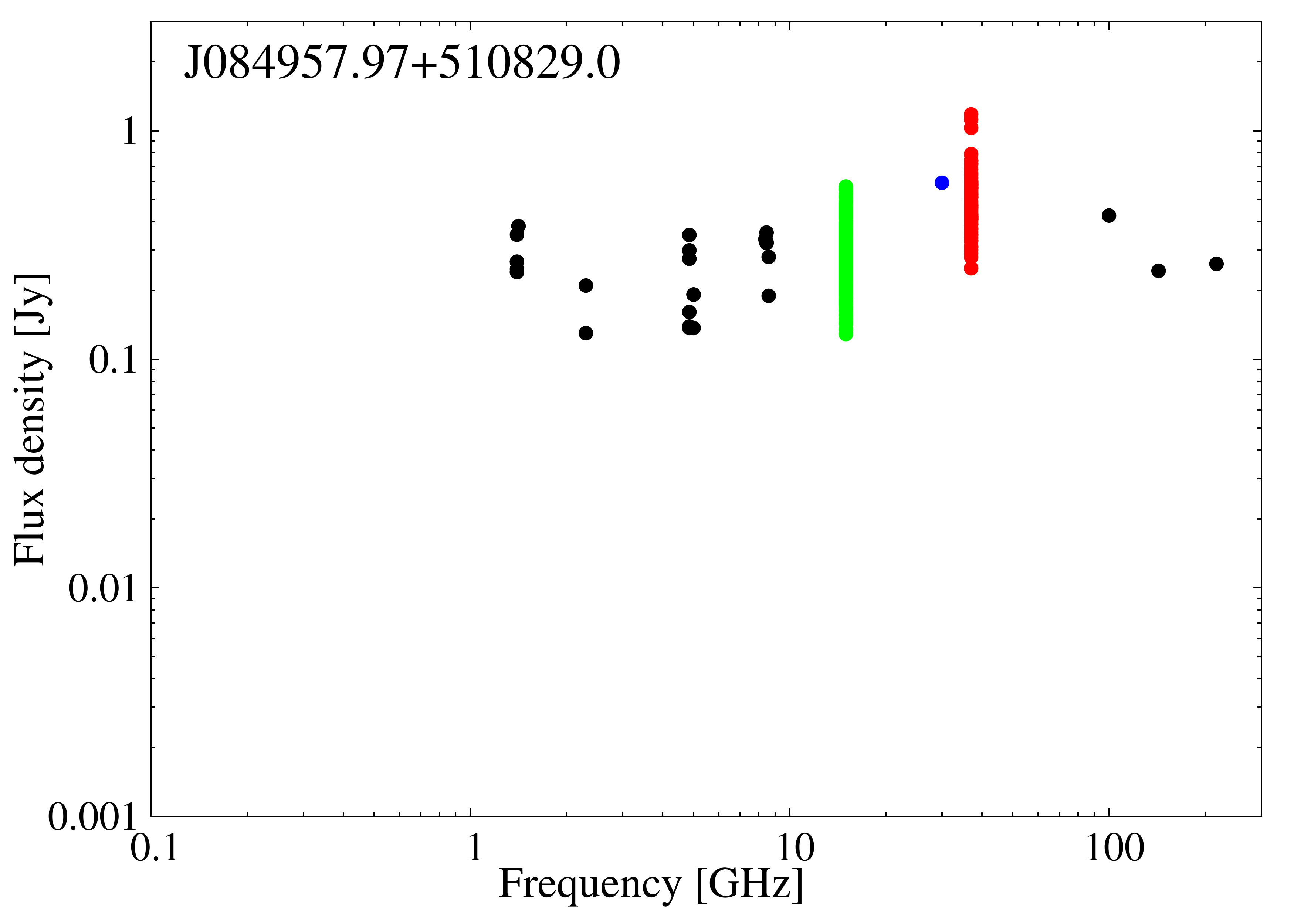}
\caption{Radio spectrum of J084957.97+510. Only detections. Colours as in Fig.~\ref{fig:spec1h0323}.} \label{fig:specj084957}
\end{minipage}
\begin{minipage}{0.47\textwidth}
\centering
\includegraphics[width=1\textwidth]{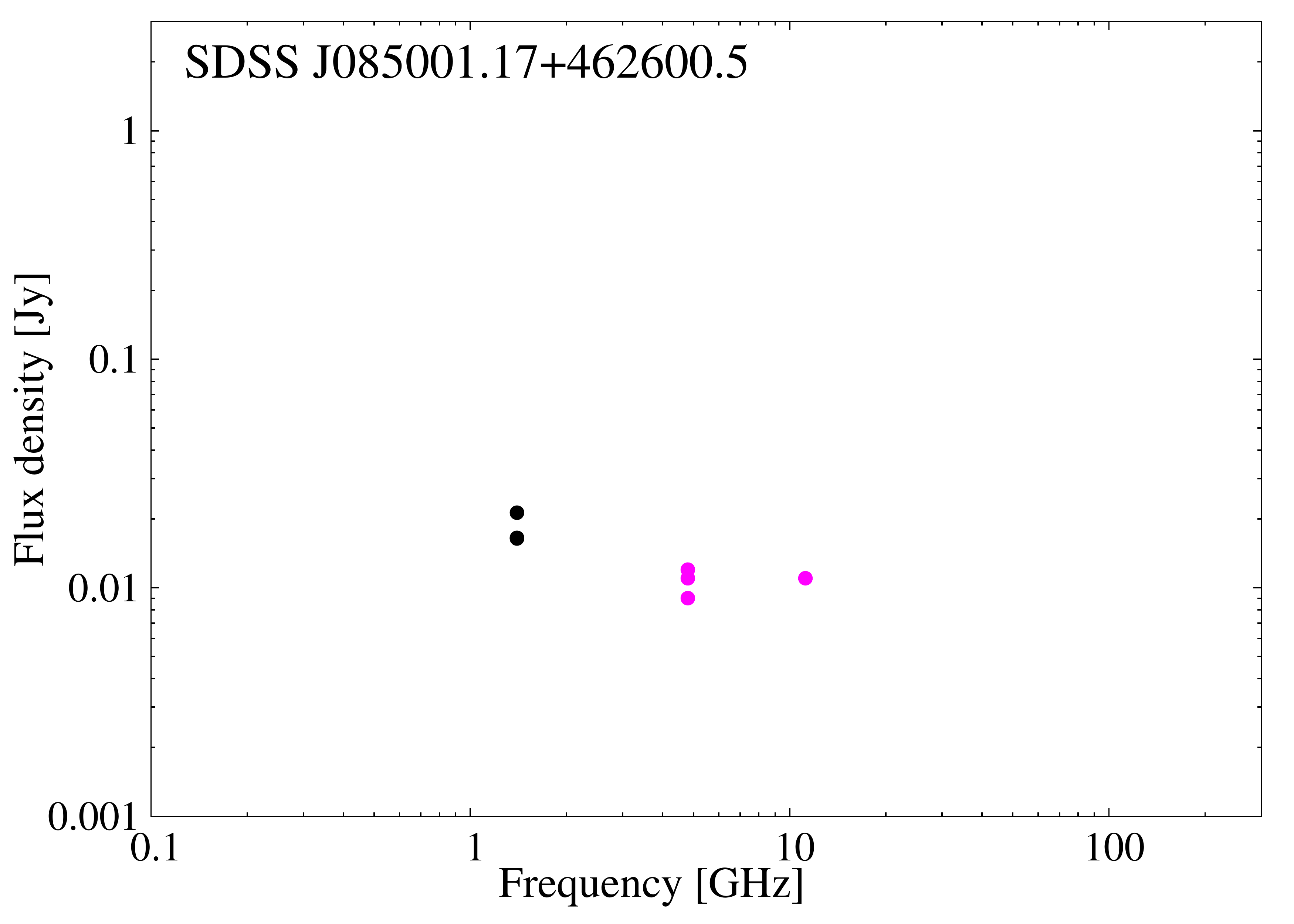}
\caption{Radio spectrum of J085001.17+462. Only detections. Colours as in Fig.~\ref{fig:spec1h0323}.} \label{fig:specj085001}
\end{minipage}\hfill
\end{figure*}
 
\begin{figure*}[ht!]
\centering
\begin{minipage}{0.47\textwidth}
\centering
\includegraphics[width=1\textwidth]{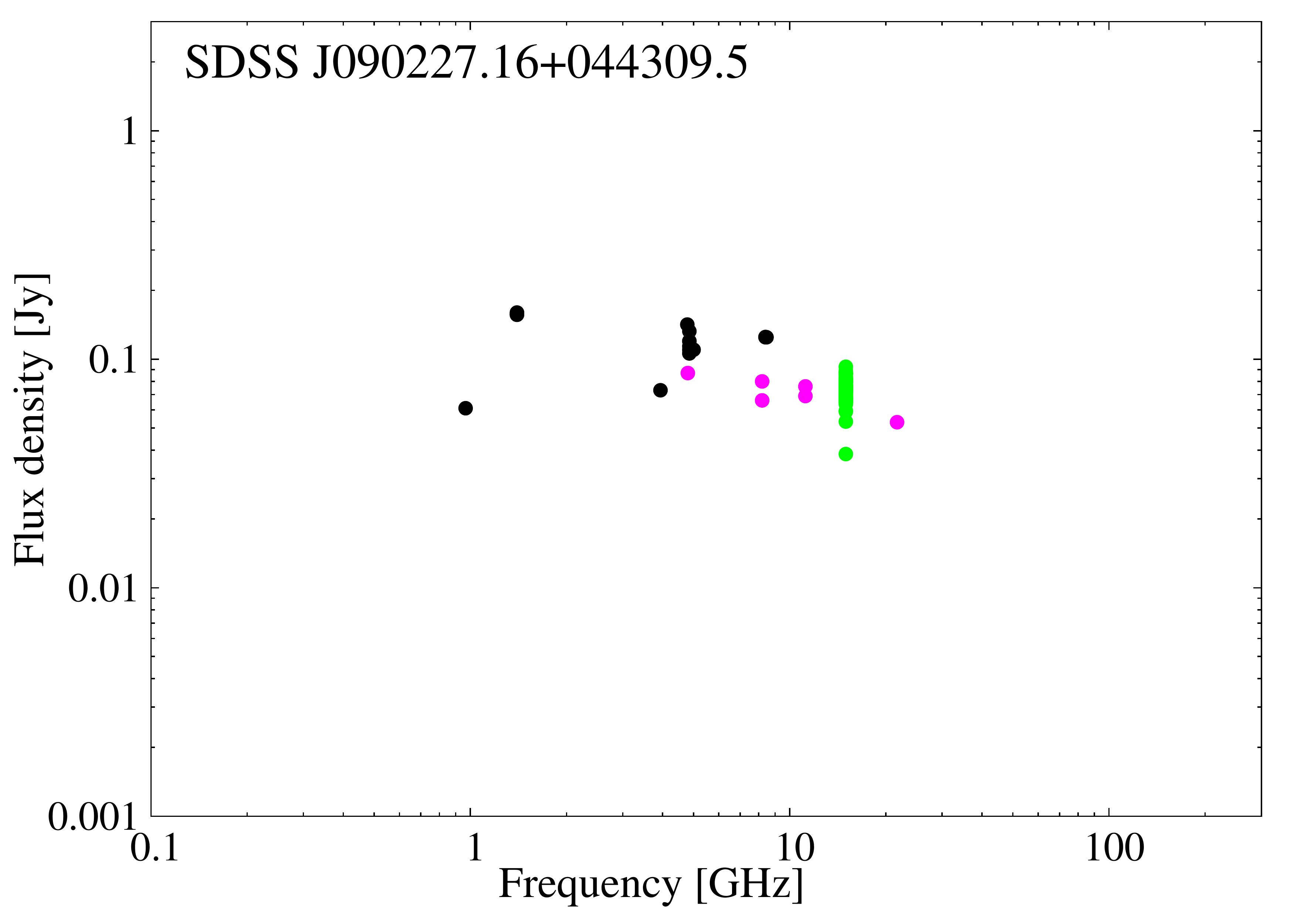}
\caption{Radio spectrum of J090227.16+044. Only detections. Colours as in Fig.~\ref{fig:spec1h0323}.} \label{fig:specj090227}
\end{minipage}
\begin{minipage}{0.47\textwidth}
\centering
\includegraphics[width=1\textwidth]{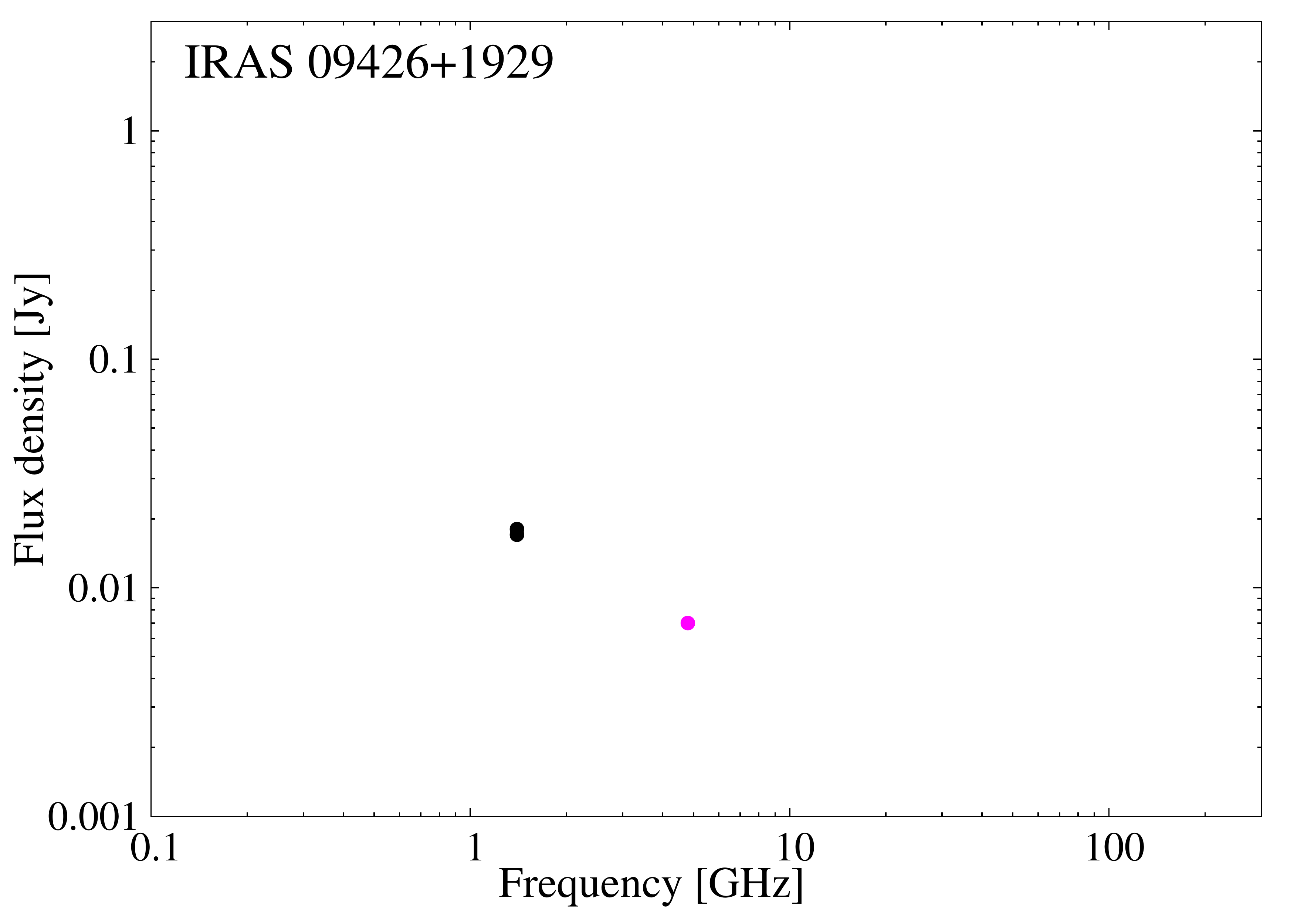}
\caption{Radio spectrum of IRAS09426+1929. Only detections. Colours as in Fig.~\ref{fig:spec1h0323}.} \label{fig:speciras09426}
\end{minipage}\hfill
\end{figure*}

\begin{figure*}[ht!]
\centering
\begin{minipage}{0.47\textwidth}
\centering
\includegraphics[width=1\textwidth]{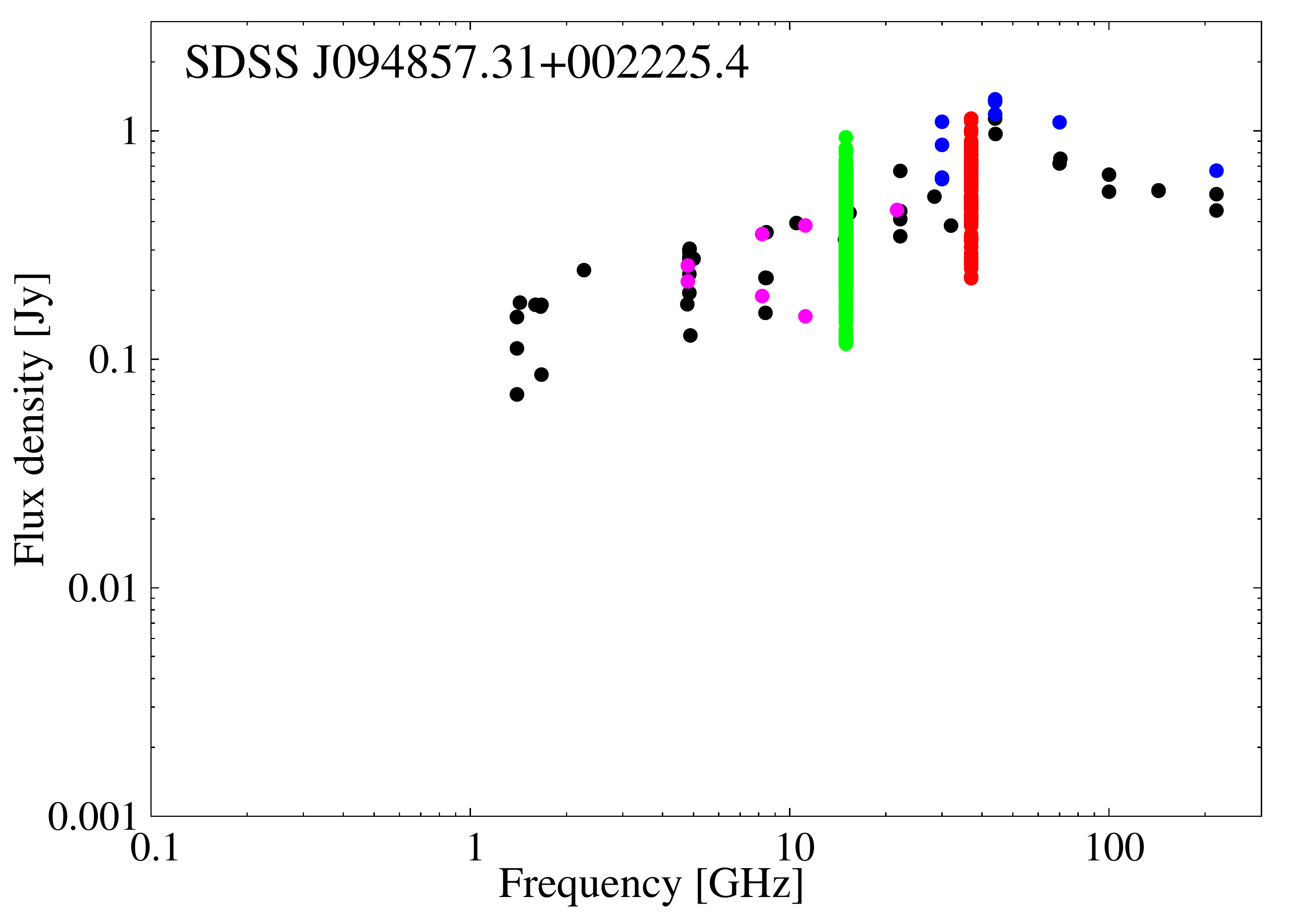}
\caption{Radio spectrum of J094857.31+002. Only detections. Colours as in Fig.~\ref{fig:spec1h0323}.} \label{fig:specj094857}
\end{minipage}
\begin{minipage}{0.47\textwidth}
\centering
\includegraphics[width=1\textwidth]{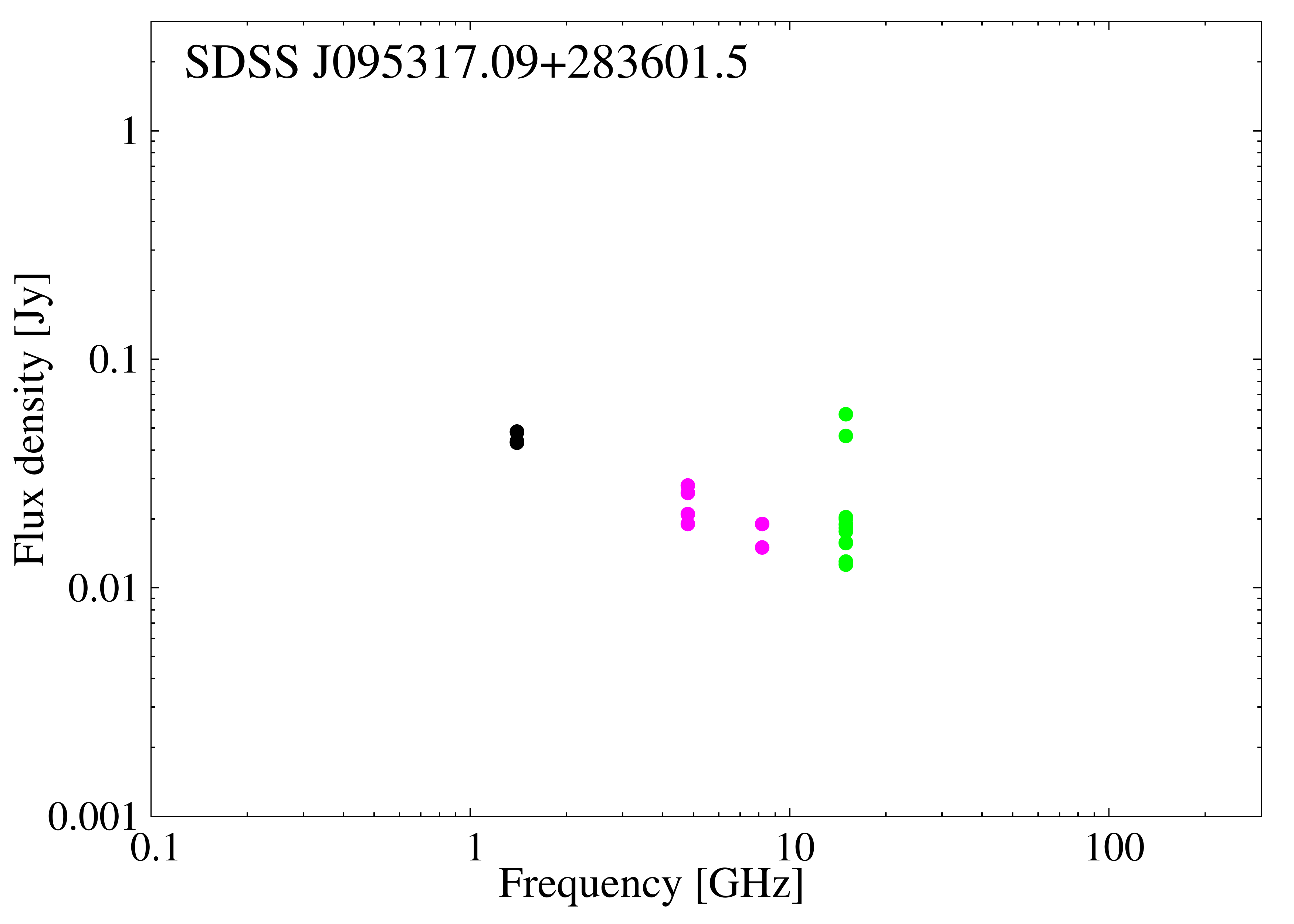}
\caption{Radio spectrum of J095317.09+283. Only detections. Colours as in Fig.~\ref{fig:spec1h0323}.} \label{fig:specj095317}
\end{minipage}\hfill
\end{figure*}

\clearpage

\newpage

\begin{figure*}[ht!]
\centering
\begin{minipage}{0.47\textwidth}
\centering
\includegraphics[width=1\textwidth]{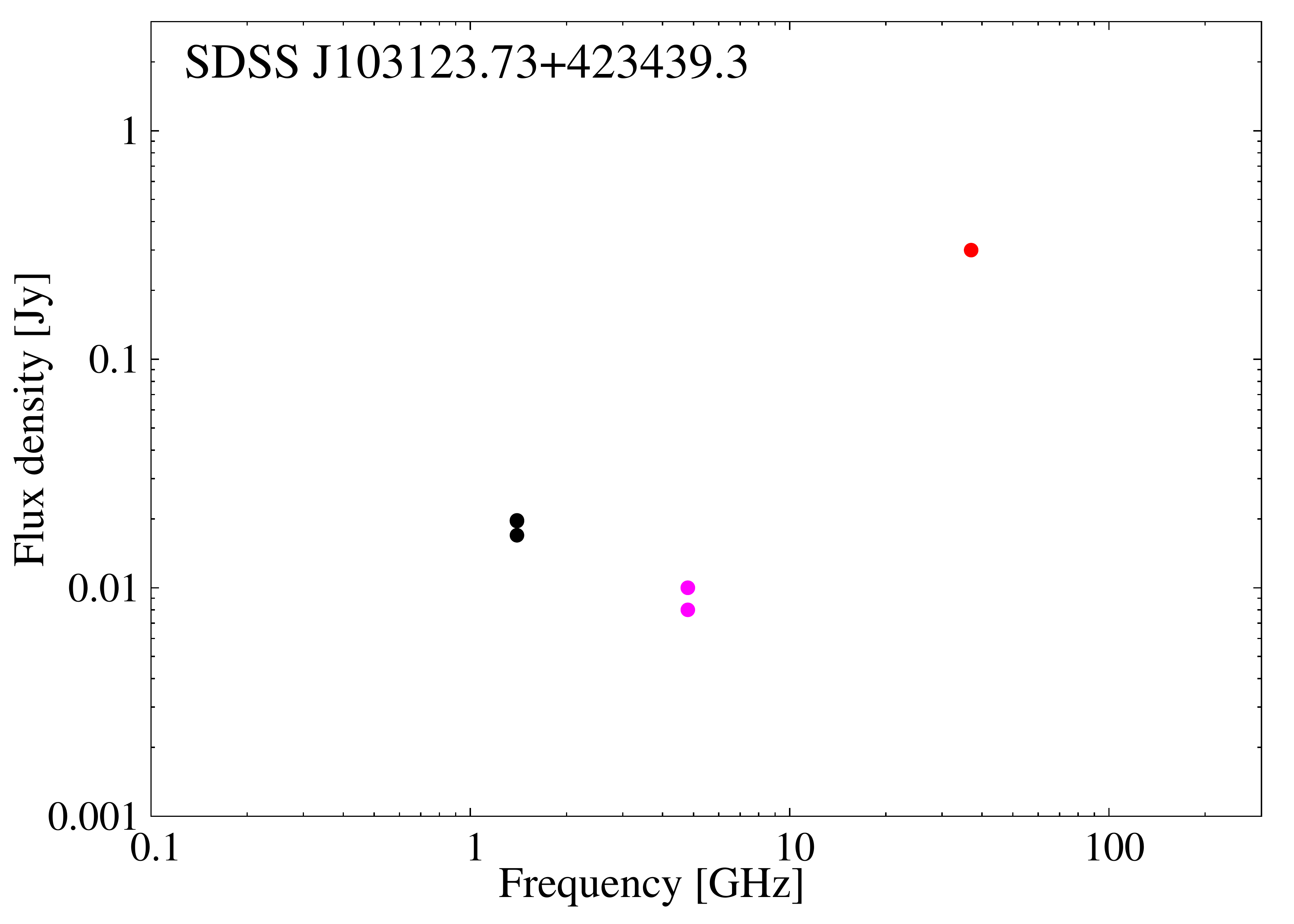}
\caption{Radio spectrum of J103123.73+423. Only detections. Colours as in Fig.~\ref{fig:spec1h0323}.} \label{fig:specj103123}
\end{minipage}
\begin{minipage}{0.47\textwidth}
\centering
\includegraphics[width=1\textwidth]{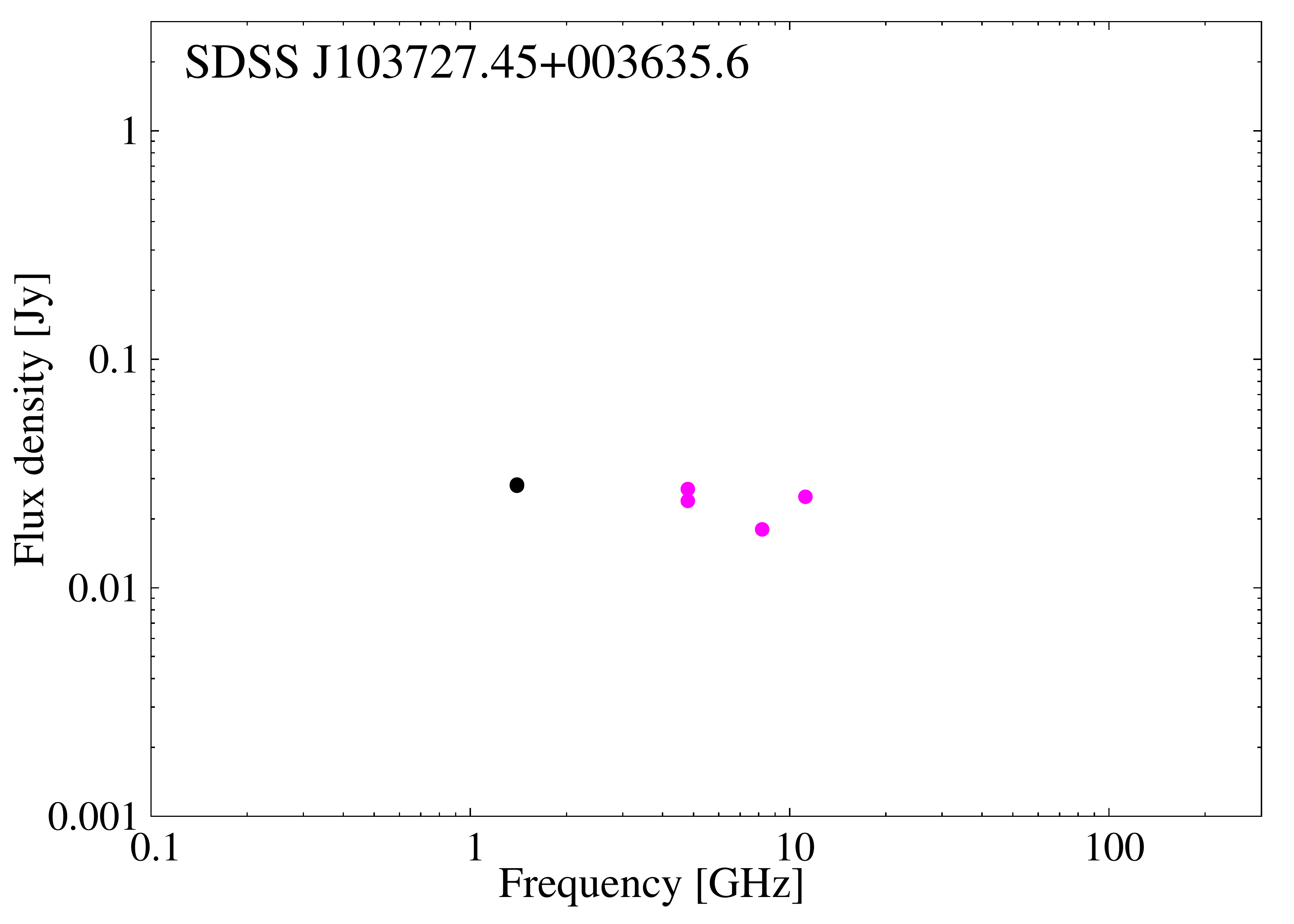}
\caption{Radio spectrum of J103727.45+003. Only detections. Colours as in Fig.~\ref{fig:spec1h0323}.} \label{fig:specj103727}
\end{minipage}\hfill
\end{figure*}

\begin{figure*}[ht!]
\centering
\begin{minipage}{0.47\textwidth}
\centering
\includegraphics[width=1\textwidth]{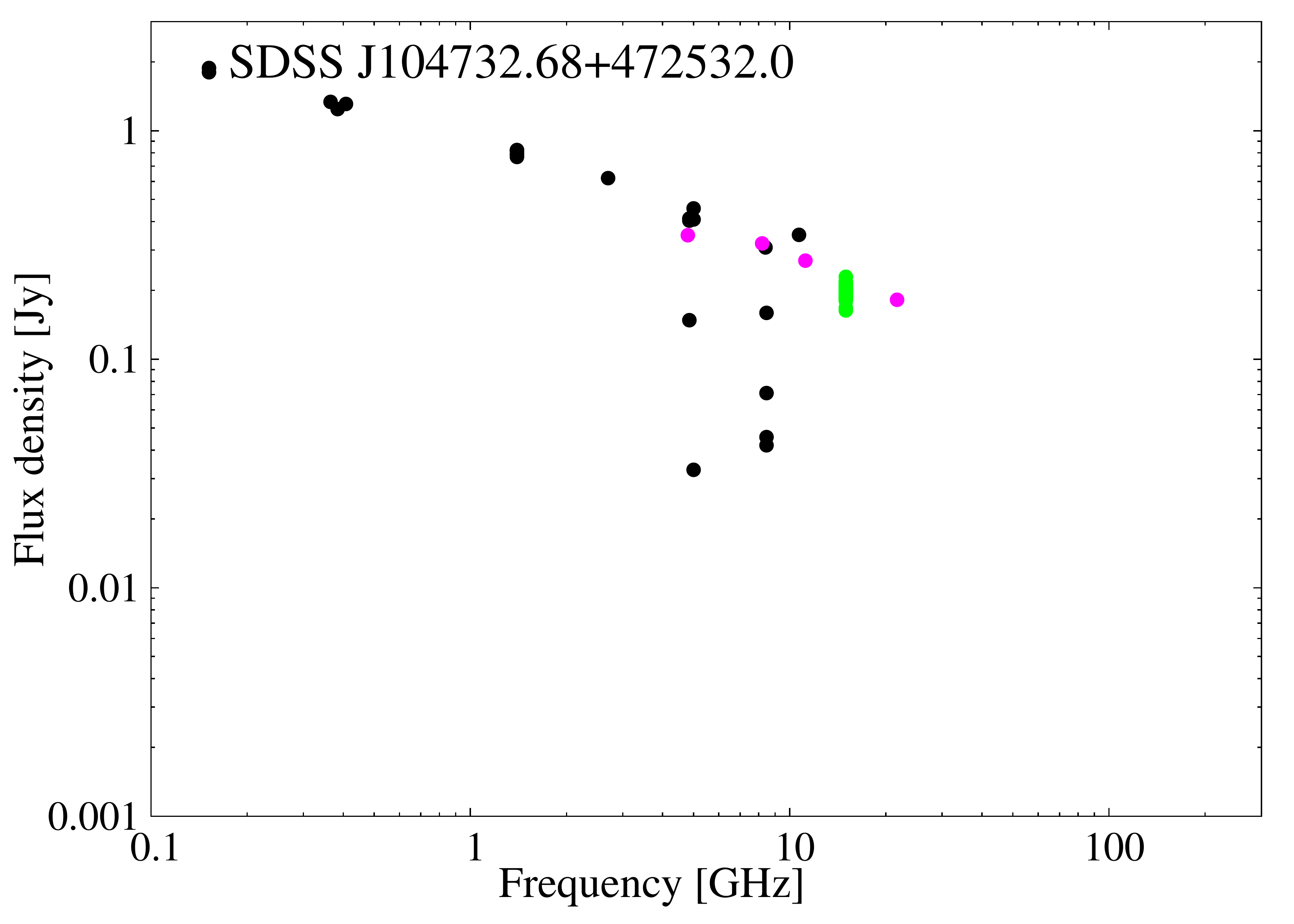}
\caption{Radio spectrum of J104732.68+472. Only detections. Colours as in Fig.~\ref{fig:spec1h0323}.} \label{fig:specj104732}
\end{minipage}
\begin{minipage}{0.47\textwidth}
\centering
\includegraphics[width=1\textwidth]{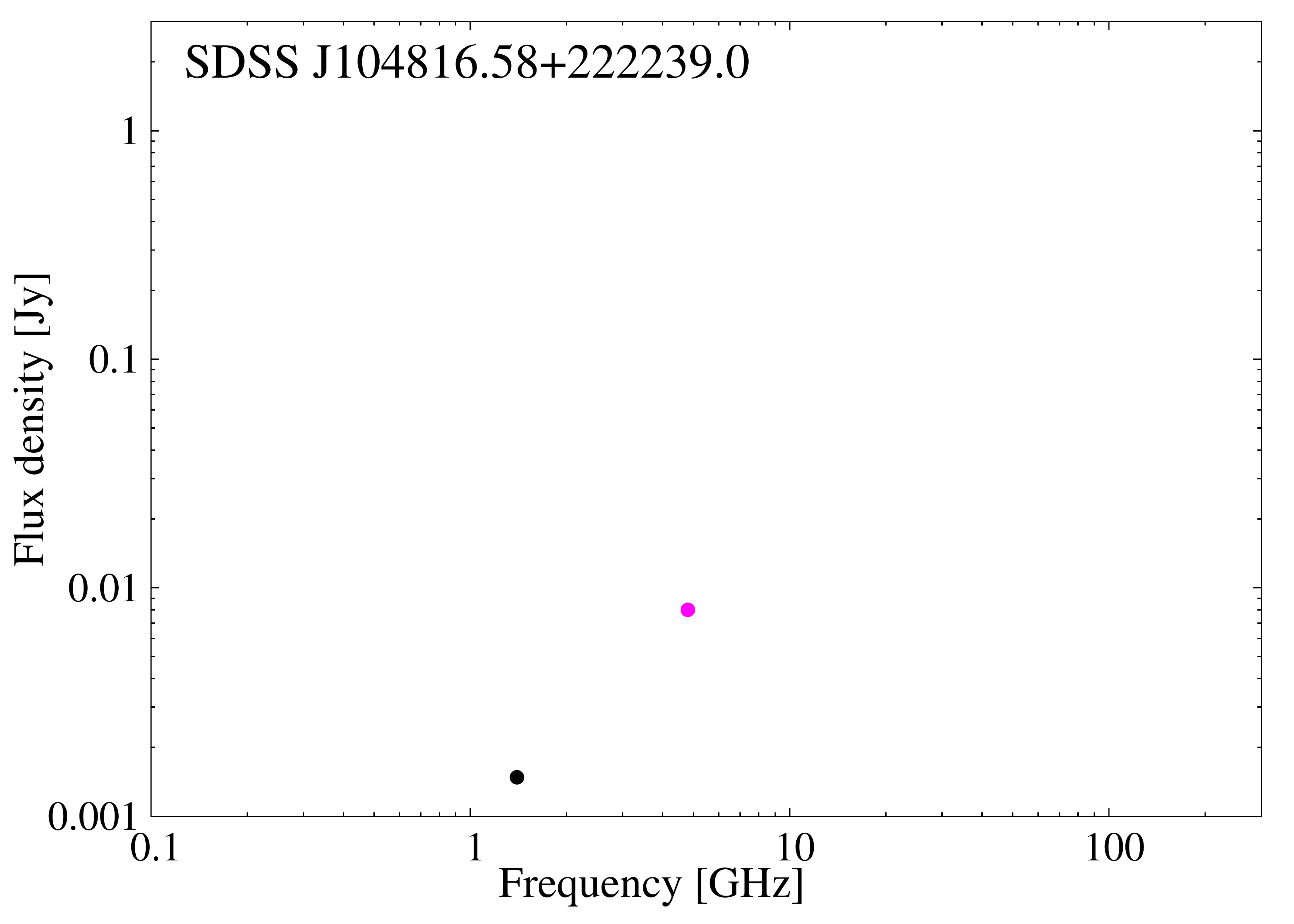}
\caption{Radio spectrum of J104816.58+222. Only detections. Colours as in Fig.~\ref{fig:spec1h0323}.} \label{fig:specj104816}
\end{minipage}\hfill
\end{figure*}

\begin{figure*}[ht!]
\centering
\begin{minipage}{0.47\textwidth}
\centering
\includegraphics[width=1\textwidth]{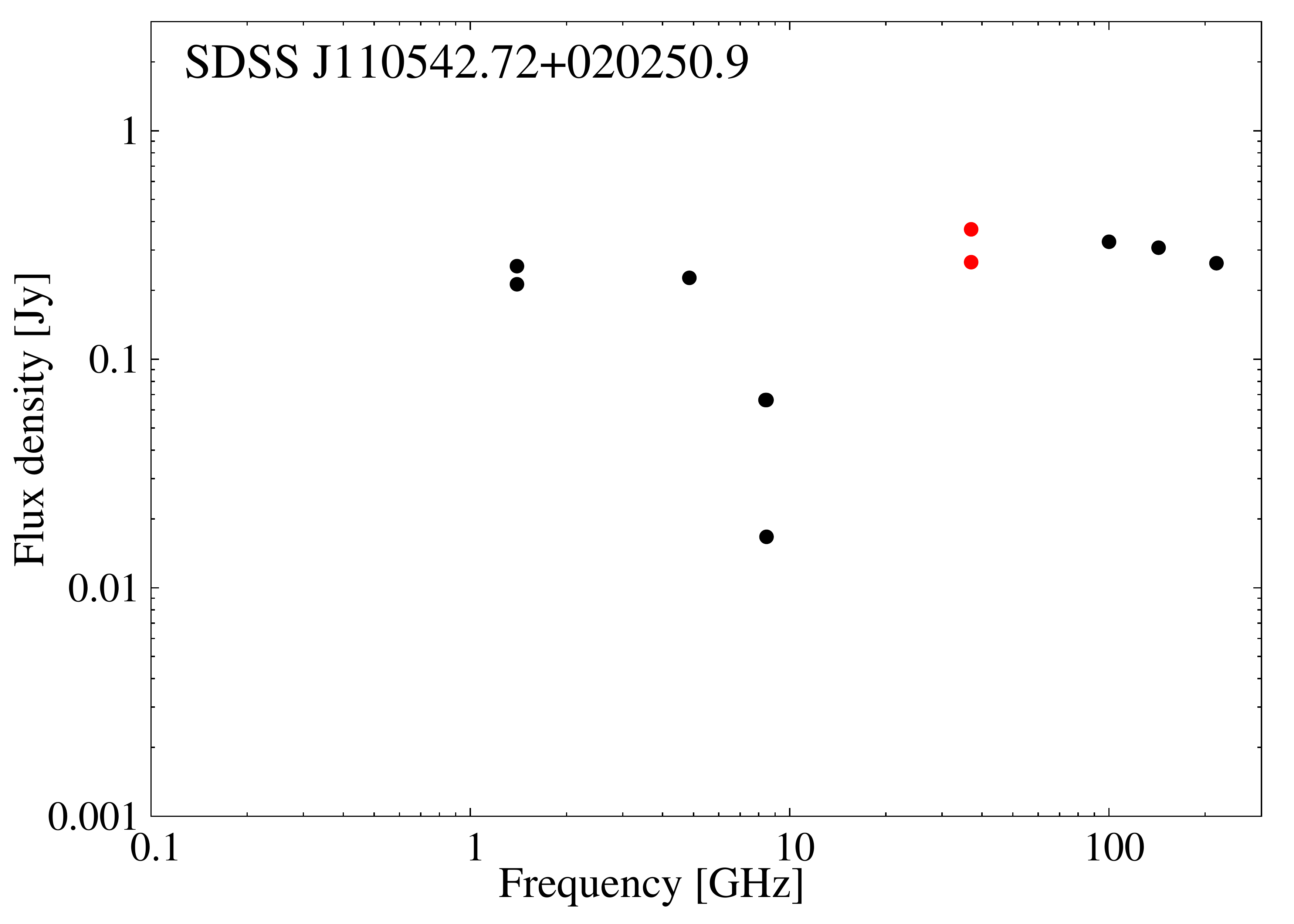}
\caption{Radio spectrum of J110542.72+020. Only detections. Colours as in Fig.~\ref{fig:spec1h0323}.} \label{fig:specj110542}
\end{minipage}
\begin{minipage}{0.47\textwidth}
\centering
\includegraphics[width=1\textwidth]{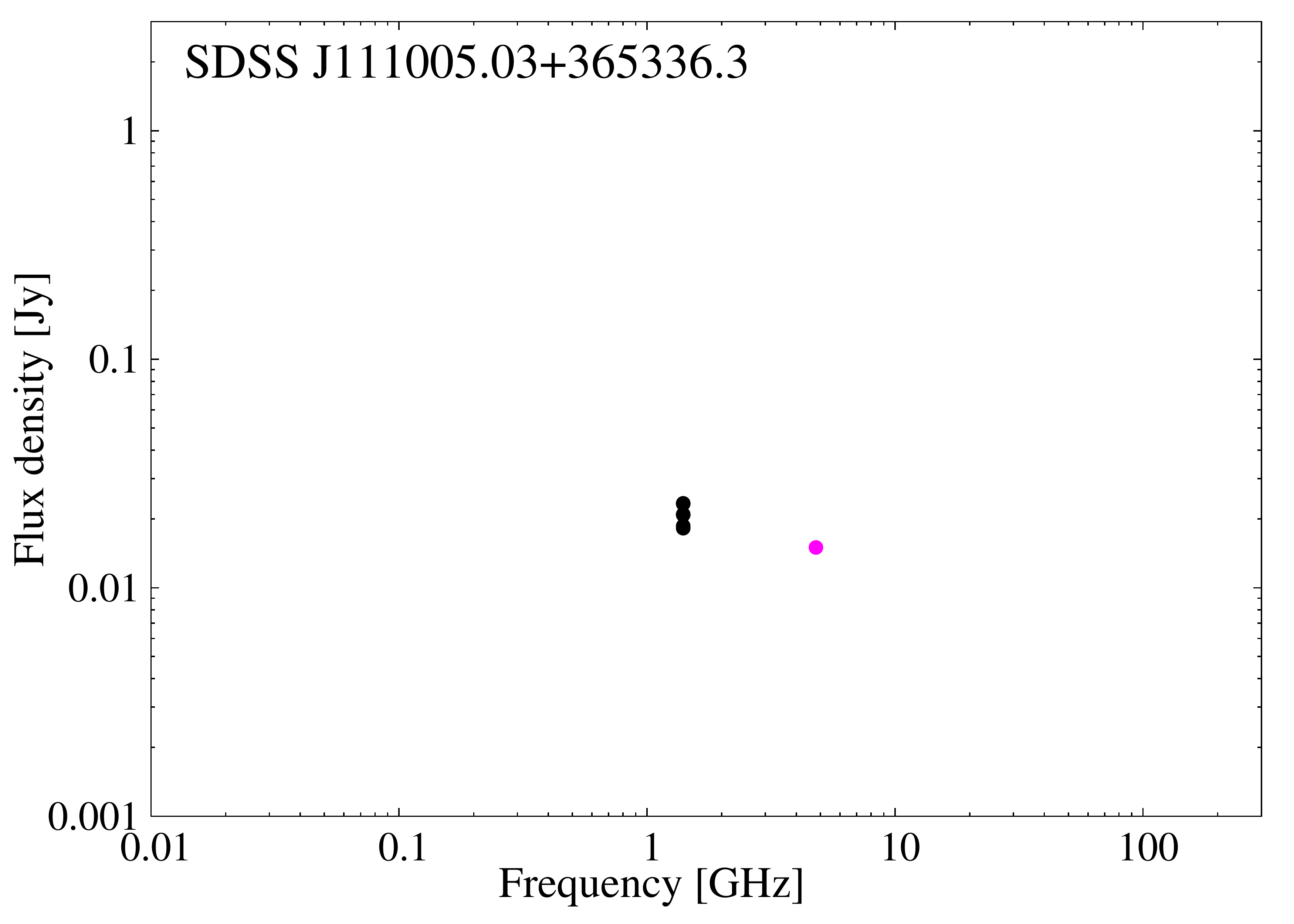}
\caption{Radio spectrum of J111005.03+365. Only detections. Colours as in Fig.~\ref{fig:spec1h0323}.} \label{fig:specj111005}
\end{minipage}\hfill
\end{figure*}

\clearpage

\newpage

\begin{figure*}[ht!]
\centering
\begin{minipage}{0.47\textwidth}
\centering
\includegraphics[width=1\textwidth]{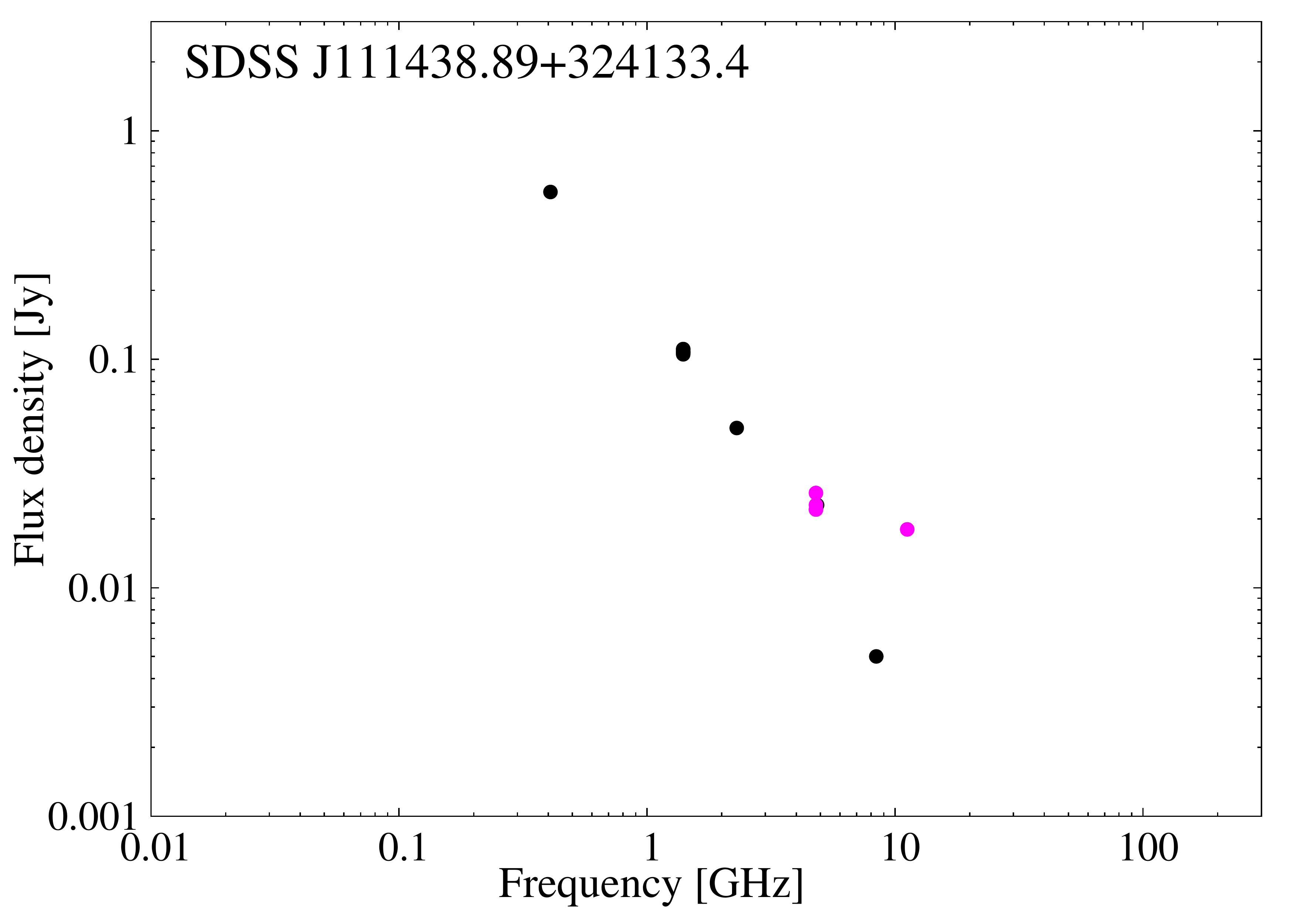}
\caption{Radio spectrum of J111438.89+324. Only detections. Colours as in Fig.~\ref{fig:spec1h0323}.} \label{fig:specj111438}
\end{minipage}
\begin{minipage}{0.47\textwidth}
\centering
\includegraphics[width=1\textwidth]{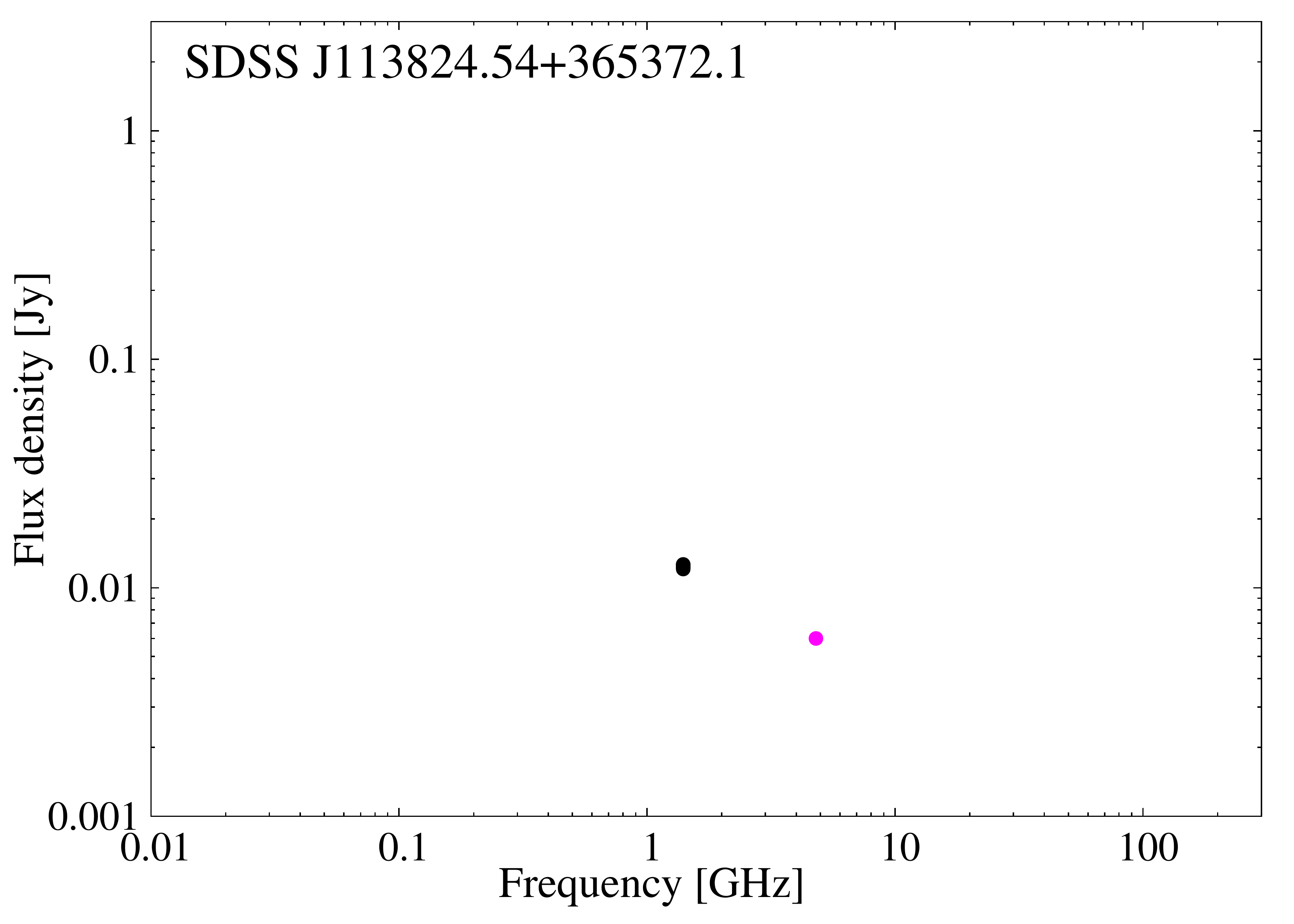}
\caption{Radio spectrum of J113824.54+365. Only detections. Colours as in Fig.~\ref{fig:spec1h0323}.} \label{fig:specj113824}
\end{minipage}\hfill
\end{figure*}

\begin{figure*}[ht!]
\centering
\begin{minipage}{0.47\textwidth}
\centering
\includegraphics[width=1\textwidth]{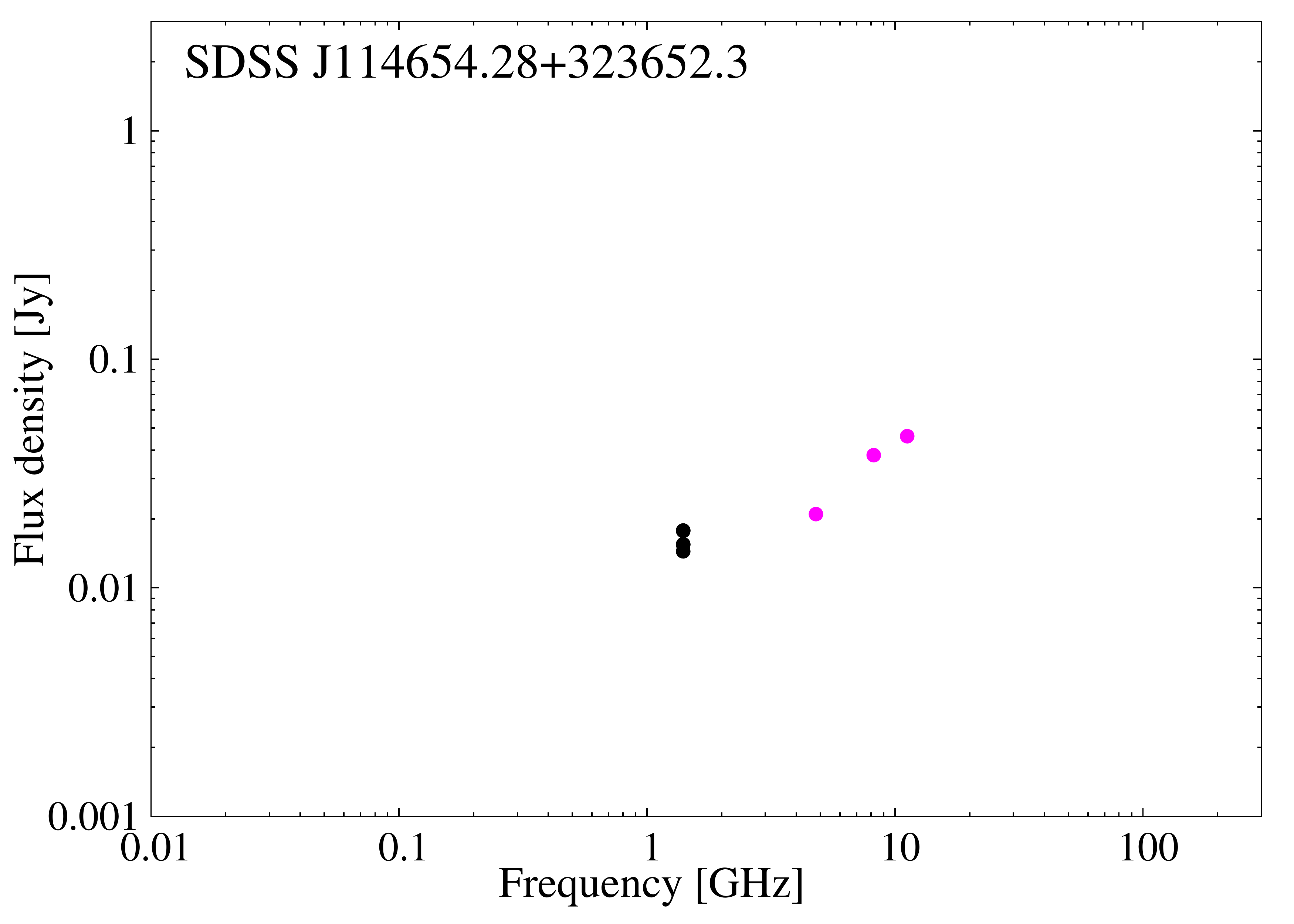}
\caption{Radio spectrum of J114654.28+323. Only detections. Colours as in Fig.~\ref{fig:spec1h0323}.} \label{fig:specj114654}
\end{minipage}
\begin{minipage}{0.47\textwidth}
\centering
\includegraphics[width=1\textwidth]{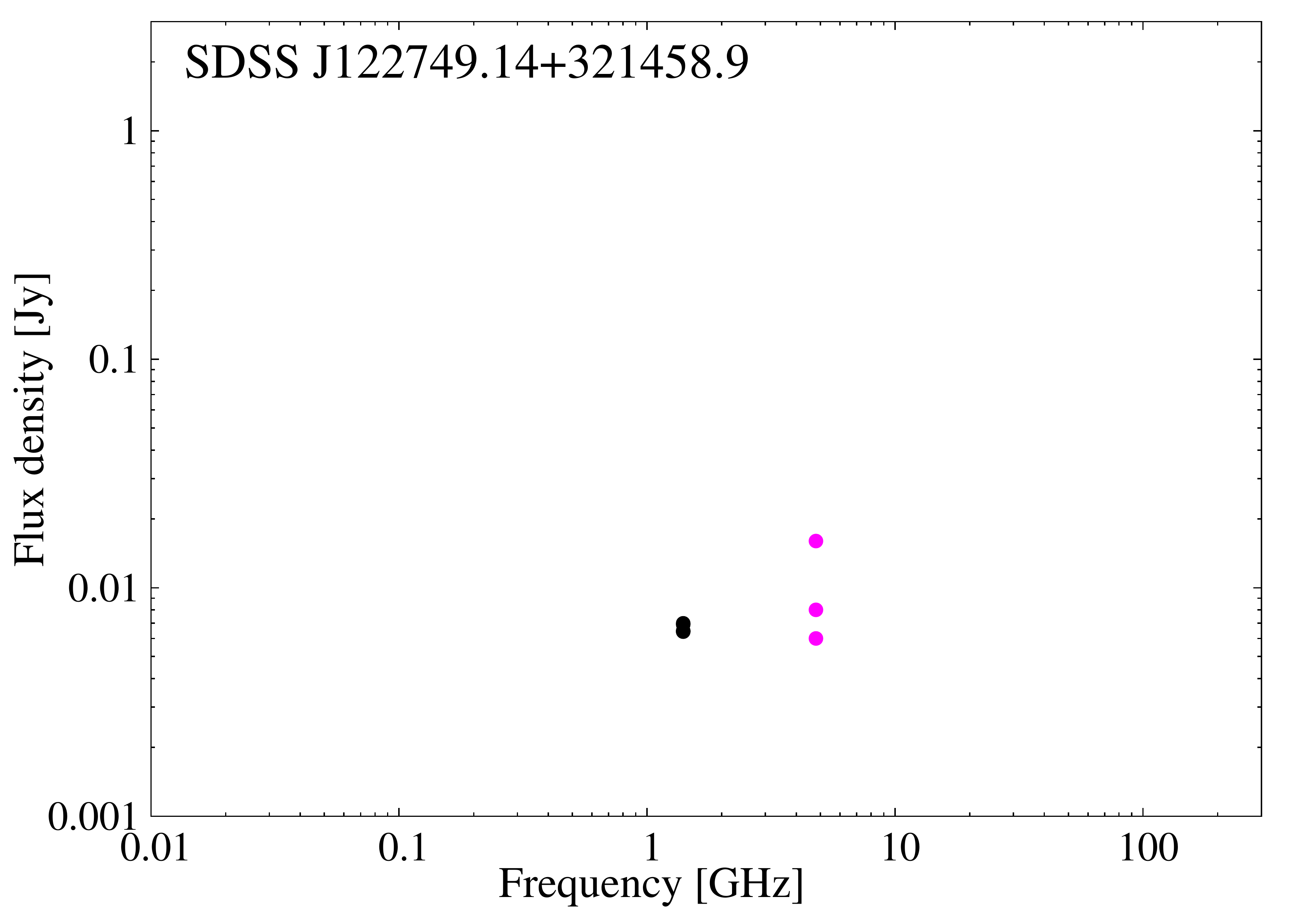}
\caption{Radio spectrum of J122749.14+321. Only detections. Colours as in Fig.~\ref{fig:spec1h0323}.} \label{fig:specj122749}
\end{minipage}\hfill
\end{figure*}

\begin{figure*}[ht!]
\centering
\begin{minipage}{0.47\textwidth}
\centering
\includegraphics[width=1\textwidth]{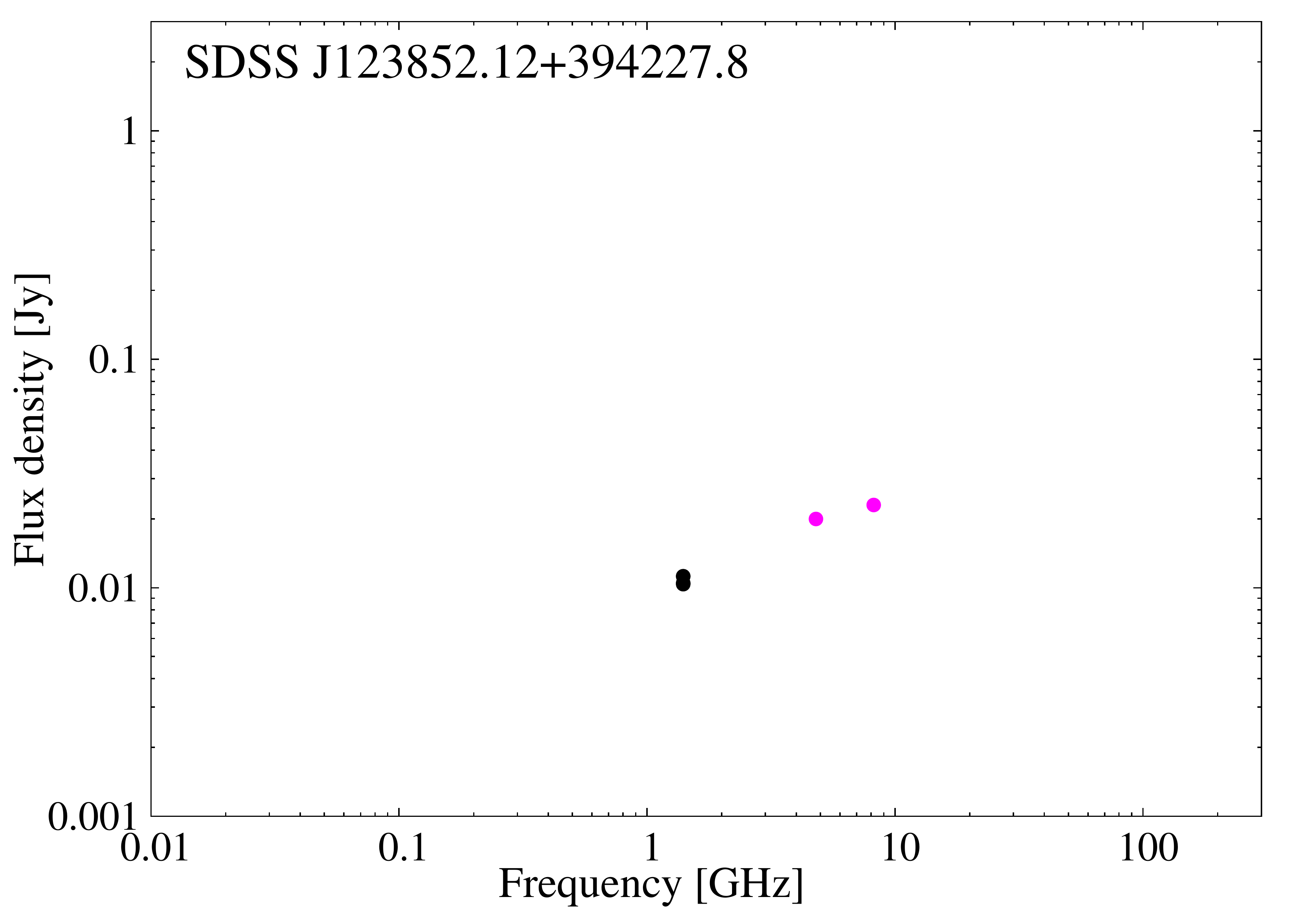}
\caption{Radio spectrum of J123852.12+394. Only detections. Colours as in Fig.~\ref{fig:spec1h0323}.} \label{fig:specj123852}
\end{minipage}
\begin{minipage}{0.47\textwidth}
\centering
\includegraphics[width=1\textwidth]{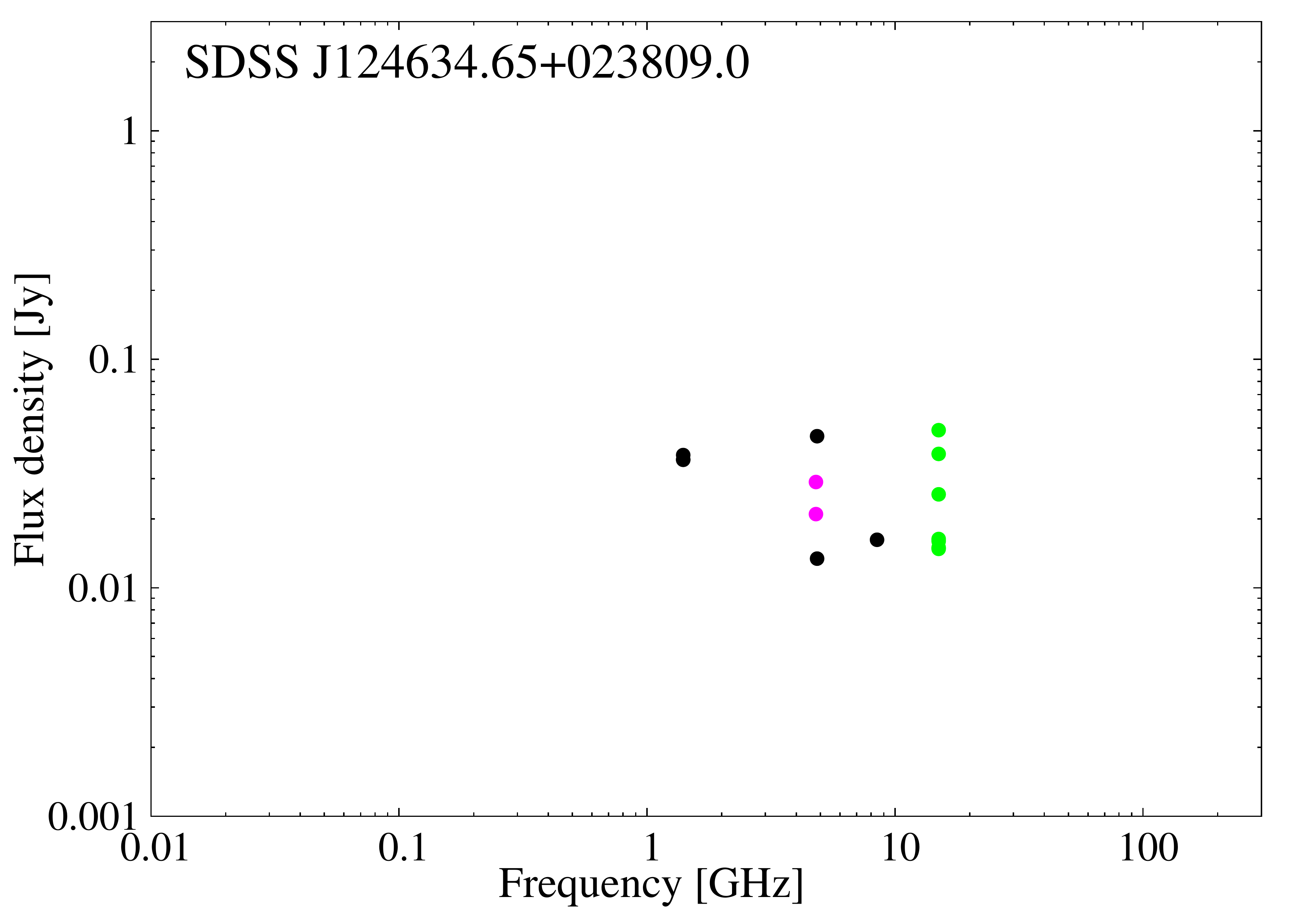}
\caption{Radio spectrum of J124634.65+023. Only detections. Colours as in Fig.~\ref{fig:spec1h0323}.} \label{fig:specj124634}
\end{minipage}\hfill
\end{figure*}

\clearpage

\newpage

\begin{figure*}[ht!]
\centering
\begin{minipage}{0.47\textwidth}
\centering
\includegraphics[width=1\textwidth]{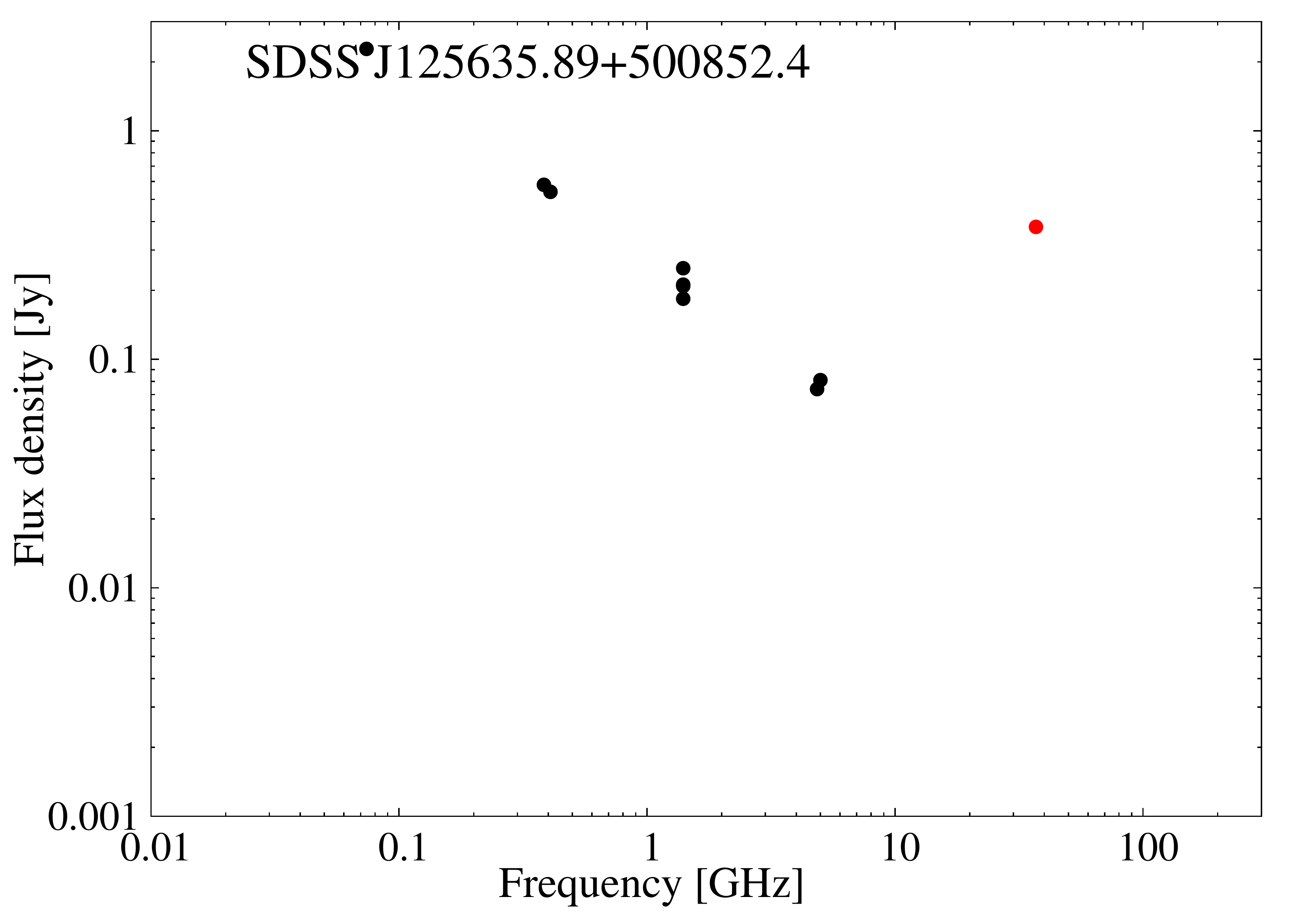}
\caption{Radio spectrum of J125635.89+500. Only detections. Colours as in Fig.~\ref{fig:spec1h0323}.} \label{fig:specj125635}
\end{minipage}
\begin{minipage}{0.47\textwidth}
\centering
\includegraphics[width=1\textwidth]{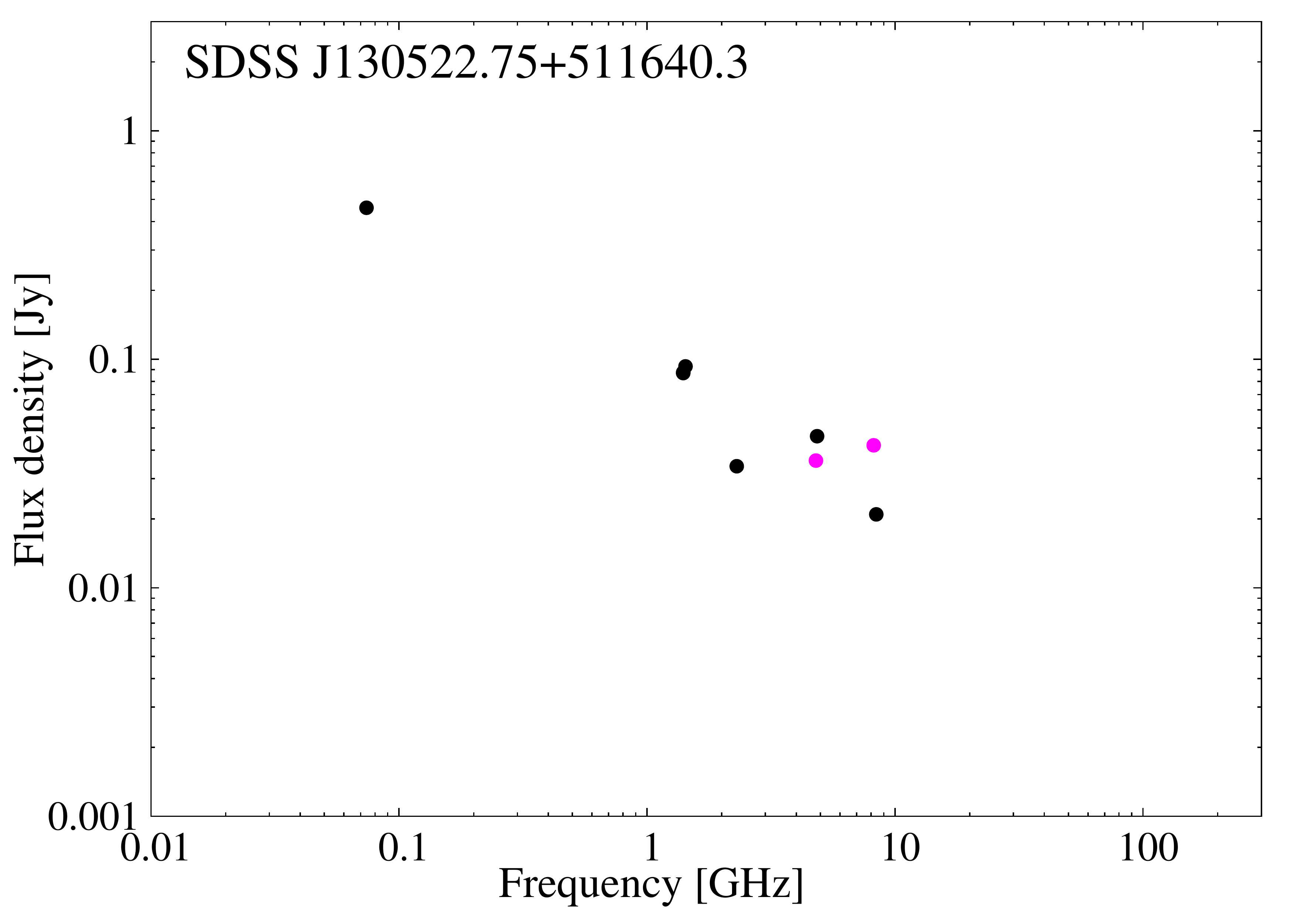}
\caption{Radio spectrum of J130522.75+511. Only detections. Colours as in Fig.~\ref{fig:spec1h0323}.} \label{fig:specj130522}
\end{minipage}\hfill
\end{figure*}

\begin{figure*}[ht!]
\centering
\begin{minipage}{0.47\textwidth}
\centering
\includegraphics[width=1\textwidth]{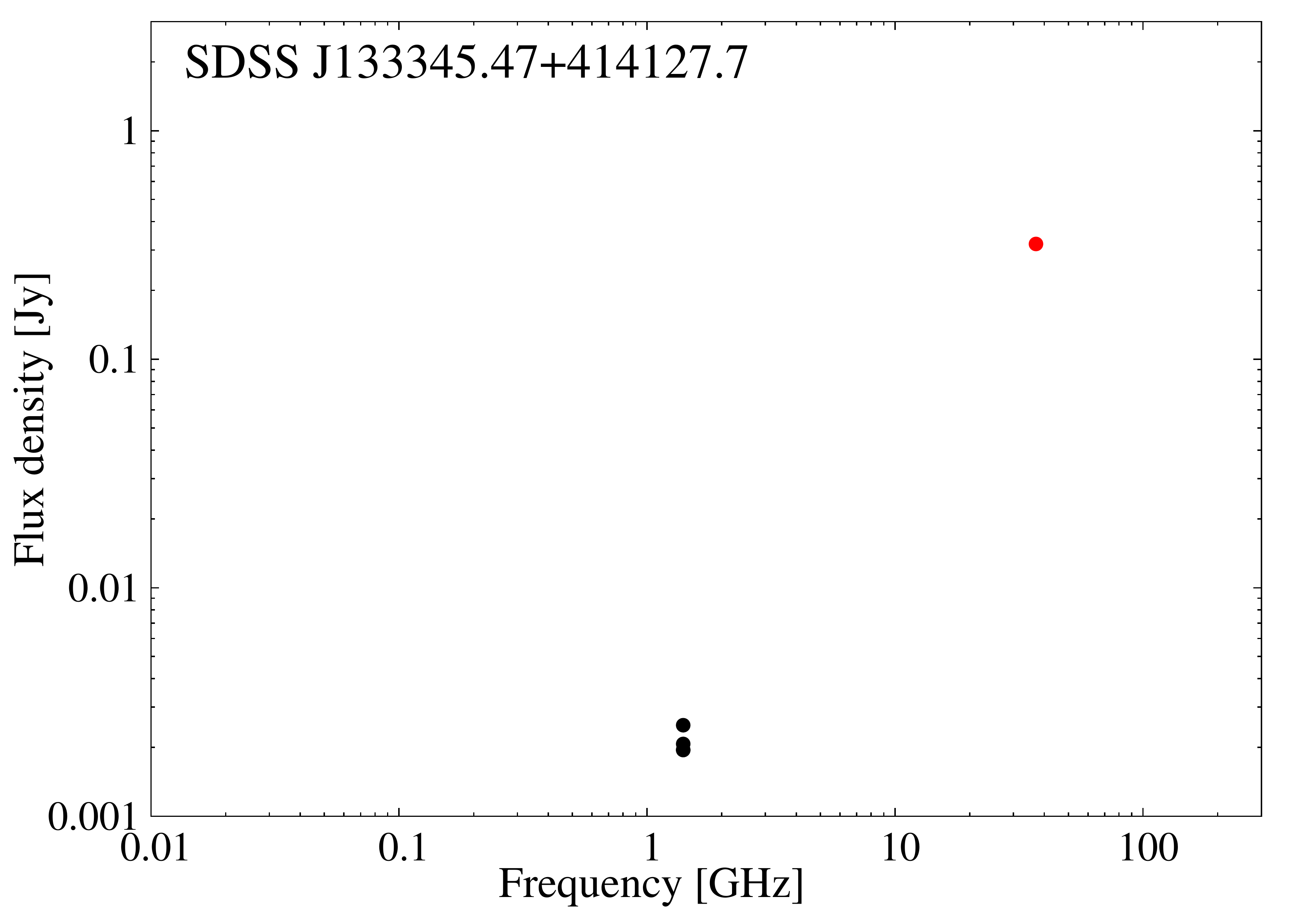}
\caption{Radio spectrum of J133345.47+414. Only detections. Colours as in Fig.~\ref{fig:spec1h0323}.} \label{fig:specj133345}
\end{minipage}
\begin{minipage}{0.47\textwidth}
\centering
\includegraphics[width=1\textwidth]{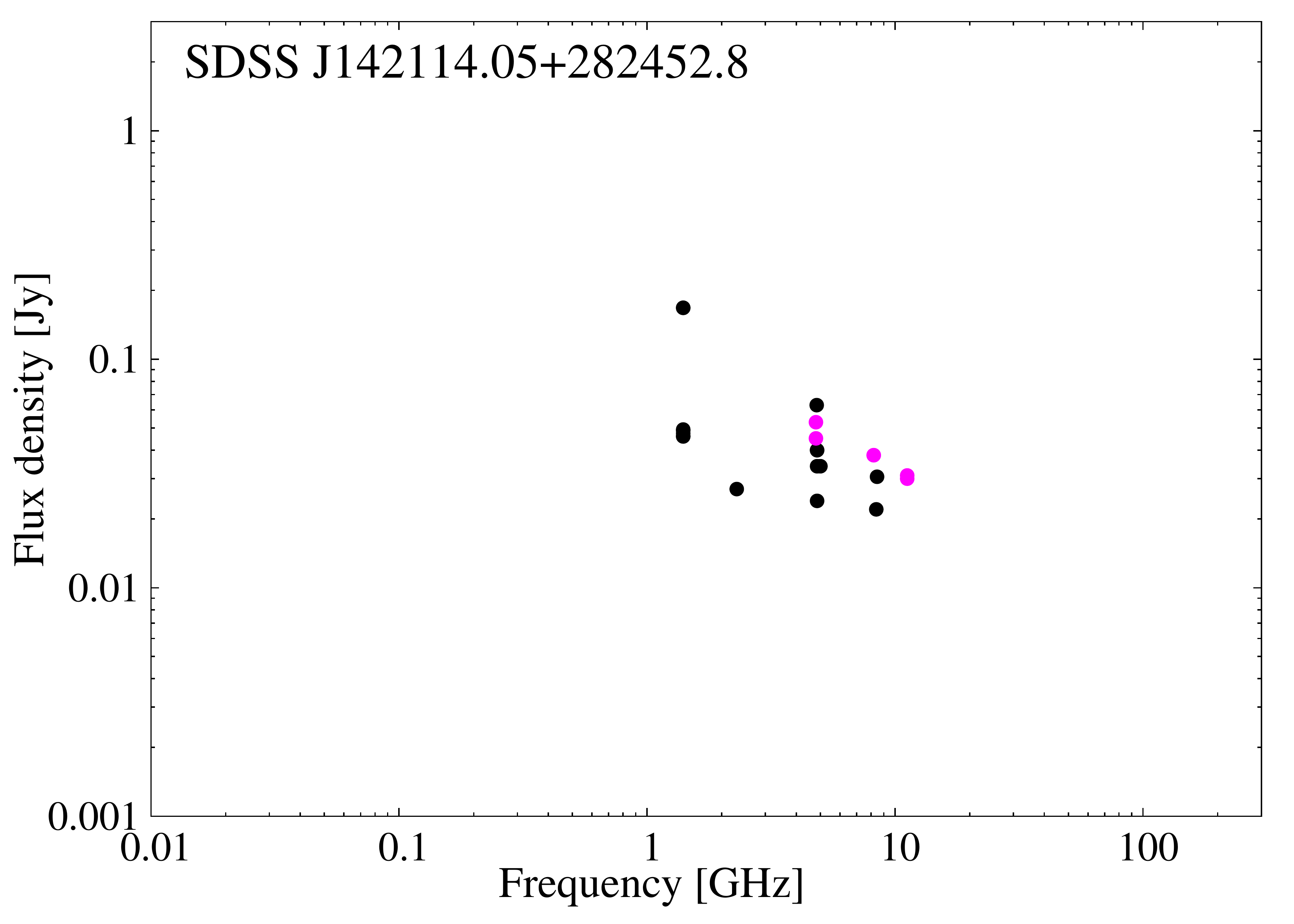}
\caption{Radio spectrum of J142114.05+282. Only detections. Colours as in Fig.~\ref{fig:spec1h0323}.} \label{fig:specj142114}
\end{minipage}\hfill
\end{figure*}

\begin{figure*}[ht!]
\centering
\begin{minipage}{0.47\textwidth}
\centering
\includegraphics[width=1\textwidth]{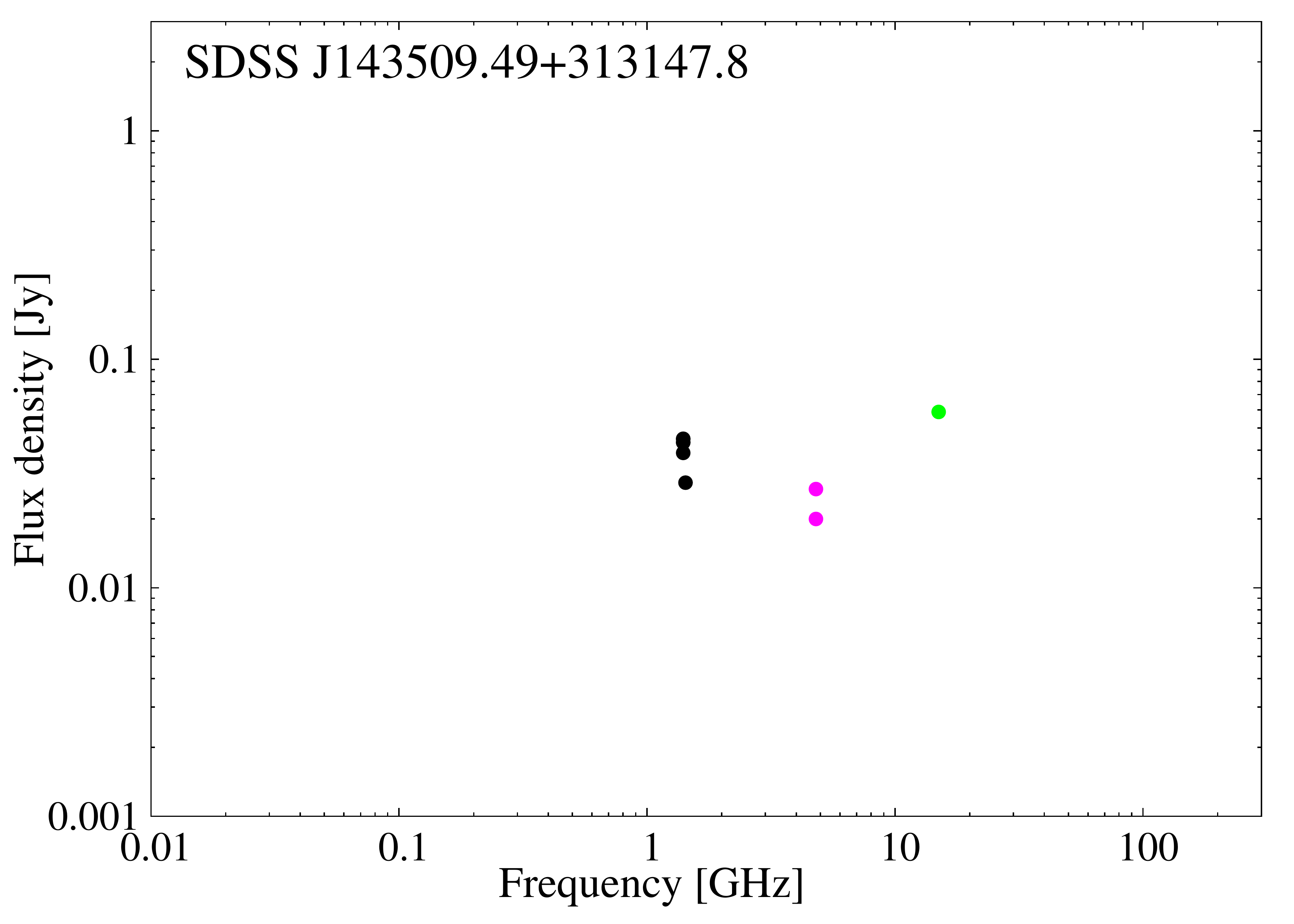}
\caption{Radio spectrum of J143509.49+313. Only detections. Colours as in Fig.~\ref{fig:spec1h0323}.} \label{fig:specj143509}
\end{minipage}
\begin{minipage}{0.47\textwidth}
\centering
\includegraphics[width=1\textwidth]{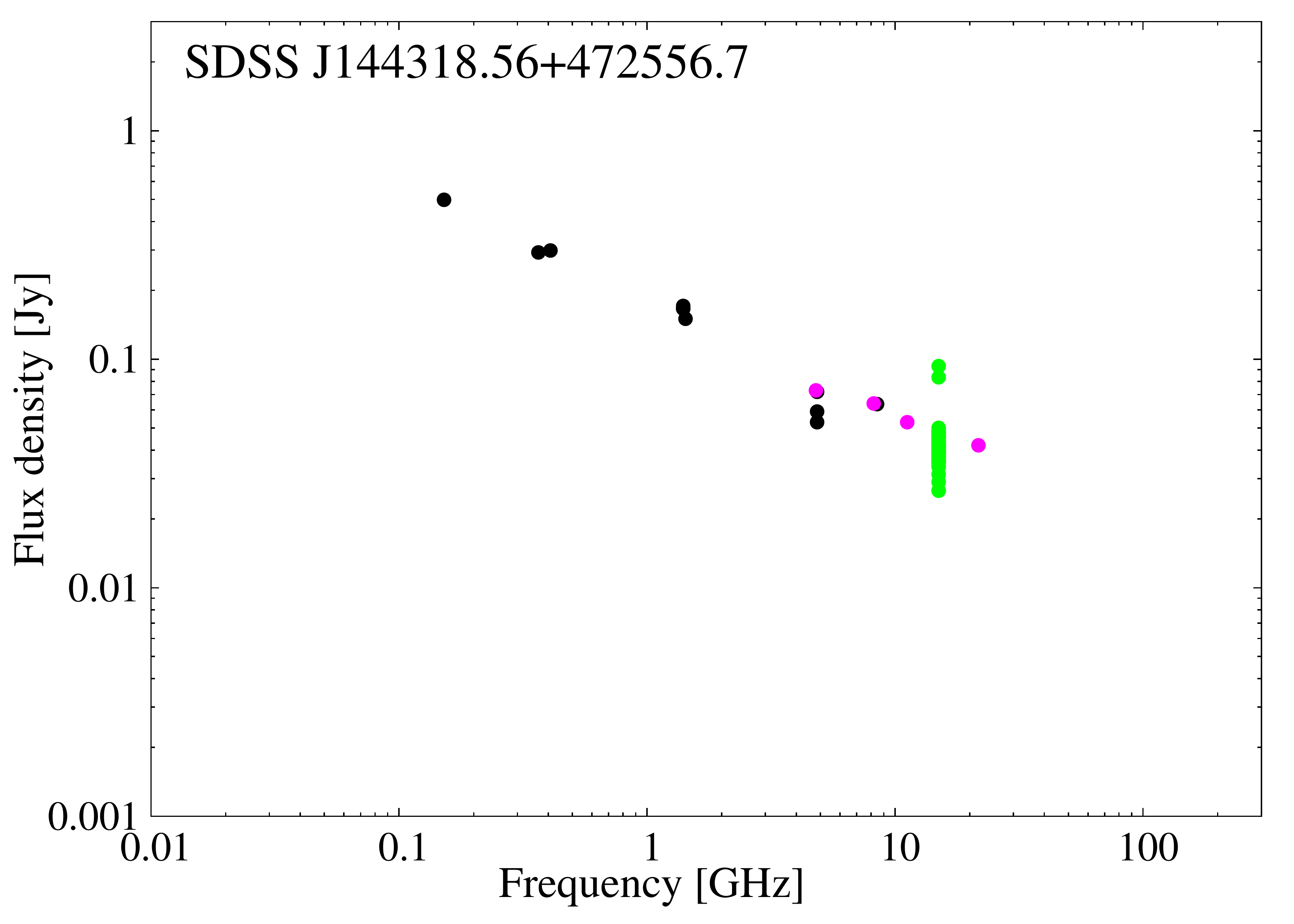}
\caption{Radio spectrum of J144318.56+472. Only detections. Colours as in Fig.~\ref{fig:spec1h0323}.} \label{fig:specj144318}
\end{minipage}\hfill
\end{figure*}

\clearpage

\newpage

\begin{figure*}[ht!]
\centering
\begin{minipage}{0.47\textwidth}
\centering
\includegraphics[width=1\textwidth]{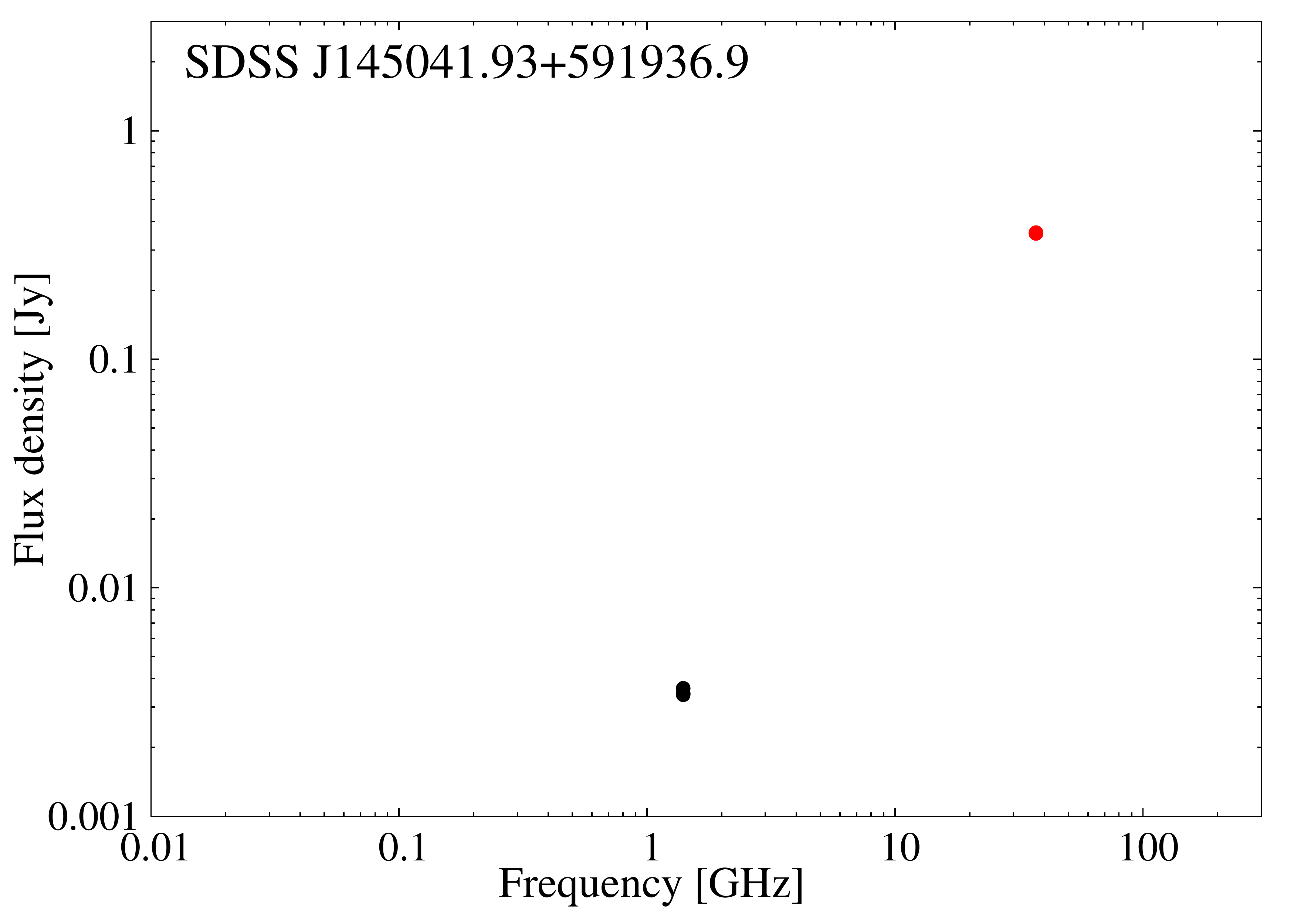}
\caption{Radio spectrum of J145041.93+591. Only detections. Colours as in Fig.~\ref{fig:spec1h0323}.} \label{fig:specj145041}
\end{minipage}
\begin{minipage}{0.47\textwidth}
\centering
\includegraphics[width=1\textwidth]{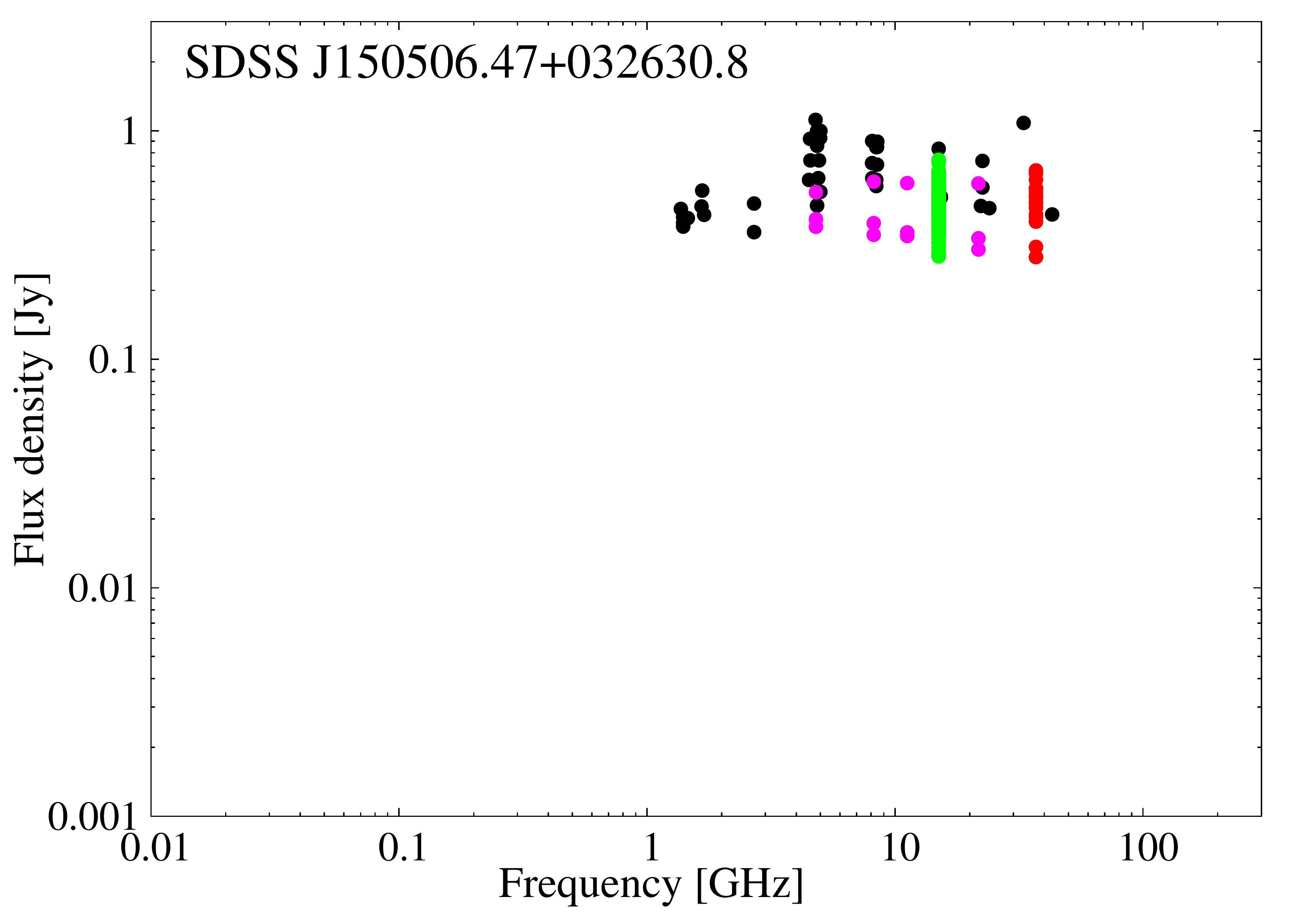}
\caption{Radio spectrum of J150506.47+032. Only detections. Colours as in Fig.~\ref{fig:spec1h0323}.} \label{fig:specj150506}
\end{minipage}\hfill
\end{figure*}

\newpage

\begin{figure*}[ht!]
\centering
\begin{minipage}{0.47\textwidth}
\centering
\includegraphics[width=1\textwidth]{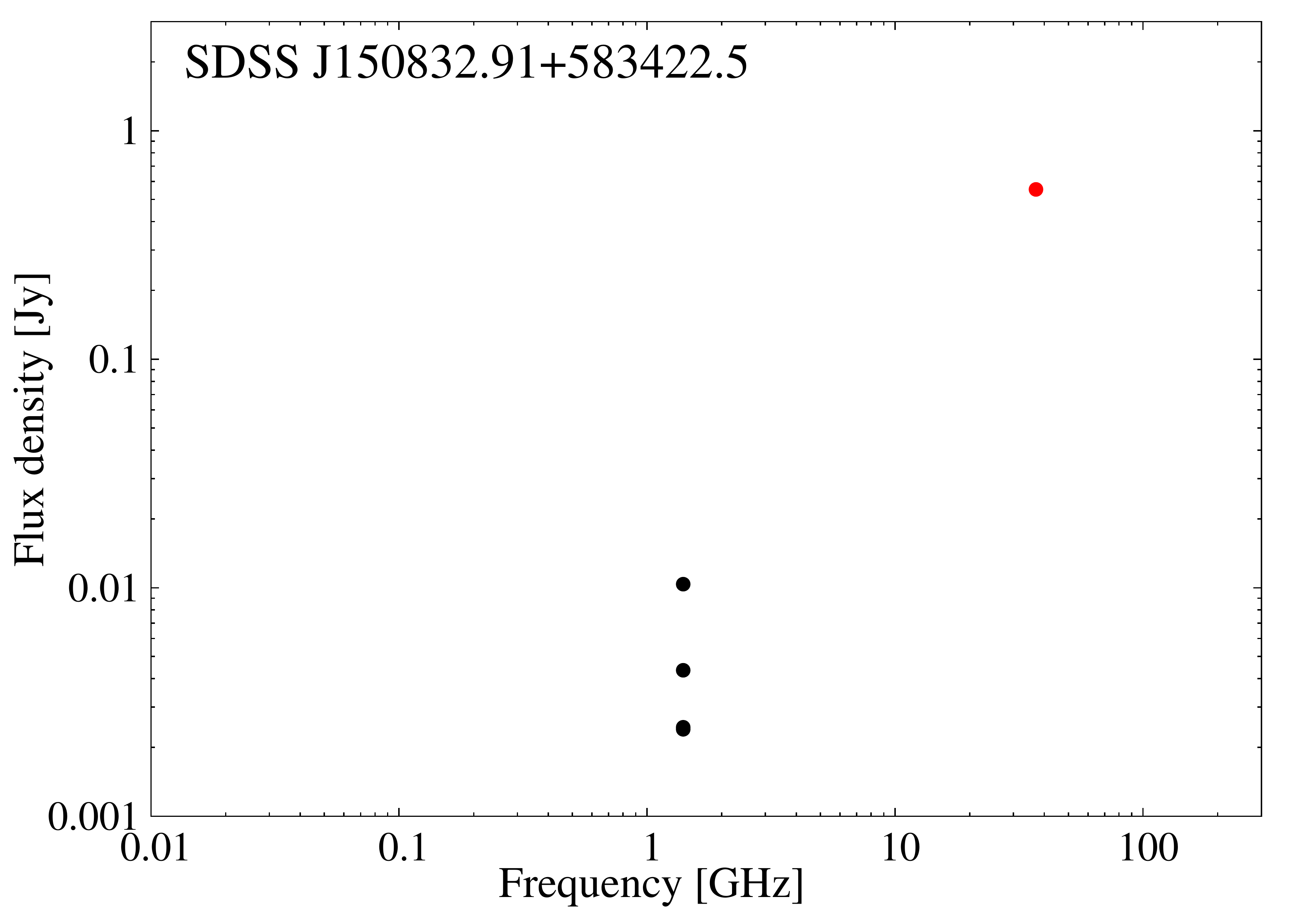}
\caption{Radio spectrum of J150832.91+583. Only detections. Colours as in Fig.~\ref{fig:spec1h0323}.} \label{fig:specj150832}
\end{minipage}
\begin{minipage}{0.47\textwidth}
\centering
\includegraphics[width=1\textwidth]{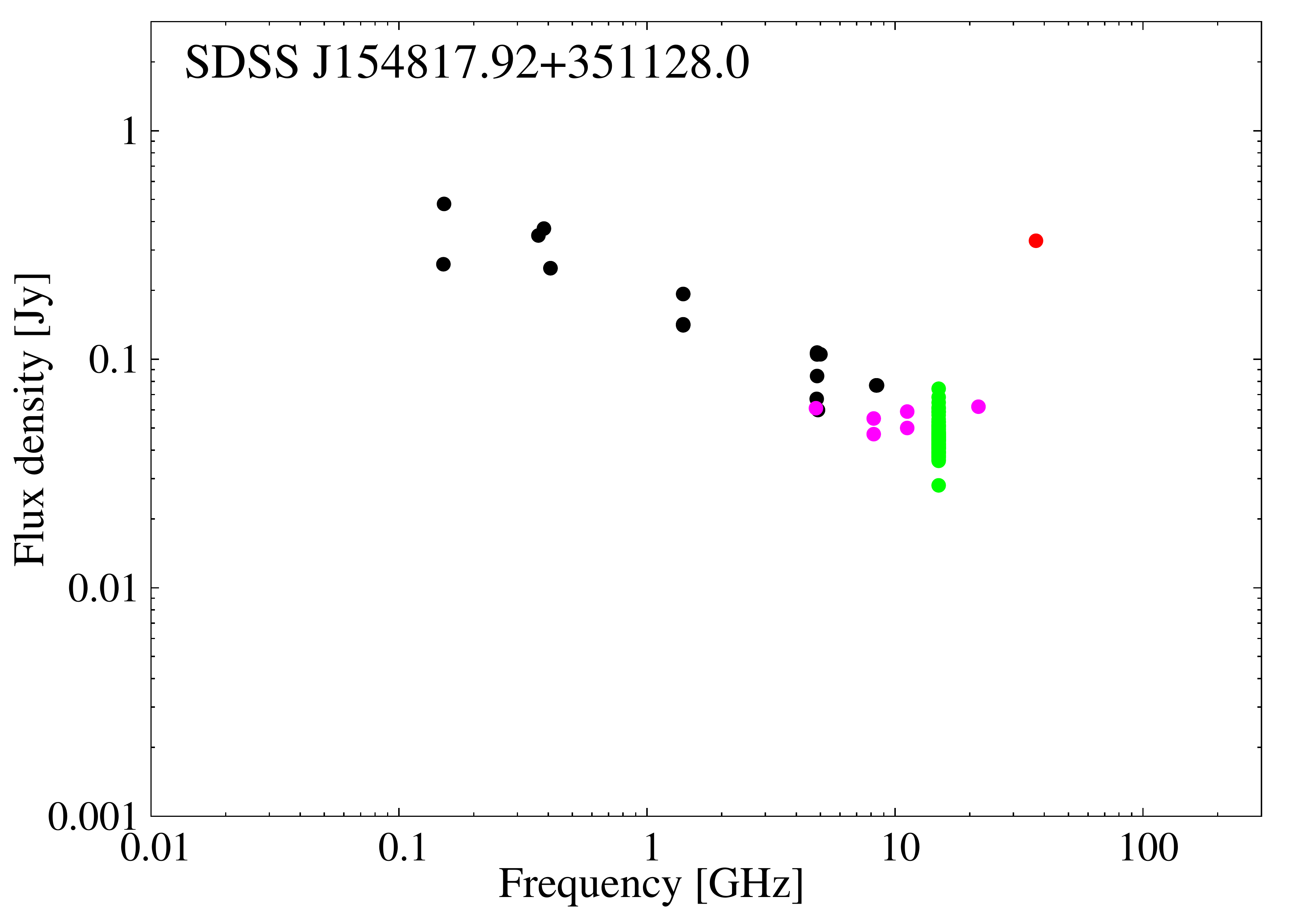}
\caption{Radio spectrum of J154817.92+351. Only detections. Colours as in Fig.~\ref{fig:spec1h0323}.} \label{fig:specj154817}
\end{minipage}\hfill
\end{figure*}

\begin{figure*}[ht!]
\centering
\begin{minipage}{0.47\textwidth}
\centering
\includegraphics[width=1\textwidth]{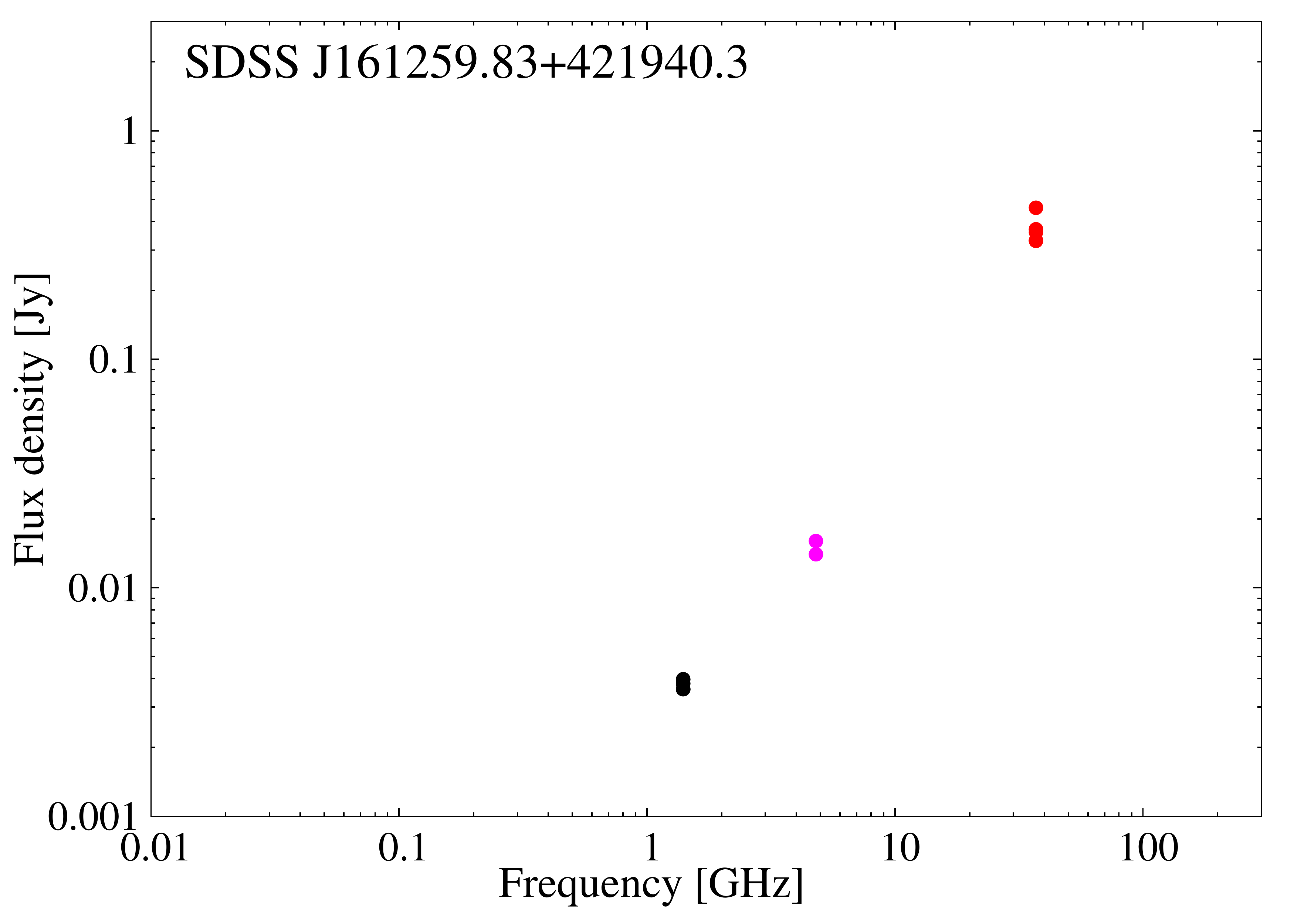}
\caption{Radio spectrum of J161259.83+421. Only detections. Colours as in Fig.~\ref{fig:spec1h0323}.} \label{fig:specj161259}
\end{minipage}
\begin{minipage}{0.47\textwidth}
\centering
\includegraphics[width=1\textwidth]{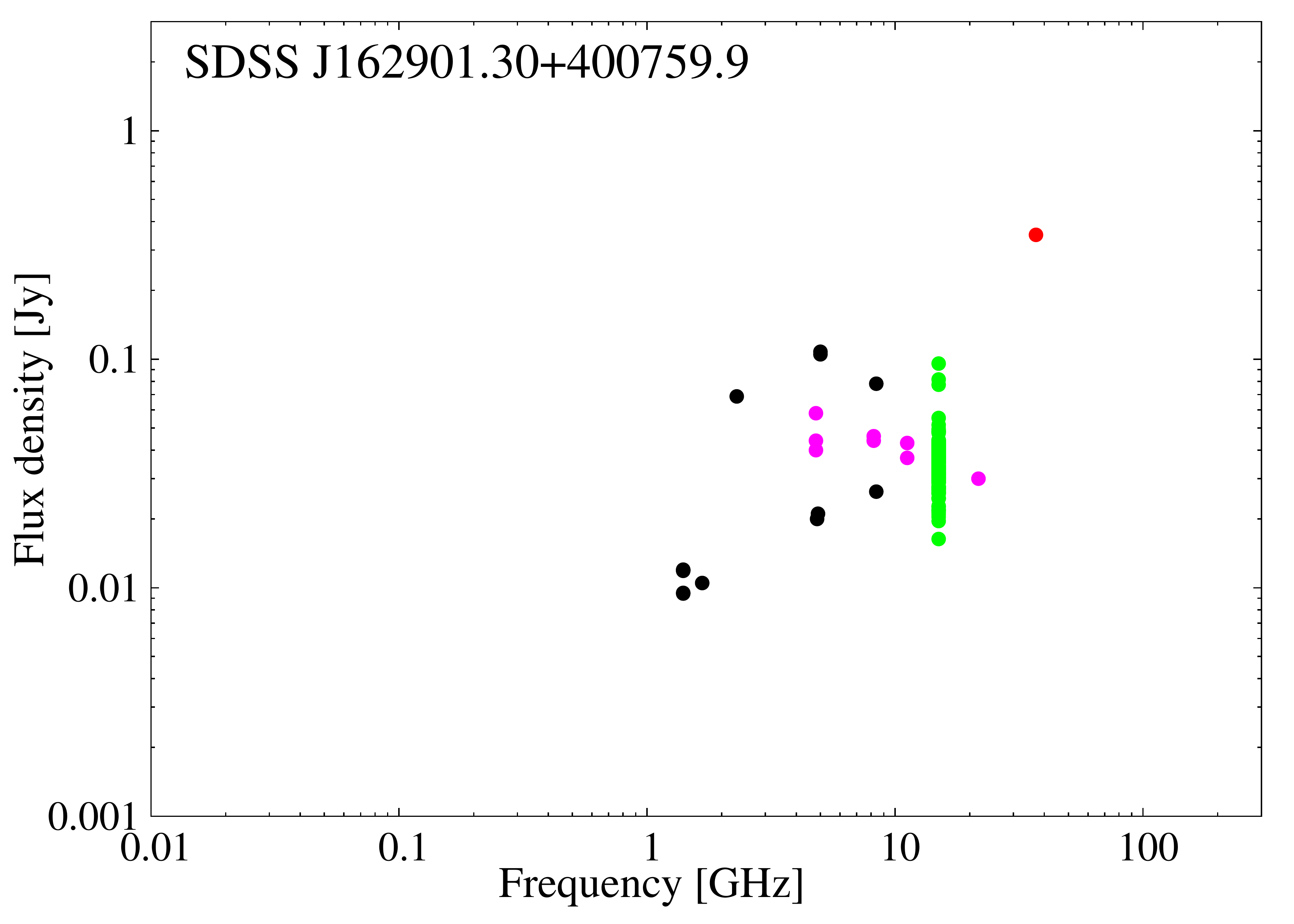}
\caption{Radio spectrum of J162901.30+400. Only detections. Colours as in Fig.~\ref{fig:spec1h0323}.} \label{fig:specj162901}
\end{minipage}\hfill
\end{figure*}

\clearpage

\newpage

\begin{figure*}[ht!]
\centering
\begin{minipage}{0.47\textwidth}
\centering
\includegraphics[width=1\textwidth]{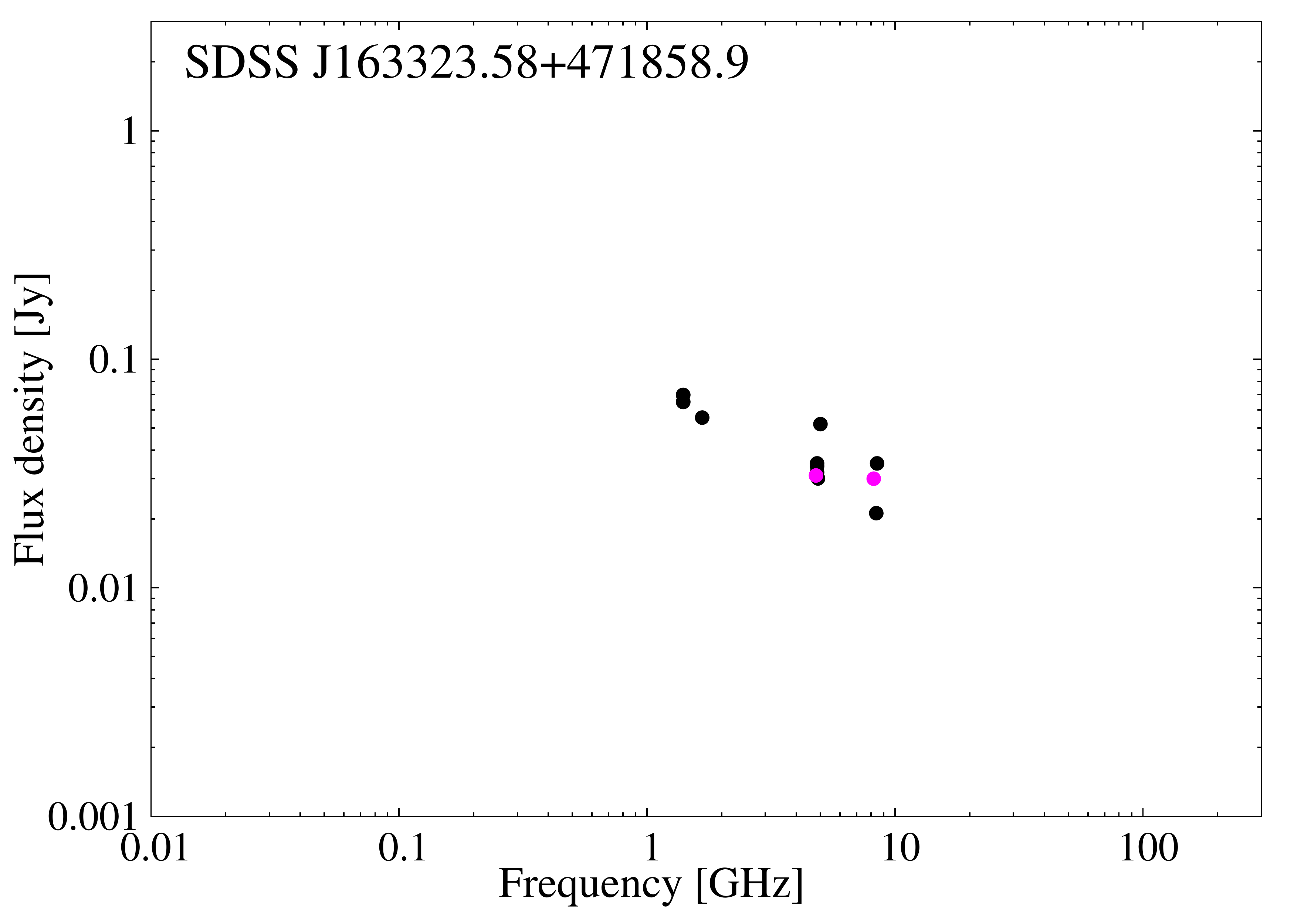}
\caption{Radio spectrum of J163323.58+471. Only detections. Colours as in Fig.~\ref{fig:spec1h0323}.} \label{fig:specj163323}
\end{minipage}
\begin{minipage}{0.47\textwidth}
\centering
\includegraphics[width=1\textwidth]{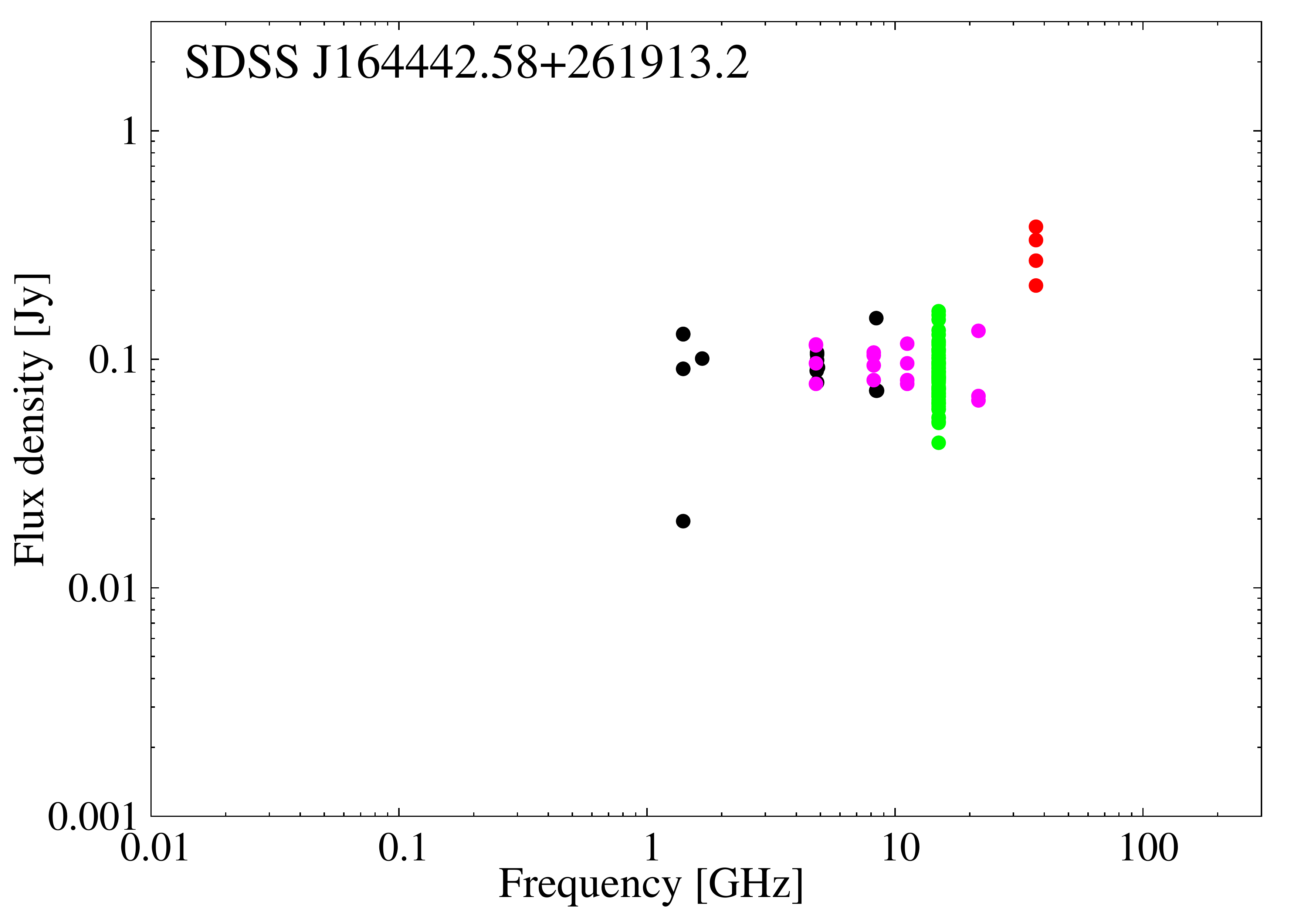}
\caption{Radio spectrum of J164442.53+261. Only detections. Colours as in Fig.~\ref{fig:spec1h0323}.} \label{fig:specj164442}
\end{minipage}\hfill
\end{figure*}

\begin{figure*}[ht!]
\centering
\begin{minipage}{0.47\textwidth}
\centering
\includegraphics[width=1\textwidth]{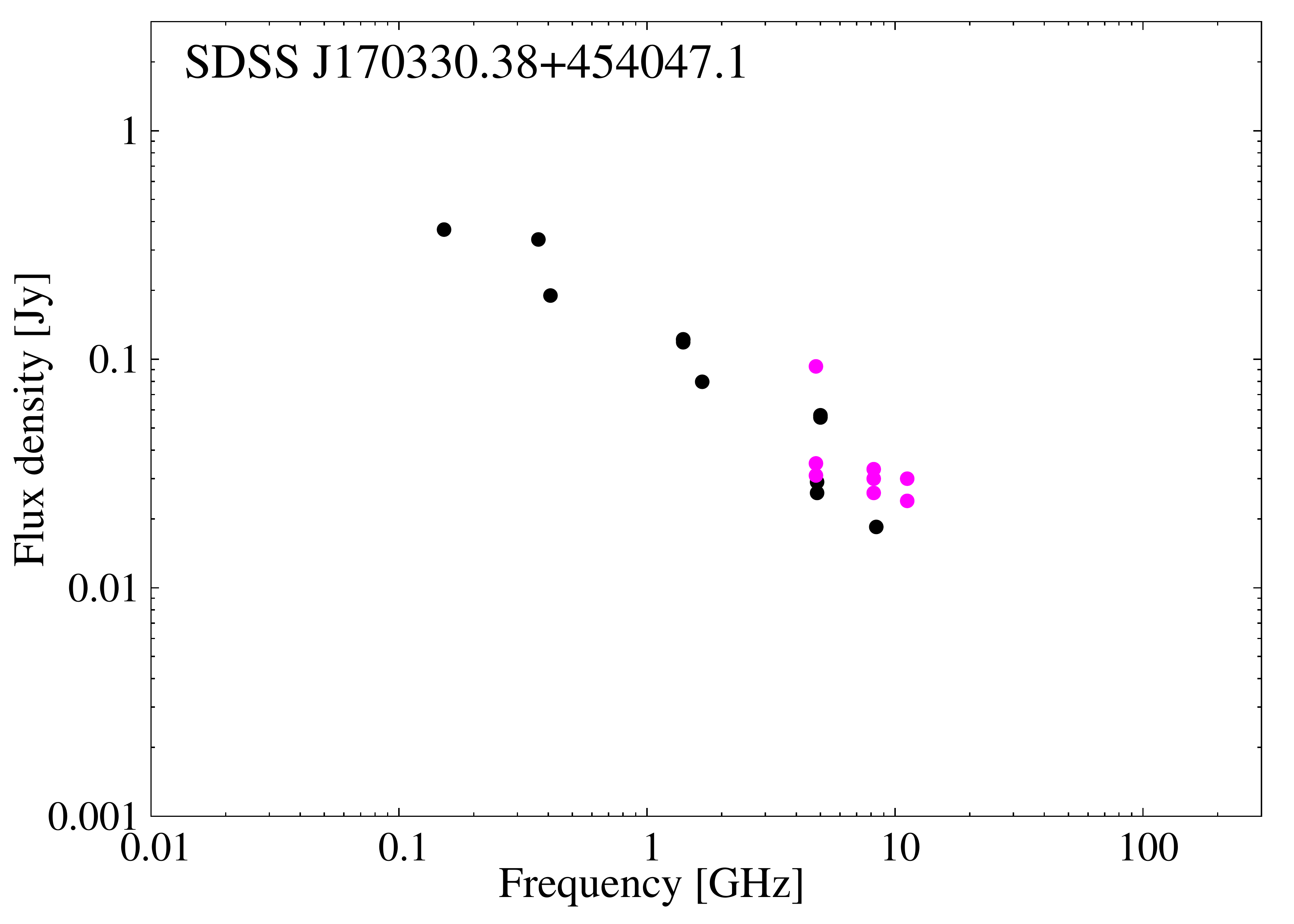}
\caption{Radio spectrum of J170330.38+454. Only detections. Colours as in Fig.~\ref{fig:spec1h0323}.} \label{fig:specj170330}
\end{minipage}
\begin{minipage}{0.47\textwidth}
\centering
\includegraphics[width=1\textwidth]{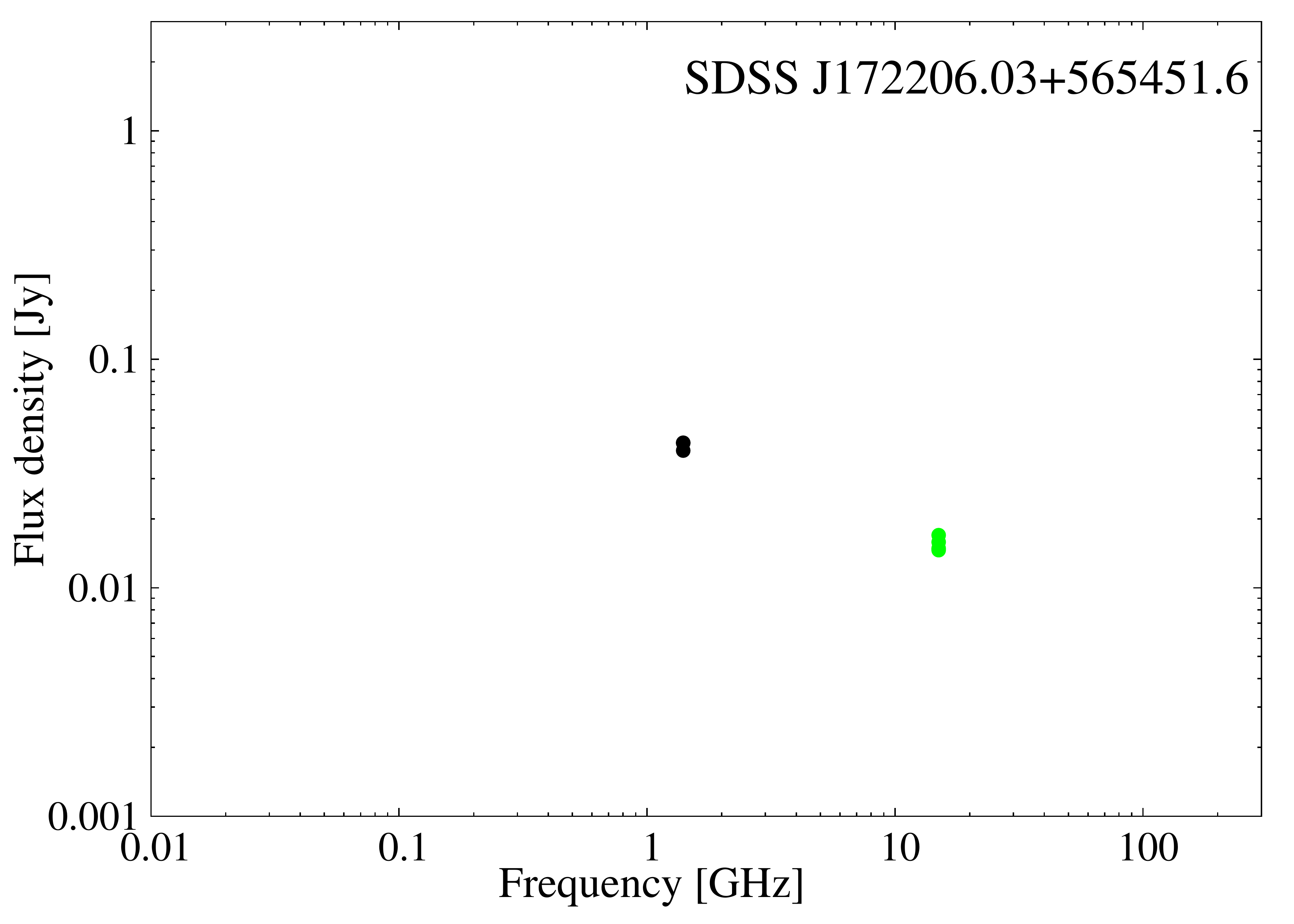}
\caption{Radio spectrum of J172206.03+565. Only detections. Colours as in Fig.~\ref{fig:spec1h0323}.} \label{fig:specj172206}
\end{minipage}\hfill
\end{figure*}

\begin{figure*}[ht!]
\begin{minipage}{0.47\textwidth}
\centering
\includegraphics[width=1\textwidth]{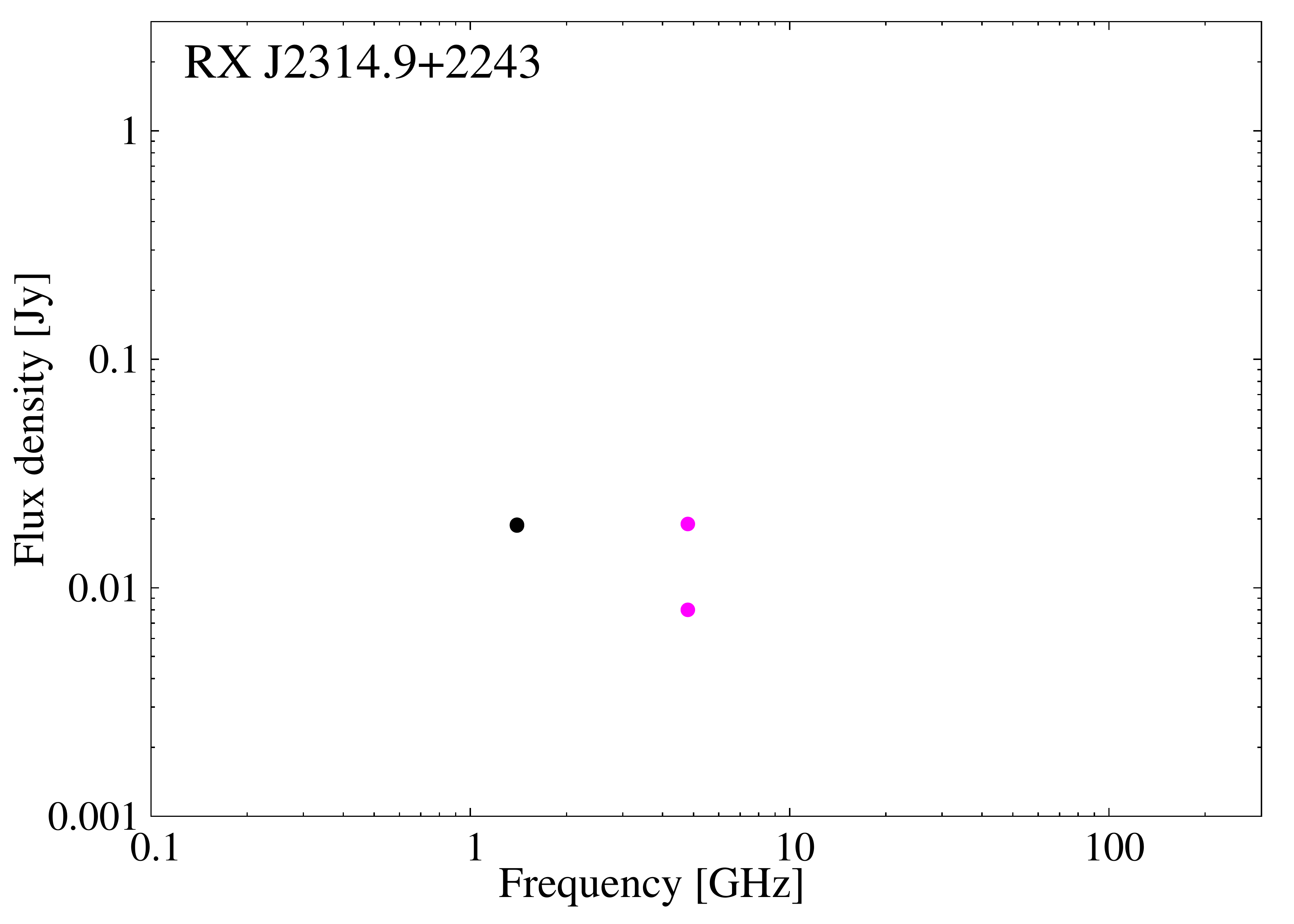}
\caption{Radio spectrum of RXJ2314.9+2243. Only detections. Colours as in Fig.~\ref{fig:spec1h0323}.} \label{fig:specrxj2314}
\end{minipage}\hfill
\end{figure*}

\end{appendix}

\end{document}